\documentclass[opre,sglanonrev]{informs4_hide}

\RequirePackage{tgtermes}
\RequirePackage{newtxtext}
\RequirePackage{newtxmath}
\RequirePackage{multirow,multicol}
\RequirePackage{xspace,dsfont}
\RequirePackage{endnotes}

\usepackage{etoc}

\DoubleSpacedXI 

\usepackage{algorithm}
\usepackage{algpseudocode}
\algrenewcommand\algorithmicrequire{\textbf{Input:}}
\algrenewcommand\algorithmicensure{\textbf{Output:}}
\usepackage{tikz}
\usepackage{xcolor}
\definecolor{dark1}{RGB}{157, 58, 103}
\definecolor{dark2}{RGB}{161, 67, 0}
\definecolor{dark3}{RGB}{115, 102, 0}
\definecolor{dark4}{RGB}{2, 120, 50}
\definecolor{dark5}{RGB}{0, 116, 122}
\definecolor{dark6}{RGB}{18, 100, 176}
\definecolor{dark7}{RGB}{116, 75, 163}
\definecolor{mid1}{RGB}{225, 119, 163}
\definecolor{mid2}{RGB}{228, 128, 77}
\definecolor{mid3}{RGB}{182, 158, 21}
\definecolor{mid4}{RGB}{87, 182, 109}
\definecolor{mid5}{RGB}{0, 181, 190}
\definecolor{mid6}{RGB}{88, 162, 242}
\definecolor{mid7}{RGB}{177, 135, 229}
\definecolor{light1}{RGB}{255, 187, 231}
\definecolor{light2}{RGB}{255, 198, 151}
\definecolor{light3}{RGB}{247, 223, 104}
\definecolor{light4}{RGB}{152, 248, 171}
\definecolor{light5}{RGB}{72, 249, 255}
\definecolor{light6}{RGB}{164, 219, 255}
\definecolor{light7}{RGB}{244, 192, 255}
\definecolor{lightyellow}{RGB}{255, 255, 204}

\newcount\Comments
\newcommand{\kibitz}[2]{\ifnum\Comments=1{\color{#1}{#2}}\fi}

\newcount\Drop  
\newcommand{\todrop}[2]{\ifnum\Drop=1{\color{#1}{#2}}\fi}
\newcommand{\drop}[1]{\todrop{mid5}{[Drop: #1]}}

\newif\ifhighlight 
\highlighttrue 

\newif\iftodo

\usepackage{color}              
\usepackage{color-edits}
\addauthor{Will}{blue}
\addauthor{Hma}{dark4}
\usepackage[colorlinks,
	citecolor = dark6,
	urlcolor = black,
	linkcolor = dark6,
    hypertexnames=false]{hyperref}
\usepackage[nameinlink]{cleveref}
\crefname{subsection}{subsection}{subsections}
\usepackage[short]{optidef}
\usepackage[font=scriptsize]{subcaption}
\newcommand{\figWidth}{0.48}

\newcommand\vtheta{{\BFtheta}}

\newcommand{\R}{\mathbb{R}}

\newcommand{\indicator}{\mathds{1}}

\newcommand{\norm}[2][]{\left\|#2\right\|_{#1}}

\newcommand{\ceil}[1]{\left\lceil #1 \right\rceil}
\newcommand{\floor}[1]{\left\lfloor #1 \right\rfloor}

\newcommand{\eps}{\varepsilon}

\newcommand{\newterm}[1]{\textit{#1}}

\newcommand{\Njob}{n}
\newcommand{\type}{\theta}
\newcommand{\typespace}{\Theta}
\newcommand{\instance}{\vtheta}
\newcommand{\sojourn}{d}
\newcommand{\potential}{p}
\newcommand{\ALG}{\mathsf{ALG}}
\newcommand{\OPT}{\mathsf{OPT}}

\newcommand{\regret}{\mathsf{Reg}}

\newcommand{\gre}{\mathsf{GRE}}
\newcommand{\PB}{\mathsf{PB}}
\newcommand{\batching}{\mathsf{BAT}}
\newcommand{\rbatching}{\mathsf{RBAT}}
\newcommand{\periodicrbatching}{\mathsf{PRBAT}}

\newcommand{\matchset}{\mathcal{M}}
\newcommand{\buffer}{A}
\newcommand{\indexf}{q}

\newcommand{\depth}{\ell}
\newcommand{\matchof}{m}

\newcommand{\reward}{r}

\newcommand{\xorigin}{O}
\newcommand{\xdestination}{D}

\newcommand{\marginalgain}{\mathsf{MG}}
\newcommand{\marginalloss}{\mathsf{ML}}

\newcommand{\dual}{\mathsf{HD}}

\newcommand{\averagedual}{\mathsf{AD}}

\usepackage{natbib}
 \bibpunct[, ]{(}{)}{,}{a}{}{,}%
 \def\bibfont{\small}%
\EquationsNumberedThrough    

\TheoremsNumberedThrough     
\ECRepeatTheorems  %

\MANUSCRIPTNO{MSOM-0001-2024.00}

\begin{document}



\RUNAUTHOR{Ma, Ma, and Romero}

\RUNTITLE{Dynamic Delivery Pooling}

\TITLE{Potential-Based Greedy Matching for Dynamic Delivery Pooling} 

\ARTICLEAUTHORS{%


\AUTHOR{Hongyao Ma}
\AFF{Graduate School of Business, Columbia University, New York, NY 10027, \EMAIL{hongyao.ma@columbia.edu}}
\AUTHOR{Will Ma}
\AFF{Graduate School of Business, Columbia University, New York, NY 10027, \EMAIL{wm2428@gsb.columbia.edu}}
\AUTHOR{Matias Romero}
\AFF{Graduate School of Business, Columbia University, New York, NY 10027, \EMAIL{mer2262@gsb.columbia.edu}}
} 

\ABSTRACT{%
We study the dynamic pooling of multiple orders into a single trip, a strategy widely adopted by online delivery platforms. When an order has to be dispatched, the platform must determine which (if any) of the available orders to pool it with, weighing the immediate efficiency gains against the uncertain, differential benefits of holding each order for future pooling opportunities. 
In this paper, we demonstrate the effectiveness of using the delivery distance as a proxy for opportunity cost via a \emph{potential-based greedy algorithm} ($\PB$). 
The algorithm is simple, pooling each departing job with the available job that maximizes the immediate savings in travel distance minus ``half its delivery distance'', which we call the \emph{potential} of the available job.
Theoretically, we show that $\PB$ achieves vanishing worst-case \emph{regret per job} as market density increases, whereas a naive greedy policy suffers constant regret.
We further show that the potential approximates the true opportunity cost of dispatching a job, in a stochastic setting with sufficient density.
Finally, we conduct extensive numerical experiments on both synthetic data and real-world data from the Meituan platform. Despite being forecast-agnostic, $\PB$ consistently outperforms greedy heuristics that rely on historical data.
Moreover, $\PB$ achieves performance comparable to computationally-intensive batching heuristics, which themselves also benefit from incorporating the potential to further improve their performance or drastically reduce computational costs.
}






\maketitle

\section{Introduction} \label{sec:intro}

On-demand delivery platforms have become an integral part of modern life, transforming how consumers search for, purchase, and receive goods from restaurants and retailers. Collectively, these platforms serve more than $1.5$ billion users worldwide, contributing to a global market valued at over \$250 billion \citep{statista2024}.
Growth has continued at a rapid pace---DoorDash in the United States reported a 19\% year-over-year increase in order volume for 2024~\citep{doordash25}, while the Chinese platform Meituan reached a peak of $150$ million daily orders in July 2025~\citep{scmp2025meituan}.
Managing this ever-expanding stream of orders poses significant operational challenges, particularly as consumers demand increasingly faster deliveries, and fierce competition compels platforms to continuously improve both operational efficiency and cost-effectiveness.

\begin{figure}[hpbt]
    \centering
    \includegraphics[width=0.9\linewidth]{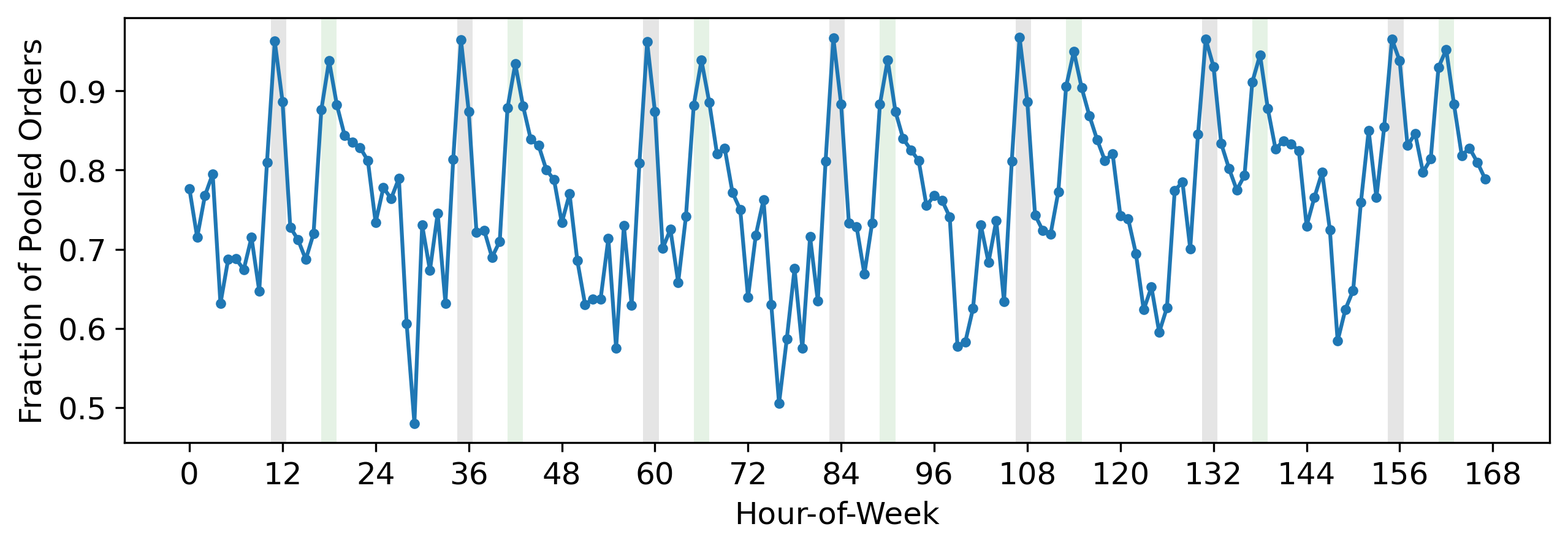}
    \caption{Fraction of pooled orders by hour-of-week in Meituan data. 
    The dataset includes $8$ days of order-level data from one city (see \Cref{sec:sim_meituan} for more details). The gray shade indicates peak lunch hours (10:30am-1:30pm), while the green shade indicates peak dinner hours (5pm-8pm). 
    }
    \label{fig:meituan_pooled_orders_per_hour}
\end{figure}

One strategy for improving efficiency and reducing labor costs is to \emph{pool} multiple orders from the same or nearby restaurants into a single trip.
This is referred to as ``stacked'', ``grouped'', or ``batched'' orders, and is advertised to drivers as opportunities for increasing earnings and efficiency~\citep{doordash2024batched}.
The strategy has been generally successful and very widely adopted~\citep{uberEatsMultiple}.
For example, data from Meituan, made public by the 2024 INFORMS TSL Data-Driven Research Challenge \citep{tsl_meituan_2024}, show that more than 85\% of orders are pooled, rising to 95\% during peak hours and consistently remaining above 50\% at all other times (see \Cref{fig:meituan_pooled_orders_per_hour}).

The delivery pooling approach introduces a highly complex decision-making problem, as orders arrive dynamically 
and need to be dispatched within a few minutes---orders in the Meituan data, for example, are offered to delivery drivers an average of around five minutes after being placed, often before meal preparation is complete (see \Cref{fig:meituan_order_to_first_dispatch_how}).
During this limited window of each order, the platform must determine which (if any) of the available orders to pool with, carefully weighing immediate efficiency gains against the uncertain differential benefits of holding each order for future pooling opportunities.
Since a single city may observe over 12 thousand orders per hour during peak times (see \Cref{fig:meituan_orders_per_hour}), a practically-viable algorithm also needs to be efficient and robust, as computational challenges are a primary bottleneck for Meituan \citep{tsl_meituan_2024}.

Similar problems have been studied in the context of various marketplaces such as ride-sharing platforms and kidney exchange programs, in a large body of work known as dynamic matching. However, we argue that the delivery pooling problem exhibits a fundamentally different \emph{reward structure}, which can be exploited to design simple yet effective algorithms.
First, many problems studied in the literature involve connecting two sides of a market, e.g. ridesharing platforms matching drivers to riders. 
In such bipartite settings, the presence of a large number of agents of identical or similar types (e.g., many riders requesting trips originating from the same area) typically leads to congestion and less efficient outcomes. 
%
The kidney exchange problem is not bipartite, but long queues of ``hard-to-match'' patient-donor pairs can still form, when many pairs share the same combination of incompatible tissue types and blood types.

In stark contrast, it is unlikely for long queues to build up for delivery \emph{pooling}, especially in dense markets, in that given enough orders, two of them will have closely situated origins and destinations, resulting in a good match. 
Illustrating further, it is actually desirable to have many customers request deliveries from similar origins to similar destinations, which we find to be the case in practice, with most requests originating from popular restaurants in busy city centers and ending in residential neighborhoods (see \Cref{appx:meituan_spatial_distribution}).
Pooling or ``stacking'' these close-by orders together (before assigning them to a courier) drastically reduces total travel distance, effectively increasing the platform's delivery capacity.

In this work, we focus on the pooling problem, and exploit the distinct \emph{reward topology} of delivery pooling where the best-case scenario for a job is to be pooled with a second job 
with identical origin and destination.
In this best case, the travel-distance-saved for each of the two jobs is half of the job's distance, which we refer to as the \emph{potential} of the job.
We demonstrate the effectiveness of using the potential as the opportunity cost of dispatching each job, via our \emph{potential-based greedy
algorithm} ($\PB$). 
When a job must be dispatched, the $\PB$ algorithm pools it with a currently available job that maximizes the adjusted pooling reward: the travel distance saved minus the potential the selected match.

On the theoretical front, we prove that $\PB$ significantly improves upon a naive greedy approach in terms of worst-case performance: as the density of the market increases, the \textit{regret per job} under $\PB$ vanishes whereas it remains constant under naive greedy. 
We further show that the potential approximates the marginal cost of dispatching each job in stochastic settings with sufficient density. 
We also conduct extensive numerical experiments on both synthetic data and real-world data from Meituan.
Despite 
being simple and forecast-agnostic, our notion of potential consistently yields better performance in comparison to dual-based marginal cost estimates derived from forecasts using historical data. 
$\PB$ dominates all greedy dispatch policies we consider, and achieves performance comparable to computationally expensive rolling batching policies.
Furthermore, we show that the potential can also be easily incorporated into the batching frameworks to either further improve their performance, or drastically reduce computation costs for equivalent outcomes.

\subsection{Model Description}

Different models of arrivals and departures have been proposed in the dynamic matching literature (see \Cref{sec:dynMatchModels}), and we believe our insight about potential is relevant across all of them.
However, in this paper we focus on a single model well-suited for the delivery pooling application.

To elaborate, we consider a dynamic non-bipartite matching problem, with a total of $\Njob$ jobs arriving sequentially to be matched.
Jobs are characterized by (potentially infinite) types $\theta\in\Theta$ representing features (e.g. locations of origin and destination) that determine the reward for the platform.
Given the motivating application to pooled deliveries, we always model jobs' types as belonging to some metric space.
Following \citet{ashlagi2019edge}, we assume that an unmatched job must be dispatched after $\sojourn$ new arrivals, which we interpret as a known \textit{internal} deadline imposed by the platform to incentivize timely service.
The platform is allowed to make a last-moment matching decision before the job leaves, termed as the job becoming \newterm{critical}.
At this point, the platform must decide either to match the critical job to another available one, collecting reward $r(\theta,\theta')$ where $\theta,\theta'$ are the types of the matched jobs and $r$ is a known reward function, or to dispatch the critical job on its own for zero reward.
The objective is to maximize the total reward collected from matching the $\Njob$ jobs.
We believe this to be an appropriate model for delivery platforms (cf. \Cref{sec:dynMatchModels}) because deadlines are known upon arrival and sudden departures (order cancellations) are rare.

Dynamic matching models can also be studied under different forms of information about the future arrivals.
Some papers \citep[e.g.][]{kerimov2024dynamic,aouad2020dynamic,eom2023batching,wei2023constant} develop sophisticated algorithms to leverage stochastic information, which is often necessary to derive theoretical guarantees under general matching rewards.
In contrast, our heuristic does not require any knowledge of future arrivals, and our theoretical results hold in an ``adversarial'' setting where no stochastic assumptions are made.
In our experiments, we consider arrivals generated both from stochastic distributions and real-world data, and find that our heuristic can outperform even algorithms that are given the correct stochastic distributions, under our specific reward topology.

\subsection{Main Contributions}

\paragraph{Notion of potential.}
As mentioned above, we design a simple greedy-like algorithm that is based purely on topology and reward structure, which we term potential-based greedy ($\PB$). More precisely, $\PB$ defines the \newterm{potential} of a job type $\theta$ to be $p(\theta)=\sup_{\theta'\in\Theta} r(\theta,\theta')/2$, measuring the highest-possible reward obtainable from matching type $\theta$.  The potential acts as an opportunity cost, and the relevance of this notion arises in settings where the potential is heterogeneous across jobs. To illustrate, let $\Theta=[0,1]$ and $r(\theta,\theta')=\min\{\theta,\theta'\}$, a reward function used for delivery pooling as we will justify in \Cref{sec:model}.  Under this reward function, a job type (which is a real number) can never be matched for reward greater than its real value, which means that jobs with higher real value have greater potential.  Our $\PB$ algorithm matches a critical job (with type $\theta$) to the available job (with type $\theta'$) that maximizes
\begin{align} \label{eqn:potentialIntro}
    r(\theta,\theta')-p(\theta')=\min\{\theta,\theta'\}-\frac12 \theta',
\end{align}
being dissuaded to use up job types $\theta'$ with high real values. Notably, this 
does not require any forecast or partial information of future arrivals, but 
assumes full knowledge of the universe of possible job types.

\paragraph{Theoretical results.}
Our theoretical results assume $\Theta=[0,1]$.  We compare $\PB$ to the naive greedy algorithm $\gre$, which selects $\theta'$ to maximize $r(\theta,\theta')$, instead of $r(\theta,\theta')-p(\theta')$ as in~\eqref{eqn:potentialIntro}.  We first consider an offline setting ($d=\infty$), showing that under reward function $r(\theta,\theta')=\min\{\theta,\theta'\}$, our algorithm $\PB$ achieves regret $O(\log n)$, whereas $\gre$ suffers regret $\Omega(n)$ compared to the optimal matching.
Our analysis of $\PB$ is tight, i.e.\ it has $\Theta(\log n)$ regret.
Building upon the offline analysis, we next consider the online setting ($d< n$), showing that $\PB$ has regret $\Theta(\frac nd\log d)$, which improves as $d$ increases. This can be interpreted as there being $\frac nd$ ``batches'' in the online setting, and our algorithm achieving a regret of $O(\log d)$ per batch.  Alternatively, it can be interpreted as the \newterm{regret per job} of our algorithm being $O(\frac {\log d}{d})$. By contrast, we show that the naive greedy algorithm $\gre$ suffers a total regret of $\Omega(n)$, regardless of $d$.

For comparison, we also analyze two other reward topologies, still assuming $\Theta=[0,1]$.
\begin{enumerate}
\item We consider the classical min-cost matching setting \citep{reingold1981greedy} where the goal is to minimize total match distance between points on a line, represented in our model by the reward function $r(\theta,\theta')=1-|\theta-\theta'|$.
For this reward function, both $\PB$ and $\gre$ have regret $\Theta(\log n)$; in fact, they are the same algorithm because all job types have the same potential. Like before, this translates into both algorithms having regret $\Theta(\frac nd \log d)$ in the online setting.
\item We consider reward function $r(\theta,\theta')=|\theta-\theta'|$, representing an opposite setting in which it is worst to match two jobs of the same type.  For this reward function, we show that any index-based matching policy must suffer regret $\Omega(n)$ in the offline setting, and that both $\PB$ and $\gre$ suffer regret $\Omega(n)$ (irrespective of $d$) in the online setting.
\end{enumerate}

Our theoretical results are summarized in \Cref{table:results}.
As our model is a special case of \citet{ashlagi2019edge}, their 1/4-competitive randomized online edge-weighted matching algorithm can be applied, which translates in our setting to an $O(n)$ upper bound for regret (under any definition of reward).
However, their regret does not improve with $d$, while we show that for specific reward topologies, the regret is $O(n\frac{\log d}d)$ and hence does improve with $d$, using a completely different algorithm and analysis.
We now outline how to prove our two main technical results, \Cref{thm:potential_log_upper_bound,thm: dynamic}.
\begin{table}[!t]
\centering
\begin{tabular}{|c|c|c|c|}
\hline
\updown $\theta,\theta'\in[0,1]$ & $r(\theta,\theta')=\min\{\theta,\theta'\}$ & $r(\theta,\theta')=1-|\theta-\theta'|$ & $r(\theta,\theta')=|\theta-\theta'|$ \\
\hline
\up\multirow{4}{*}{Regret of $\PB$} & Offline: $\Theta(\log n)$ & & \\
\down & (\Cref{thm:potential_log_upper_bound}, \Cref{prop:loglowerboundOffline}) & & \\
\up & Online: $\Theta(\frac nd \log d)$ & Offline: $\Theta(\log n)$ & Offline: $\Omega(n)$ \\
\down & (\Cref{thm: dynamic}, \Cref{prop:loglowerboundOnline}) & Online: $\Theta(\frac nd \log d)$ & Online: $\Omega(n)$ \\
\cline{1-2}
\up\multirow{4}{*}{Regret of $\gre$} & Offline: $\Omega(n)$ & (\Cref{sec:reward2}) & (\Cref{sec: reward 3}) \\
\down & (\Cref{prop:greedy_linear_lower_bound}) & & \\
\up & Online: $\Omega(n)$ & & \\
\down & (\Cref{prop:greedy_linear_lower_bound_dynamic}) & & \\
\hline
\end{tabular}
\caption{
Summary of theoretical results under different reward functions $r:[0,1]^2\to \R$.  The total number of jobs is denoted by $n$, and the batch size in the online setting is approximately $d$.
}
\label{table:results}
\end{table}
\paragraph{Proof techniques.}
We establish an upper bound on the regret of $\PB$ by comparing its performance to the sum of the potential of all jobs.
Under reward function $\reward(\type,\type')=\min\{\type,\type'\}$, the key driver of regret is the sum of the distances between jobs matched by $\PB$.
In \Cref{thm:potential_log_upper_bound}, we study this quantity by analyzing the intervals induced by matched jobs in the offline setting ($\sojourn=\infty$).
We show that the intervals formed by the matching output of $\PB$ constitute a \emph{laminar set family}; that is, every two intervals are either disjoint or one fully contains the other.
We then prove that more deeply nested intervals must be exponentially smaller in size. Finally, we use an LP to show that the sum of interval lengths remains bounded by a logarithmic function of the number of intervals.

In \Cref{thm: dynamic}, we partition the set of jobs in the online setting into roughly $\frac{\Njob}{d}$ ``batches'' and analyze the sum of the distances between matched jobs within a single batch.
Our main result is to show that we can use \Cref{thm:potential_log_upper_bound} to derive an upper bound for an arbitrary batch.
However, a direct application of the theorem on the offline instance defined by the batch would not yield a valid upper bound, because the resulting matching of $\PB$ in the offline setting could be inconsistent with its online matching decisions.
To overcome this challenge, we carefully construct a modified offline instance that allows for the online and offline decisions to be coupled, while ensuring that the total matching distance did not go down.  This allows us to upper-bound the regret per batch by $O(\log d)$, for a total regret of $O(\frac nd \log d)$.

\paragraph{Simulations on synthetic data.} We test $\PB$ on random instances in the setting of our theoretical results, except that job destinations are drawn uniformly at random from $[0,1]$ (instead of adversarial). 
We benchmark its performance against $\gre$, as well as more sophisticated algorithms, including (i) forecast-aware heuristics that use historical data to estimate the opportunity costs of different job types and (ii) batching-based heuristics that compute optimal outcomes based on the set of currently available jobs. 
We show that $\PB$ consistently outperforms \emph{all} benchmarks starting from relatively low market densities, achieving over 95\% of the hindsight optimal solution that has full knowledge of arrivals.
This result is robust to different distributions of job types, and also two-dimensional locations with a common origin.

\paragraph{Simulations on real data.}

Finally, we test the practical applicability of $\PB$ via extensive numerical experiments using order-level data from the Meituan platform, made available from the 2024 INFORMS TSL Data-Driven Research Challenge \citep{tsl_meituan_2024}.
The key information we extract from this dataset is the exact timestamps of each delivery order, and the geographic coordinates of pick-up and drop-off locations.
These aspects differ from our theoretical setting in that (i) requests may not be available to be pooled for a fixed number of new arrivals before being dispatched, and (ii) locations are two-dimensional with delivery orders having heterogeneous origins.
To address (i), we assume that the platform sets a fixed time window after which an order becomes critical, corresponding to each job being able to wait for at most a fixed sojourn time before being dispatched.
For (ii), we extend our definition of reward (that captures the travel distance saved) to two-dimensional heterogeneous origins.
Under this new reward definition, the main insight from our theoretical model remains true: longer deliveries have higher potential reward from being pooled with other jobs in the future, and thus $\PB$ aims to keep them in the system.

In contrast to our simulations with synthetic data, the real-life delivery locations may now be correlated and exhibit time-varying effects.
Regardless, we find that $\PB$ improves significantly upon the naive greedy and forecast-aware greedy policies, achieving close to 90\% of the maximum possible travel distance saved from pooling. 
Although batching-based policies can achieve even better performance (over 90\%), this comes at a computational cost two orders of magnitudes larger than that for $\PB$.
Moreover, we show that the potential can also be incorporated as the opportunity cost into the batching frameworks  (which already implicitly capture near-term opportunity costs by optimizing the current state) to either further improve their performance, or drastically reduce computation costs for equivalent outcomes.

\paragraph{Explanation for the effectiveness of the potential as the opportunity cost.}
The robust performance of $\PB$, and the effectiveness of incorporating the \emph{potential} as the opportunity cost across diverse settings, warrants explanation. 
In \Cref{sec: interpretation}, we provide intuition for this phenomenon within a stylized framework. 
Theoretically, we prove that as market size $n$ increases, the ``correct'' shadow prices (i.e. the marginal value and marginal cost of each job) converge to our notion of potential at a rate of $O\left(1/n\right)$. 
Through simulations, we also illustrate that the shadow prices exhibit strong concentration around the potential, even at moderate densities. 

\subsection{Further Related Work}

\subsubsection{Dynamic matching.} \label{sec:dynMatchModels}

Dynamic matching problems have received growing attention from different communities in economics, computer science, and operations research. We discuss different ways of modeling the trade-off between matching now vs.\ waiting for better matches, depending on the application that motivates the study.

One possible model \citep{kerimov2024dynamic,kerimov2025optimality,wei2023constant} 
considers a setting with jobs that arrive stochastically in discrete time and remain in the market indefinitely, but use a notion of \newterm{all-time regret} that evaluates a matching policy, at every time period, against the best possible decisions until that moment, to disincentivize algorithms from trivially delaying until the end to make all matches.

A second possible model, motivated by the risk of cancellation in ride-sharing platforms, is to consider sudden departures modeled by jobs having heterogeneous \textit{sojourn times} representing the maximum time that they stay in the market \citep{aouad2020dynamic}. These sojourn times are unknown to the platform, and delaying too long risks many jobs being lost without a chance of being matched.
\citet{aouad2020dynamic} formulates an MDP with jobs that arrive stochastically in continuous time, and leave the system after an exponentially distributed sojourn time. They propose a policy that achieves a multiplicative factor of the hindsight optimal solution. 
%

%
A third possible model \citep{huang2018match} also considers jobs that can leave the system at any period, but allows the platform to make a last-moment matching decision right before a job leaves, termed as the job becoming \newterm{critical}.  In this model, one can without loss assume that all matching decisions are made at times that jobs become critical.

Finally, the model we study also makes all decisions at times that jobs become critical, with the difference being that the sojourn times are known upon the arrival of a job, as studied in \citet{ashlagi2019edge, eom2023batching}. Under these assumptions, \citet{eom2023batching} assume stationary stochastic arrivals, while \citet{ashlagi2019edge} allow for arbitrary arrivals.
Our research focuses on deterministic greedy-like algorithms that can achieve good performance as the market thickness increases.
Moreover, our work differs from these papers in the description of the matching value. While they assume arbitrary matching rewards, we consider specific reward functions known to the platform in advance, and use this information to derive a simple greedy-like algorithm with good performance.
This aligns with a stream of literature on online matching, that captures more specific features into the model to get stronger guarantees \citep[see][]{chen2023feature,balkanski2023power}.

We should note that our paper also relates to a recent stream of work studying the effects of batching and delayed decisions in online matching \citep[e.g.][]{feng2024batching}, as well as works studying the relationship between market thickness and quality of online matches \citep[e.g.][]{ashlagi2021kidney}.

\subsubsection{Delivery operations.}
On-demand delivery routing has been widely studied using approximate dynamic programming techniques \citep[e.g.][]{reyes2018meal,ulmer2021restaurant} to tackle complex mathematical programs, often lacking theoretical performance guarantees and offering limited interpretability.
Closer to our research goal, \citet{chen2024courier} analyze the optimal dispatching policy on a stylized queueing model representing a disk service area centered at one restaurant, while
\citet{cachon2023fast} study the interplay between the number of couriers and platform efficiency, assuming a one-dimensional geography with one single origin, which is also the primary model in our theoretical results.


\section{Preliminaries}
\label{sec:model}
In this section, we introduce a dynamic non-bipartite matching model for the delivery pooling problem. A total of $\Njob$ jobs arrive to the platform sequentially.
Each job $j \in [\Njob] = \{1, \ldots, \Njob\}$, indexed in the order of arrival, has a \emph{type} $\type_j \in \typespace$. 
We adopt the criticality assumption from \citet{ashlagi2019edge}, in the sense that each job remains available to be matched for $\sojourn\ge 1$ new arrivals, after which it becomes \textit{critical}. The parameter $\sojourn$ can also be interpreted as the market density.
When a job becomes critical, the platform decides whether to (irrevocably) match it with another available job, in which case they are dispatched together (i.e. \emph{pooled})
and the platform collects a \emph{reward} given by a known function $\reward:\typespace^2\to\R$. If a critical job is not pooled with another job, it has to be dispatched by itself for zero reward.

\subsection{Reward Topology}
Our modeling approach directly imposes structure on the type space, as well as the reward function that captures the benefits from pooling delivery orders together. Similar to previous numerical work on pooled trips in ride-sharing platforms \citep{eom2023batching, aouad2020dynamic}, we assume that the platform's goal is to reduce the total distance that needs to be traveled to complete all deliveries.
The reward of pooling two orders together is therefore the travel distance saved when they are delivered by the same driver in comparison to delivered separately. 
Formally, we consider a \newterm{linear city model} where job types $\type \in \typespace = [0,1]$ represent destinations of the delivery orders, and assume that all orders need to be served from the origin 0 (similar to that analyzed in \citet{cachon2023fast}).
If a job of type $\type$ is dispatched by itself, the total travel distance from the origin 0 is exactly $\type$. If it is matched with another job of type $\type'$, they are pooled together on a single trip to the farthest destination $\max\{\type,\type'\}$. Thus, the distance saved by pooling is 
\begin{equation}
  \reward(\type,\type') = \type + \type' - \max\{\type,\type'\} = \min\{\type,\type'\}.
  \label{eq:defn_reward_A} 
\end{equation}

Although our main theoretical results leverage the structure of this reward topology, our proposed algorithm can be applied to other reward topologies. 
In particular, we derive theoretical results for two other reward structures for $\typespace = [0,1]$. First, we consider 
\begin{equation}
    \reward(\type,\type') = 1 - |\type - \type'|  \label{eq:defn_reward_B}
\end{equation}
to capture the commonly-studied spatial matching setting~\citep[e.g.][]{kanoria2021dynamic,balkanski2023power}, where the reward is larger if the distance $|\theta-\theta'|$ is smaller. 
We also consider 
\begin{equation}
    \reward(\type,\type') = |\type - \type'| \label{eq:defn_reward_C}
\end{equation}
with the goal of matching types that are far away from each other, contrasting the other two reward functions.
In addition, we perform numerical experiments for delivery pooling in two-dimensional (2D) space, where the reward function corresponds to the travel distance saved in the 2D setting.

\subsection{Benchmark Algorithms}

Given market density $\sojourn$ and reward function $\reward$, if the platform had full information of the sequence of arrivals $\instance \in \typespace^\Njob$, the hindsight optimal $\OPT(\instance,\sojourn)$ can be computed by the integer program (IP) defined in \eqref{eq: OPT}. We denote the hindsight optimal matching solution as $\matchset_{\OPT} = \{(j,k):x_{j,k}^{\ast}=1\}$, where $x^\ast$ is an optimal solution of \eqref{eq: OPT}.
\begin{maxi}
    {x}{ \sum_{j,k : j\neq k, |j-k|\le \sojourn} x_{jk} \reward(\type_j,\type_k)}
    {\label{eq: OPT}}{\OPT(\instance,\sojourn) =}
    \addConstraint{ \sum_{k:j\neq k} x_{jk}}{\le 1,}{j\in [\Njob]}
    \addConstraint{ x_{jk}}{\in\{0,1\},}{j,k\in [\Njob], j\neq k.}
\end{maxi}

An online matching algorithm operates over an instance $\instance\in\typespace^\Njob$ sequentially: when job $j\in [\Njob]$ becomes critical, the types of future arrivals $k > j + d$ are unknown, and any matching decision has to be made based on the currently available information.
Given an algorithm $\ALG$, we denote by $\ALG(\instance,\sojourn)$ the total reward collected by an algorithm on such instance.
We analyze the performance of algorithms via \emph{regret}, as follows:
\[
    \regret_\ALG(\instance,\sojourn) = \OPT(\instance,\sojourn)-\ALG(\instance,\sojourn).
\]

We study a class of online matching algorithms that we call \newterm{index-based greedy matching algorithms}.
Each index-based greedy matching algorithm is specified by an \emph{index function} $\indexf:\typespace^2\to\R$, and only makes matching decisions when some job becomes critical (in our model, it is without loss of optimality to wait to match).
A general pseudocode is provided in \Cref{alg:dynamic}.
When a job $j \in [\Njob]$ becomes critical, the algorithm observes the set of available jobs in the system $A(j) \subseteq [\Njob]$ that have arrived but are not yet matched (this is the set $\buffer\setminus\{j\}$ in \Cref{alg:dynamic}), and chooses a match $\matchof(j) \in A(j)$ that maximizes the index function (even if negative), collecting a reward $\reward(\type_j,\type_{\matchof(j)})$. If there are multiple jobs that achieve the maximum, the algorithm breaks the tie arbitrarily.
The algorithm makes a set of matches $\matchset = \{ (j,m(j)) : j \in C \}$, where $C$ denotes the set of jobs that are matched when they become critical, and its total reward is $\ALG(\instance,\sojourn) = \sum_{j\in C} \reward(\type_j,\type_{\matchof(j)})$.
Note that $\matchof(j)$ is undefined if $j$ is either unmatched, or was not critical at the time it was matched.

\begin{algorithm}
\caption{Index-based Greedy Matching Algorithm}\label{alg:dynamic}
\begin{algorithmic}[1]
\Require Instance $\instance$, density $\sojourn$, index function $\indexf$, reward function $\reward$
\Ensure $C$, $m:C\to[n]$
\State Initialize set of available jobs $\buffer=\emptyset$, and jobs that are critical when matched $C=\emptyset$
\For{job $t=1,\ldots,\Njob+\sojourn+1$}
    \If{$t\le\Njob$}
        \State $\buffer=\buffer\cup \{t\}$ \Comment{job $t$ arrives}
    \EndIf         
    \If {$t-d\in \buffer$} 
        \State $j=t-d$ \Comment{job $j=t-d$ becomes critical}
        \If{$\buffer \setminus\{j\} \neq\emptyset$}
            \State Choose $m(j) \in \argmax_{k \in \buffer \setminus\{j\} } \indexf(\type_{j}, \type_{k})$ \Comment{Ties are broken arbitrarily}
            \State $C\gets C\cup\{j\}$
            \State $\buffer \gets \buffer\setminus \{j,\matchof(j)\} $ \Comment{jobs $j,\matchof(j)$ are dispatched together}
        \Else
            \State $\buffer \gets \buffer\setminus \{j\} $ \Comment{job $j$ is dispatched by itself}
        \EndIf
    \EndIf
\EndFor
\State\Return $C, \{\matchof(j)\}_{j\in C}$
\end{algorithmic}
\end{algorithm}

As an example, the \newterm{naive greedy} algorithm ($\gre$) uses index function $\indexf_{\gre}(\type,\type') = \reward(\type,\type')$. When 
a job becomes critical, the algorithm matches it with an available job to maximize the (instant) reward.
To illustrate the behavior of this algorithm under \drop{reward function}%
$\reward(\type,\type') = \min\{\type,\type'\}$, we first make the following observation, that the algorithm always chooses to match each critical job with a higher type job when possible. 
\begin{remark}\label{deliverygreedy}
    When a job $j \in [\Njob]$ becomes critical, if the set $A_+(j) = \{k\in A(j):\type_k \ge \type_j\}$ is nonempty, then under the naive greedy algorithm $\gre$, we have $\matchof(j)\in A_+(j)$ and $\reward(\type_j,\type_{\matchof(j)}) = \type_j $. 
\end{remark}

\section{Potential-Based Greedy Algorithm}
\label{sec:PB} 

We introduce in this section the potential-based greedy algorithm and prove that it substantially outperforms the naive greedy approach in terms of worst case regret.

The \newterm{potential-based greedy} algorithm ($\PB$) is an index-based greedy matching algorithm (as defined in \Cref{alg:dynamic}) whose index function is specified as
\begin{equation}
    \indexf_{\PB}(\type,\type') = \reward(\type, \type') - \potential(\type'),
\end{equation}
where $\potential(\type)$ is the \newterm{potential} of a job of type $\type$, formally defined as follows
\begin{equation}
    \potential(\type)=\frac{1}{2}\sup_{\type'\in\typespace} \reward(\type,\type'). \label{eq:defn_potential}
\end{equation} 

This new notion of potential can be interpreted as an optimistic measure of the marginal value of holding on to a job that could be later matched with a job that maximizes instant reward.
Intuitively, this ``ideal'' matching outcome appears as the optimal matching solution when the density of the market is arbitrarily large. We provide a formal statement of this intuition in \Cref{sec: interpretation}.
Note that this definition of potential is quite simple in the sense that it relies on only in the knowledge of the reward topology: reward function and type space.
In particular, for our topology of interest, $\reward(\type,\type')=\min\{\type,\type'\}$ on $\typespace=[0,1]$, the ideal scenario is achieved when two identical jobs are matched and the potential $\potential(\type)=\type/2$ is simply proportional to the length of a solo trip, capturing that longer deliveries have higher potential reward from being pooled with other jobs in the future.

We begin by giving a straightforward interpretation for this algorithm under the linear city model and our reward of interest, as we did for the naive greedy algorithm $\gre$ in \Cref{deliverygreedy}.

\begin{remark}\label{deliverypotential}
    Under the 1-dimensional type space $\Theta = [0,1]$ and the reward function $\reward(\type,\type')=\min\{\type,\type'\}$, 
    the $\PB$ algorithm always matches each critical job to an available job that's the closest in space, since 
    \begin{align*}
        \matchof(j) 
        \in \argmax_{k\in A(j) } \indexf_\PB(\type_j,\type_k) 
        = \argmin_{k\in A(j) } |\type_j-\type_k|. 
    \end{align*}    %
    This follows from the fact that $ |\type_j-\type_k| = \type_j + \type_k - 2\min\{\type_j, \type_k\} = \type_j - 2\indexf_\PB(\type_j,\type_k) $.
\end{remark}

In the rest of this section, we first assume $d = \infty$ and study the \textit{offline} performance of various index-based greedy matching algorithms under the reward function $\reward(\type,\type') = \min\{\type,\type'\}$.
The performance of greedy procedures for offline matching is a fundamental question in its own right, but this analysis will also later help us analyze the algorithms' \emph{online} performances when $d<\infty$.
%

Results for the two alternative reward structures defined in \eqref{eq:defn_reward_B} and \eqref{eq:defn_reward_C} are reported in \Cref{appx:alternative_reward_topologies}.

\subsection{Offline Performance under Reward Function $\reward(\type,\type') = \min\{\type,\type'\}$}\label{sec: static}

Suppose $\sojourn=\infty$, meaning that all jobs are available to be matched when the first job becomes critical. We study the \emph{offline} performance of both the naive greedy algorithm $\gre$ and potential-based greedy algorithm. We show that the regret of $\gre$ grows linearly with the number of jobs, while the regret of $\PB$ is logarithmic.

\begin{proposition}[proof in \Cref{pf:greedy_linear_lower_bound}]\label{prop:greedy_linear_lower_bound}
Under reward function $\reward(\type,\type') = \min\{\type,\type'\}$, when the number of jobs $\Njob$ is divisible by 4, there exists an instance $\instance \in [0,1]^\Njob$ for which $\regret_\gre(\instance,\infty) \ge n/4$.
\end{proposition}

In particular, the \newterm{regret per job} of $\gre$, i.e.\ dividing the regret by $\Njob$, is constant in $\Njob$.
In contrast, we prove in the following theorem that $\PB$ performs substantially better. In fact, its regret per job gets better for larger market sizes $\Njob$.

\begin{theorem}
\label{thm:potential_log_upper_bound}
    Under reward function $\reward(\type,\type') = \min\{\type,\type'\}$, we have $\regret_\PB(\instance,\infty) \le 1 + \log_2(\Njob/2+1)/2$, for any $\Njob$, and any instance $\instance \in [0,1]^\Njob$.
\end{theorem}
\proof{Proof.}
Let $(C,\matchof(\cdot))$ be the output of $\PB$ on $\instance$ (generated as in \Cref{alg:dynamic}).
In particular, 
$\PB(\instance,\infty) = \sum_{j \in C} \reward(\type,\type')$.
On the other hand, since $\reward(\type,\type') \le \potential(\type) + \potential(\type')$ for all $\type,\type'\in[0,1]$ and $\potential(\type)=\type/2$, we have 
\begin{align*}
    \OPT(\instance,\infty) 
    = \sum_{(j,k) \in \matchset_\OPT} \reward(\type_j,\type_k)  
    \le \sum_{j=1}^\Njob \potential(\type_j) 
    \le
    \sup_{\type\in \Theta}\potential(\type) + \sum_{j\in C} \potential(\type_j) + \potential(\type_{\matchof(j)}) 
    =
    \frac12 + \sum_{j\in C} \frac{\type_j}{2} + \frac{\type_{\matchof(j)}}{2}
    ,
\end{align*}
where the last inequality comes from the fact that $\PB$, and in fact any index-based matching algorithm, leaves at most one job unmatched (when $\Njob$ is odd).
Hence,
\begin{align}
&\regret_\PB(\instance,\infty)
=\OPT(\instance,\infty)-\PB(\instance,\infty)\nonumber \\
&\qquad \le \frac12 + \sum_{j \in C} \frac{\type_j}{2} + \frac{\type_{\matchof(j)}}{2} - \min\{\type_j,\type_{m(j)}\} \nonumber\\
&\qquad = \frac12 + \sum_{j \in C} \frac{|\type_j - \type_{\matchof(j)}|}{2}\label{eq: offline_regret_bound}. 
\end{align}
%
To bound the right-hand side, we first prove the following. For each $j\in C$, consider the interval $I_j=( \min\{\type_j,\type_{\matchof(j)}\},\max\{\type_j,\type_{\matchof(j)}\})\subseteq[0,1]$ and $\depth(j)=|\{k : I_j \subseteq I_k\}|$. We argue that
\begin{equation}\label{eq: distancebound}
    |\type_j - \type_{\matchof(j)}| \le 2^{1-\depth(j)}.
\end{equation}

First, note that $\{I_j\}_{j\in C}$ is a \newterm{laminar set family}: for every $j,k\in C$, the intersection of $I_j$ and $I_k$ is either empty, or equals $I_j$, or equals $I_k$. Indeed, without loss of generality, assume $j<k$. Then, when job $j$ becomes critical we have $\matchof(j),k,\matchof(k) \in A(j) $. Therefore, from \Cref{deliverypotential}, $\matchof(j)$ is the closest to $j$ and thus $k,\matchof(k)\notin I_j$, i.e.\ either $I_j \cap I_k = \emptyset$, or $I_j \cap I_k = I_j$.
Moreover, if $I_j\subsetneq I_k$ (i.e. $I_j\subsetneq I_k$ and $I_j\neq I_k$), then necessarily $j < k$, since otherwise $k$ becomes critical first and having $j,\matchof(j) \in A(k)$ closer in space, $\PB$ would not have chosen $\matchof(k)$.
Now, we can prove \eqref{eq: distancebound} by induction. If $\depth(j)=1=|\{j\}|$, clearly $|\type_j-\type_{\matchof(j)}|\le 1$. Suppose that the statement is true for $\depth(j)=\depth \ge 1$. If $\depth(j)=\depth+1$ then there exists $k$ such that $I_j\subsetneq I_k$ and $\depth(k)=\depth$. By the previous property, $j<k$ and since when job $j$ becomes critical we had $k,\matchof(k)\in A(j)$, it must be the case that
\begin{align*}
    |\type_j-\type_{\matchof(j)}| 
    &<\max\{ \min\{\type_j,\type_{m(j)}\} - \min\{\type_k,\type_{m(k)}\},\max\{\type_k,\type_{m(k)}\}-\max\{\type_j,\type_{m(j)}\}\}\\
    &\le \min\{\type_j,\type_{m(j)}\} - \min\{\type_k,\type_{m(k)}\} + \max\{\type_k,\type_{m(k)}\} - \max\{\type_j,\type_{m(j)}\} \\
    &= |\type_k-\type_{\matchof(k)}| - |\type_j-\type_{\matchof(j)}|.
\end{align*}
Thus, $|\type_j-\type_{\matchof(j)}|\le|\type_k-\type_{\matchof(k)}|/2\le 2^{1-(\depth+1)}$,
completing the induction.

Having established \eqref{eq: distancebound}, we use it to derive an upper bound for $\sum_{j\in C}|\type_j-\type_{\matchof(j)}|$. Note that since $\depth(j)\in \{1,\ldots,\floor{\Njob/2}\}$, we have
\begin{align} \label{eq: distanceboundbylayer}
\sum_{j \in C} |\type_j - \type_{\matchof(j)}| = \sum_{\depth = 1}^{\floor{\Njob/2}}\sum_{j\in C:\depth(j) = \depth} |\type_j - \type_{\matchof(j)}|.
\end{align}
For every $\depth$, we have $\sum_{j\in C:\depth(j) = \depth} |\type_j - \type_{\matchof(j)}|\le \min\{1,2^{1-\depth}|\{j:\depth(j)=\depth\}|\}$ by the laminar property and \eqref{eq: distancebound}.
Let $z_\depth$ denote $2^{1-\depth}|\{j:\depth(j)=\depth\}|$, where we note that 
\begin{align*}\label{eq: numofintervalsbound}
    \sum_{\depth=1}^{\floor{\Njob/2}} 2^{\depth-1}z_\depth = \sum_{\depth=1}^{\floor{\Njob/2}} |\{j:\depth(j)=\depth\}| = |C| \le \Njob/2.
\end{align*}
We can then use the following LP to upper-bound the value of~\eqref{eq: distanceboundbylayer} under an adversarial choice $\{z_\depth:\depth=1,\ldots,\lfloor n/2\rfloor\}$:
\begin{maxi}
    {z}{ \sum_{\depth=1}^{\floor{\Njob/2}} z_\depth }
    {\label{eq: knapsack}}{}
    \addConstraint{ \sum_{\depth=1}^{\floor{\Njob/2}} 2^{\depth-1}z_\depth }{\le \Njob/2}
    \addConstraint{0\le z_\depth }{\le 1,\quad }{ \depth = 1,\ldots,\floor{\Njob/2}.}
\end{maxi}
This is a fractional knapsack problem, whose optimal value is at most $\log_2({\Njob/2+1})$, and thus
\begin{equation}\label{dist_bound}
    \sum_{j \in C} |\type_j - \type_{\matchof(j)}| \le \log_2(\Njob/2+1).
\end{equation}
Substituting back into \eqref{eq: offline_regret_bound} yields $\regret_\PB(\instance,\infty) \le 1/2 + \log_2(\Njob/2+1)/2$, completing the proof.
\Halmos\endproof

The following result shows that this analysis of $\PB$ is tight, up to constants.

\begin{proposition}[proof in \Cref{pf:loglowerboundOffline}]
\label{prop:loglowerboundOffline}
Under reward function $\reward(\type,\type') = \min\{\type,\type'\}$, when the number of jobs is $\Njob=2^{k+3} - 4$ for some integer $k\ge 0$, there exists an instance $\instance \in [0,1]^\Njob$ for which $\regret_\PB(\instance,\infty) \ge (\log_2(n+4)-3)/4$.
\end{proposition}
%


\subsection{Online Performance under Reward Function $\reward(\type,\type') = \min\{\type,\type'\}$}\label{sec:dynamic}

We now study the performance of algorithms in the online setting, i.e. when $\sojourn<\Njob$, and derive informative performance guarantees conditional on $\sojourn$, the number of new arrivals before a job becomes critical.
In fact, the regret per job of $\gre$ remains constant, whereas the one of $\PB$ decreases with $\sojourn$.
To build upon our results proved in \Cref{sec: static} for the offline setting, we partition the set of jobs into $b=\lceil\Njob/(\sojourn + 1) \rceil$ \newterm{batches}.
To be precise, we define the $t$-th batch as $B_{t} = \{(t-1)(\sojourn+1)+1,\ldots, t(\sojourn+1)\} \cap [\Njob]$, for each $t = 1, \ldots, b$.
Scaling $\sojourn$ while keeping the number of batches $b$ fixed can be interpreted as increasing the density of the market.
We first show that the regret under $\gre$ remains linear in the number of jobs $\Njob$, independent of $\sojourn$, which can be understood as suffering a regret of order $\Omega(\sojourn)$ per batch.

\begin{proposition}[proof in \Cref{pf:greedy_linear_lower_bound_dynamic}]
\label{prop:greedy_linear_lower_bound_dynamic}
    Under reward topology $\reward(\type,\type') = \min\{\type,\type'\}$, if $(\sojourn+1)$ is divisible by 4, then for any number of jobs $\Njob$ divisible by $(\sojourn+1)$, there exists an instance $\instance \in [0,1]^\Njob$ for which $\regret_\gre(\instance,\sojourn) \ge n/4$.
\end{proposition}

In contrast, the following theorem shows that the performance of $\PB$ improves as market density increases.

\begin{theorem}
\label{thm: dynamic}
Under reward topology $\reward(\type,\type') = \min\{\type,\type'\}$, we have $\regret_\PB(\instance,\sojourn) \le 1/2 + (\frac{\Njob}{d+1}+1)(1+\log(\sojourn+2))/2$ for any $\Njob$, and any instance $\instance\in[0,1]^\Njob$.
\end{theorem}
\proof{Proof.}
Let $\sojourn\ge 1$ and $(C,\matchof(\cdot))$ be the output of $\PB$ on $\instance$.
Similar to the proof of \Cref{thm:potential_log_upper_bound}, we have
\begin{align}
\regret_\PB(\instance,\sojourn)
& \le 
\frac12 + \frac12 \sum_{t=1}^{b} \left(\sum_{j\in  C\cap B_t} |\type_j-\type_{\matchof(j)}| \right), \label{eq:online_dist}
\end{align}
where we split the sum into the $b = \lceil \Njob/(\sojourn + 1) \rceil$ batches, defined as $B_{t} = \{(t-1)(\sojourn+1)+1,\ldots, t(\sojourn+1)\} \cap [\Njob]$ for $t = 1, \ldots, b$. We analyze the term in large parentheses for an arbitrary $t$. Intuitively, we would like to consider an offline instance consisting of jobs $(\type_j)_{j\in C \cap B_t}$ and their matches $(\type_{\matchof(j)})_{j\in C\cap B_t}$, and apply \Cref{thm:potential_log_upper_bound} on the offline instance which has size $2|C\cap B_t|\le 2(\sojourn+1)$.
However, since the offline setting allows jobs to observe the full instance, as opposed to only the next $\sojourn$ arrivals, the resulting matching of $\PB$ on the offline instance could be inconsistent with its output on the online instance. In particular, executing $\PB$ on the offline instance could match jobs with indices more than $d$ apart, which is not possible in the online setting. Thus, to derive a proper upper bound on regret, we need to construct a modified offline instance for which each resulting match of $\PB$ coincides with the matching output in the online counterpart, and in which matching distances were not decreased.

To construct this modified offline instance, first note that if $\matchof(j)\in B_t$ for all $j\in C\cap B_t$, then no modification is needed, since all matched jobs have indices less than $\sojourn$ apart.
However, if $\matchof(j)\in B_{t+1}$ for some $j\in C\cap B_t$, then $\matchof(j)$ can be available in the offline instance for some job $k<j$ with $|\matchof(j)-k|>\sojourn$ and be matched to $k$ in the offline instance even though this would not be possible in the online instance (see \Cref{fig:thm3_exm}).
Thus, in the modified offline instance, we ``move'' job $m(j)$ ensure job $k$ would not choose $m(j)$ for its match.
To do so, we distinguish three cases.
For each $j\in C\cap B_t$, let $A(j)$ be the set of jobs that the online algorithm could have chosen from to match with $j$, i.e.~$A_j$ is the set $A\setminus\{j\}$ in \Cref{alg:dynamic} right after job $j$ becomes critical.
If $\matchof(j)\in A(j)\cap B_t$, then both jobs $j,\matchof(j)$ are included in the offline instance we construct.
In the second case, if $\matchof(j)\in A(j)\cap B_{t+1}$ and $A(j)\cap B_t\neq\emptyset$, then we modify $\type_{m(j)}$ to be equal to the best available matching candidate within batch i.e.\ we ``move'' job $m(j)$ to location $\argmin \{ |\type_j - \type| : \type = \type_k, \ k\in A(j)\cap B_t \}$.
In the third case, if $\matchof(j)\in A(j)\cap B_{t+1}$ and $A(j)\cap B_t=\emptyset$, then we don't consider job $j$ or its match (if any) in the offline instance we construct. This can happen at most once per batch, since any other job $k<j$ would have $j\in A(k)$. 

\begin{figure}[H]
  \centering
  \subcaptionbox{Batch $B_t$ of the online instance $\instance$.
    \label{fig:thm3_exm_original}
    }[0.4 \textwidth]
    {
    \begin{tikzpicture}

\fill[lightyellow] (1.8, 0) rectangle (6.2, 3.2);

\draw[->] (-0.5, 0) -- (8.5, 0) node[right] {$j$};
\draw[->] (0, -0.1) -- (0, 3.3) node[above] {$\type_j$};

\foreach \x in {1,...,8}
    \draw (\x, -0.1) -- (\x, 0.1);

\node[below] at (1, 0) {\tiny $5t$};
\node[below] at (2, 0) {\tiny $5t+1$};
\node[below] at (3, 0) {\tiny $5t+2$};
\node[below] at (4, 0) {\tiny $5t+3$};
\node[below] at (5, 0) {\tiny $5t+4$};
\node[below] at (6, 0) {\tiny $5t+5$};
\node[below] at (7, 0) {\tiny $5t+6$};
\node[below] at (8, 0) {\tiny $5t+7$};

\foreach \y in {0.1,0.2,0.3,0.4,0.5,0.6,0.7,0.8,0.9,1}
    \draw (-0.1, \y*3) -- (0.1, \y*3) node[left, xshift=-0.2cm] {\tiny \y};

\foreach \x/\y in {1/0.9, 2/0.1, 3/0.7, 4/0.3, 5/0.4, 6/1, 7/0.2, 8/0.8} 
{
    \fill[black] (\x, \y*3) circle (0.1);
    \draw[gray, dashed] (0, \y*3) -- (\x, \y*3);
    }

\draw[thick] (1, 0.9*3) -- (3, 0.7*3);     
\draw[thick] (2, 0.1*3) -- (4, 0.3*3);
\draw[thick] (5, 0.4*3) -- (7, 0.2*3);     
\draw[thick] (6, 1*3) -- (8, 0.8*3);  

\end{tikzpicture}
    }
  \hfill
  \subcaptionbox{Modified offline instance $\hat{\instance}$. 
    \label{fig:thm3_exm_phantom}
    }[0.4 \textwidth]
    {
    \begin{tikzpicture}

\draw[->] (-0.5, 0) -- (4.5, 0) node[right] {$j$};
\draw[->] (0, -0.1) -- (0, 3.3) node[above] {$\type_j$};

\foreach \x in {1,...,4}
    \draw (\x, -0.1) -- (\x, 0.1);

\node[below] at (1, 0) {\tiny $5t+1$};
\node[below] at (2, 0) {\tiny $5t+3$};
\node[below] at (3, 0) {\tiny $5t+4$};
\node[below] at (4, 0) {\tiny $5t+6$};

\foreach \y in {0.1,0.2,0.3,0.4,0.5,0.6,0.7,0.8,0.9,1}
    \draw (-0.1, \y*3) -- (0.1, \y*3) node[left, xshift=-0.2cm] {\tiny \y};

\foreach \x/\y in {1/0.1, 2/0.3, 3/0.4, 4/1} 
{
    \fill[black] (\x, \y*3) circle (0.1);
    \draw[gray, dashed] (0, \y*3) -- (\x, \y*3);
    }

\draw[thick] (1, 0.1*3) -- (2, 0.3*3);     
\draw[thick] (3, 0.4*3) -- (4, 1*3);

\end{tikzpicture}
    }
\caption{Illustration of the construction of the modified offline instance, with $\sojourn=4$. In \Cref{fig:thm3_exm_original}, job $5t+1$ chooses job $5t+3$ in the online execution of $\PB$, because only jobs up to $5t+5$ would have arrived when job $5t+1$ becomes critical.  However, in the offline execution of $\PB$, job $5t+1$ would choose job $5t+6$, because all jobs are available.  Our modified instance in \Cref{fig:thm3_exm_phantom} corrects the inconsistent decision.
}
  \label{fig:thm3_exm}
\end{figure}
To formally define the construction, fix an arbitrary batch $t$. Let
$\hat{C}=\{j\in C\cap B_t: A(j)\cap B_t\neq \emptyset\}$ and $\hat{n}= 2|\hat{C}|\le 2(\sojourn+1)$. 
$\hat{C}$ is the set of critical jobs in cases one or two above, and we include jobs $j\in\hat{C}$ and their matches $m(j)$ in the offline instance.  There are $\hat{n}$ jobs in the offline instance and the ordering of indices is consistent with the original instance.
Define modified locations in the offline instance as follows:
%
\begin{align*}
\hat{\type}_j &= \type_j &\text{for $j\in\hat{C}$,}
\\ \hat{\type}_{m(j)} &= \type_{m(j)} &\text{for $j\in\hat{C}$, if $m(j)\in A(j)\cap B_t$ (case one),}
\\ \hat{\type}_{m(j)} &= \argmin \{ |\type_j - \type| : \type = \type_k, \ k\in A(j)\cap B_t \} &\text{for $j\in\hat{C}$, if $m(j)\in A(j)\cap B_{t+1}$ (case two).}
\end{align*}

Now, we show that the output of (offline) $\PB$ on $\hat{\instance}$ provides an upper bound for the expression $\sum_{j\in C\cap B_t}|\type_j-\type_{\matchof(j)}|$ in~\eqref{eq:online_dist}.
In particular, we show that the output of $\PB$ is exactly $(\hat{C},m(\cdot))$ when executed on the modified offline instance.
First, note that the construction potentially introduces ties if there is a job $k\in C\cap B_t$ with $\matchof(k)\in B_{t+1}$. However, since \Cref{thm:potential_log_upper_bound} holds under arbitrary tie-breaking, we assume $\PB$ breaks ties on $\hat{\instance}$ to maintain consistent decisions with $\PB$ on the original instance $\instance$. Thus, from \Cref{deliverypotential}, it suffices to show that for every $j\in \hat{C}$, $\matchof(j)$ minimizes the distance among available jobs.
Indeed, note that every job in the offline instance has a type that is identical to a job in $B_t$. Then, proceeding in the same order as in the online instance, the job types of the set of available jobs in the offline instance is a subset of the original available jobs. For a match in case one, we know that it minimizes the distance among the original available jobs, and hence it minimizes the distance among the offline available jobs. Moreover, we have $|\hat{\type}_j-\hat{\type}_{\matchof(j)}| = |\type_j - \type_{\matchof(j)}|$. In the second case, the match minimizes distance among offline available jobs by construction, and moreover $|\hat{\type}_j-\hat{\type}_{\matchof(j)}| \ge |\type_j - \type_{\matchof(j)}|$.
Consequently, we can upper bound 
$    \sum_{j\in C\cap B_t } |\type_j-\type_{\matchof(j)}|
\le 1 + \sum_{j\in  \hat{C}} |\hat{\type}_j-\hat{\type}_{\matchof(j)}|$.
%
Then, recalling \eqref{dist_bound} in the proof of \Cref{thm:potential_log_upper_bound}, we have $\sum_{j\in  \hat{C}} |\hat{\type}_j-\hat{\type}_{\matchof(j)}| \le \log(\hat{n}/2 + 1) \le \log(\sojourn + 2)$, and 
\begin{align}
    \sum_{j\in C\cap B_t } |\type_j-\type_{\matchof(j)}|
\le 1 +  \log(\sojourn + 2).\label{eq: distance_dynamic}
\end{align}
Substituting back into~\eqref{eq:online_dist}, and using the fact that $b\le\frac n{d+1}+1$, we get $\regret_\PB(\instance,\sojourn)\le 1/2 + (\frac{\Njob}{d+1}+1)(1+\log(\sojourn+2))/2$, completing the proof. 
\hfill \Halmos\endproof
Lastly, we show that this online analysis of $\PB$ is also tight, up to constants.
\begin{proposition}[proof in \Cref{pf:loglowerboundOnline}]
\label{prop:loglowerboundOnline}
Under reward topology $\reward(\type,\type') = \min\{\type,\type'\}$, if $\sojourn+1 = 2^{k+3}-4$ for some $k\in\{0,1,\ldots\}$, then for any number of jobs $\Njob$ divisible by $(\sojourn+1)$, there exists an instance $\instance \in [0,1]^\Njob$ for which $\regret_\PB(\instance,\sojourn) \ge \frac{\Njob}{3(\sojourn+1)}(\log_2(\sojourn+5)-3)/4$.
\end{proposition}

\section{Numerical Experiments} \label{sec:numerical_experiments}

In this section, we evaluate our potential-based greedy algorithm ($\PB$) against benchmark algorithms via numerical simulations. 
First, using synthetic data generated for a setting aligned with our theoretical model, but considering uniformly-random instead of worst-case instances, we find that $\PB$ outperforms all benchmarks at densities as low as $d \geq 5$. 
We then apply these algorithms to real data from the Meituan platform. Although the real-world setting differs from our theoretical model in a number of ways (see \Cref{sec:sim_meituan}), $\PB$ consistently dominates all greedy dispatch algorithms. To beat $\PB$, one requires batching-based heuristics that use more computation by orders of magnitude; moreover, even these methods can benefit from incorporating our potential as an opportunity cost adjustment.
\Cref{sec: sim_results_extra} provides additional simulation results, including additional performance metrics, more complex synthetic environments (e.g. two-dimensional locations, more general spatial distributions), different reward topologies,
and a more practically feasible ``periodic'' batching algorithm.

\subsection{Benchmark Algorithms} \label{sec:sim_benchmarks}
  
We compare $\PB$ against benchmark algorithms that (i) incorporate alternative opportunity costs derived from data, and/or (ii) adopt non-greedy dispatch policies that optimize over all currently-available jobs.

\subsubsection{Alternative opportunity costs.}
Both $\PB$ and naive greedy ($\gre$) are \newterm{forecast-agnostic}, in that they do not rely on any historical data or demand 
forecast.
In practice, however, platforms possess historical data that can be used to estimate the opportunity cost of dispatching a job.
To benchmark against such \newterm{forecast-aware} heuristics, we consider algorithms that use optimal dual solutions (i.e. shadow prices) as proxies for the opportunity cost of matching each job. 
Specifically, given an instance of historical data, we can solve the dual of a linear relaxation of the IP defined in \eqref{eq: OPT} (see \Cref{sec: LP} for details).
The optimal dual variables associated with the capacity constraint of each job represents the marginal value of the job, which we use to construct two benchmark algorithms.

\paragraph{Hindsight Dual.}
If we had access to the dual optimal solution for an instance in advance, we may use them as the opportunity cost of dispatching each job.
To measure the value of this information, we simulate a \newterm{hindsight-dual} algorithm ($\dual$) that operates retrospectively on the same instance.
Formally, given an instance $\instance\in\typespace^\Njob$ and parameter $\sojourn$, let $\lambda_j=\lambda_j(\instance,\sojourn)$ be the optimal dual variable for the capacity constraint $\sum_{k:j\neq k} x_{jk} \le 1$.
The $\dual$ algorithm operates as an index-based greedy algorithm (\Cref{alg:dynamic}) where the index function is given by $\indexf_{\dual}(\type_j,\type_k) = \reward(\type_j,\type_k) - \lambda_k$, for all $j,k\in [\Njob]$, such that $j\neq k, |j-k| \le d$.
Because this requires knowledge of the entire realized instance, $\dual$ serves only as an idealized benchmark.

\paragraph{Average Dual.}
A practically more feasible approach for estimating opportunity costs is to average the hindsight duals from historical data.
This, however, yields a shadow price for every \emph{observed} type, which might not cover the entire continuous type space and can be very noisy.
Therefore, we discretize the type space, and estimate average duals for each \emph{discrete type} as the empirical average of the optimal dual variables for jobs associated to the discrete type.
To be precise, let $H$ be the set of all instances in the historical data, and consider a discretization $\{\typespace_1,\typespace_2,\ldots,\typespace_p\}$
that partitions the space into $p$ cells (these cells are mutually exclusive and jointly exhaustive).
Each cell $i=1,2,\ldots,p$ is associated with a shadow price $\Bar{\lambda}_i$ that is the average of historical jobs lying in the same cell, i.e. $\Bar{\lambda}_i$ is the average of $\{\lambda_j(\instance,\sojourn) : \instance\in H, \type_j \in \typespace_i \}$.
Then, the \newterm{average-dual} algorithm
($\averagedual$) is an index-based greedy matching algorithm that uses the shadow prices for the cells, i.e.\ uses index function
$\indexf_{\averagedual}(\type,\type') = \reward(\type,\type') - \sum_{i=1}^p\Bar{\lambda}_{i}\indicator\{\type'\in\typespace_i\}.$

\medskip

\subsubsection{Alternative dispatch policies.}
We also consider batching-based dispatch policies that compute optimal outcomes given the entire set of currently-available jobs.

\paragraph{Full Batching.} 

When an available job becomes critical, the \newterm{full batching} algorithm ($\batching$) computes the optimal pooling outcome for all available jobs, and dispatches all jobs accordingly. 
Note that under the theoretical model introduced in \Cref{sec:model}, $\batching$ clears the market every $\sojourn + 1$ arrivals, which is equivalent to the algorithm analyzed in \citet{ashlagi2019edge}. 
An obvious drawback of $\batching$ is that it dispatches non-critical jobs unnecessarily, eliminating the possibility of better matches in the future.

\paragraph{Rolling Batching.} 

To address the issue of dispatching non-critical jobs, we consider the \newterm{rolling batching} ($\rbatching$) algorithm, which also computes an optimal solution every time any available job becomes critical. 
Instead of dispatching all jobs, however, $\rbatching$ only dispatches the critical job and its match (if any) according to the optimal solution, holding the rest until the next time a job becomes critical.

\subsection{Synthetic Data}\label{sec:sim_unif_1D}

We consider the setting analyzed in our theoretical results, but now draw job destinations uniformly at random from a one-dimensional space instead of assuming worst-case sequences.
We fix the total number of jobs at $\Njob=1000$ and vary the density level from $d=5$ to $d=30$.
For each $(\Njob, \sojourn)$ pair, we generate 100 instances $\instance \in[0,1]^\Njob$ uniformly at random, \drop{apply all algorithms on each of them,} and compare in \Cref{fig:1D} the average regret and reward \newterm{ratio} $\ALG(\instance,d)/\OPT(\instance,d)$ achieved by each algorithm and benchmark.
Additional performance metrics (the fraction of jobs that are pooled and the fraction of the total distance that is reduced by pooling) are provided in \Cref{sec:match_rate_2}.
Similar results for non-uniform spatial distribution are presented in \Cref{sec: nonunif_1D}.

\subsubsection{Comparison with alternative opportunity costs.}

\begin{figure}
  \centering
  \subcaptionbox{Average regret (lower is better).%
    \label{fig:1D_greedy_regret}}[0.49 \textwidth]{\includegraphics[width = \figWidth \textwidth]{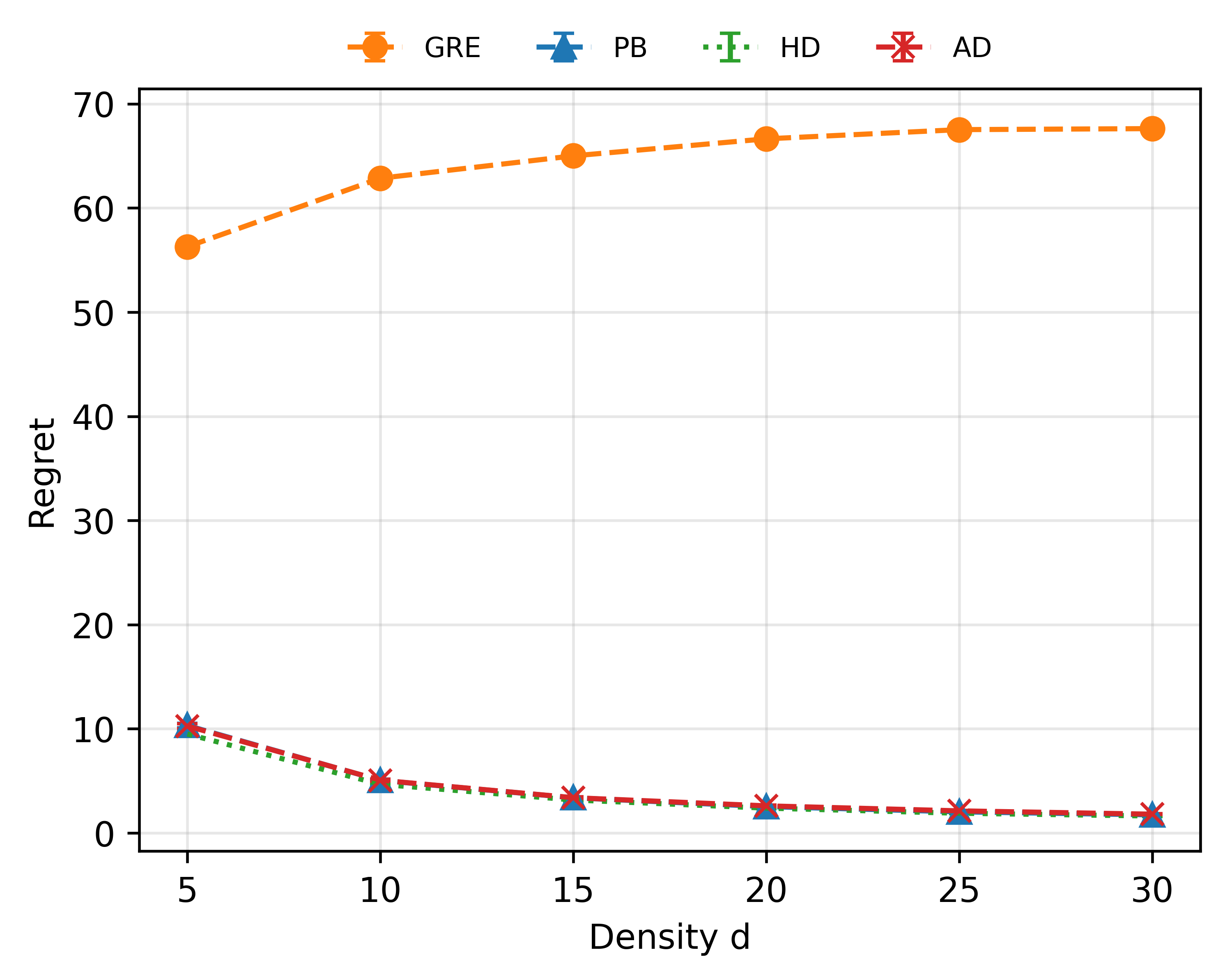}}
  \hfill
  \subcaptionbox{Average ratio (higher is better).%
    \label{fig:1D_greedy_ratio}}[0.49 \textwidth]{\includegraphics[width = \figWidth \textwidth]{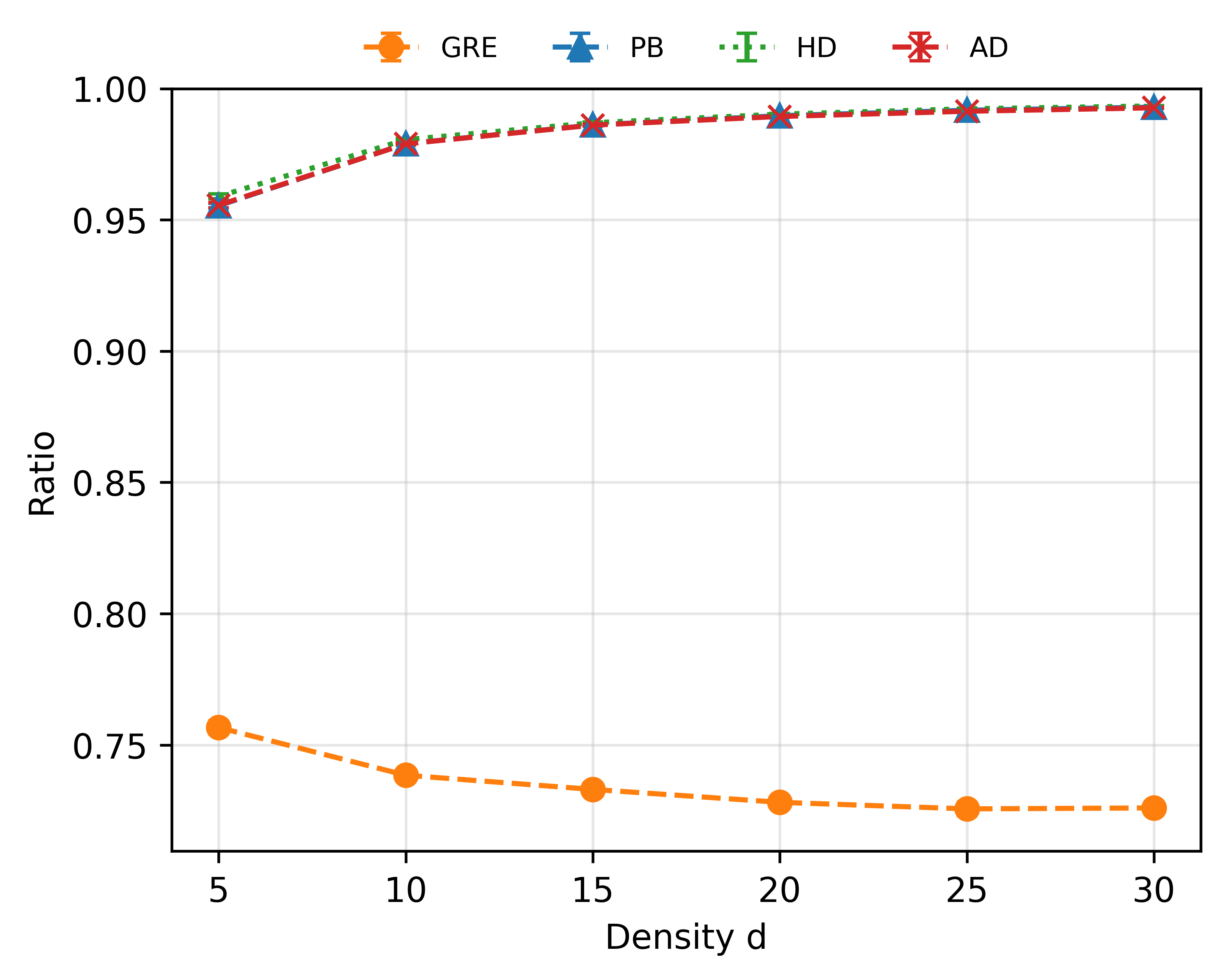}}
  \caption{Average regret and reward ratio, comparing with alternative opportunity costs, in random 1D instances.
  }
  \label{fig:1D}
\end{figure}

\Cref{fig:1D} compares potential-based greedy $\PB$ with naive greedy $\gre$ and the two forecast-aware benchmarks, $\dual$ and $\averagedual$.
We first observe that consistent with our theoretical analysis, $\PB$ substantially outperforms $\gre$.
Even at relatively low density levels, $\PB$ achieves over 95\% of $\OPT$ (the maximum possible travel distance saved from pooling), and the performance of $\PB$ consistently improves with $d$. 
In stark contrast, the performance of $\gre$ relative to $\OPT$ \textit{worsens} as the density $d$ increases, despite the fact that $\gre$ does achieve a higher pooling reward as the density increases (see \Cref{fig:1D_unif_saving_fraction} in \Cref{sec: 1D_extra}).
For the $\dual$ and $\averagedual$ benchmarks, we uniformly discretize the type space $[0,1]$ into 100 intervals.
Under $\averagedual$, the opportunity costs are estimated using the average dual variables from 400 sample instances drawn IID from the same distribution as the 100 evaluation instances.
We see that $\PB$ performs on par with $\averagedual$ and $\dual$ across all density levels, without relying on any forecast/historical data. 
This effectiveness of using the potentials as opportunity costs is explained in \Cref{sec: interpretation}, where we illustrate that the hindsight duals concentrate more tightly around the potential as $\sojourn$ grows.
Moreover, we prove that (in this stochastic setting) the marginal value of dispatching a job converges to the potential in arbitrarily dense markets.

\subsubsection{Comparison with batching-based policies.}

\begin{figure}
  \centering
  \subcaptionbox{Average regret (lower is better).%
    \label{fig:1D_regret}}[0.49 \textwidth]{\includegraphics[width = \figWidth \textwidth]{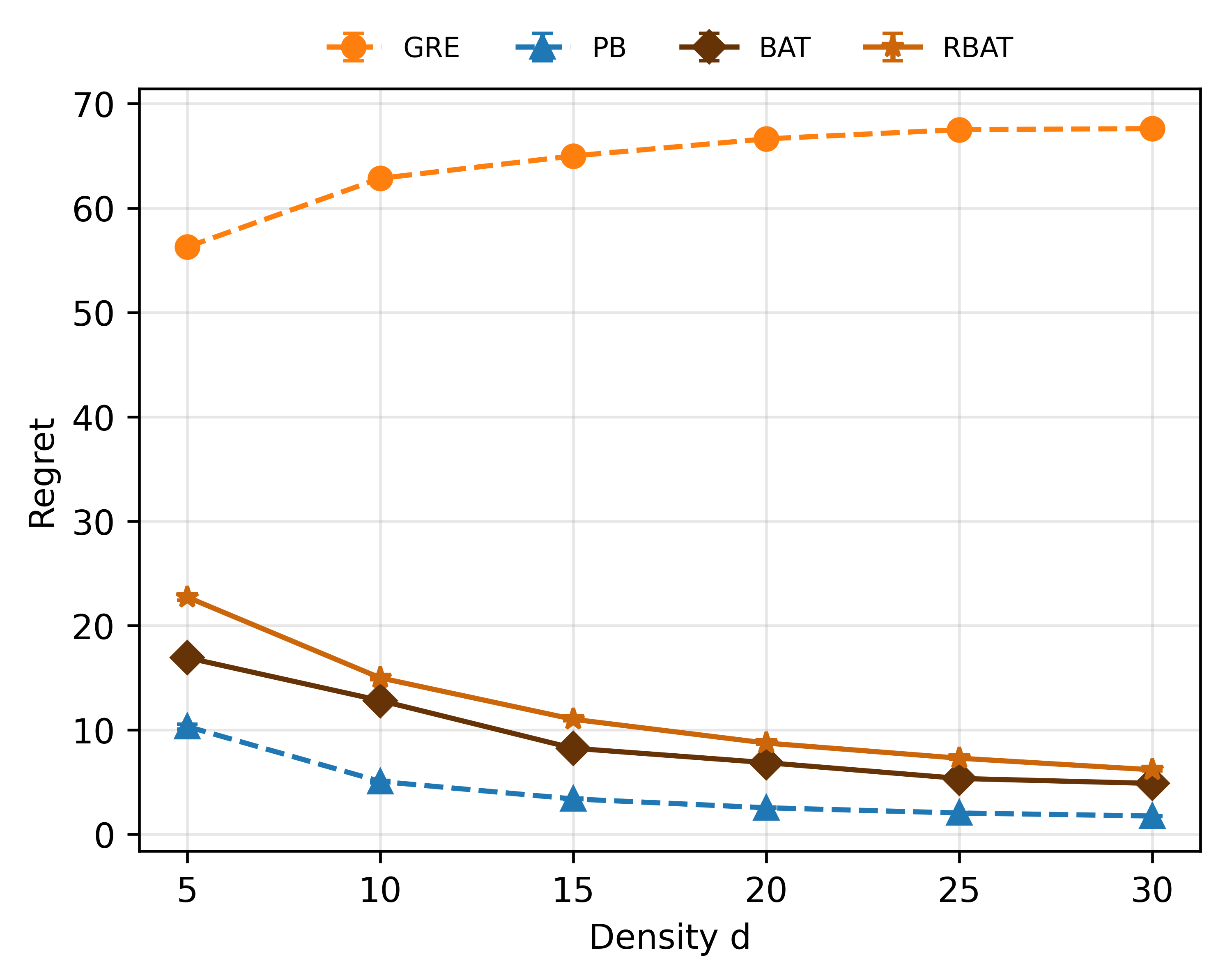}}
  \hfill
  \subcaptionbox{Average ratio (higher is better).%
    \label{fig:1D_ratio}}[0.49 \textwidth]{\includegraphics[width = \figWidth \textwidth]{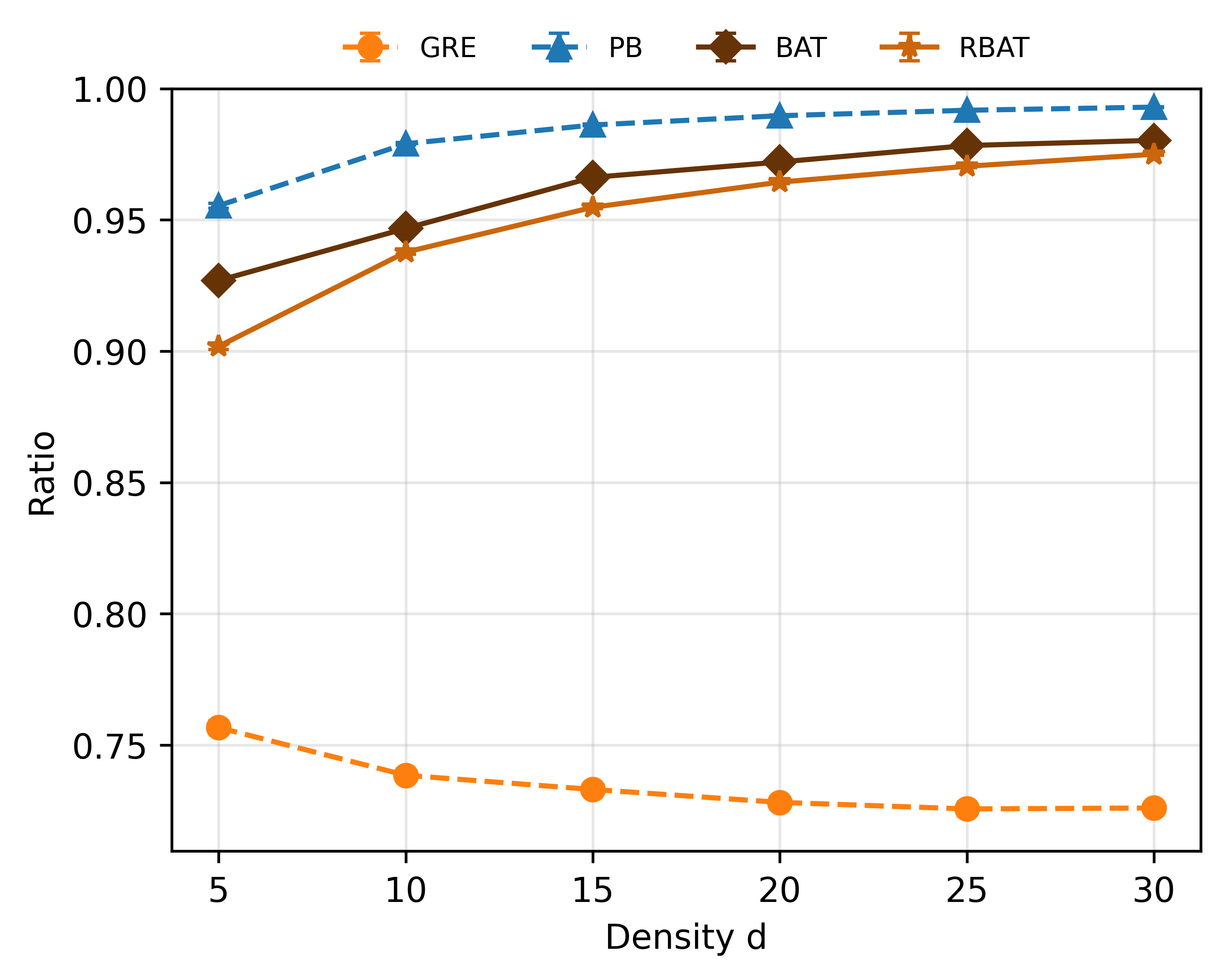}}
  \caption{Average regret and reward ratio, comparing with batching-based heuristics, in random 1D instances.
  }
  \label{fig:1D_batching}
\end{figure}

\Cref{fig:1D_batching} compares $\PB$ and $\gre$ with the non-greedy algorithms $\batching$ and $\rbatching$.
While batching heuristics implicitly capture near-term opportunity costs by optimizing for the current state (i.e., the set of all available jobs), $\PB$ surprisingly outperforms both $\batching$ and $\rbatching$ across all density levels.
This suggests that in this synthetic environment, the long-term view of future arrivals, reflected in the potential, is more important than optimizing given the current state.
Although the batching algorithms can also be extended to incorporate opportunity costs, the optimality gap of $\PB$ in this setting is already small, leaving little headroom for improvement.
In the real-data experiments that follow, however, we will see that state-dependent optimization becomes more valuable, and shadow-price augmentation yields further gains.

\subsection{Data from the Meituan Platform}
\label{sec:sim_meituan}

In this \namecref{sec:sim_meituan}, we test the algorithms and benchmarks using data from the Meituan platform, made public for the 2024 INFORMS TSL Data-Driven Research Challenge \citep{tsl_meituan_2024}
(\Cref{sec:market_dynamics} provides more details on the dataset and high-level illustrations of market dynamics over both space and time).

The dataset provides detailed information on various aspects of the platform's delivery service, covering 8 days of operations in a single city during October 2022.
For each of the 569 million delivery orders, the dataset provides precise timestamps and anonymized coordinates for pick-up and drop-off locations, allowing us to accurately reconstruct job arrivals and pooling opportunities.

Since the pick-up and drop-off locations now reside in a two-dimensional (2D) space, we first extend the definition of pooling reward and derive the potential of different job types. 
We then provide the notion of a pooling window (i.e. the sojourn time) that determines which jobs can be pooled together, and describe the data setting and the dual-estimation procedure used to construct the forecast-aware benchmarks.

\paragraph{Pooling reward.}
Each job $j$ has type $\type_j = (\xorigin_j,\xdestination_j)$, 
with $\xorigin_j,~\xdestination_j\in \R^2$ corresponding to the coordinates of the job origin and destination.
The travel distance saved by pooling two orders together is the difference between (i) the travel distance required to fulfill the jobs in two separate trips, and (ii) the minimum travel distance for fulfilling the two requests in a single, pooled trip.
The distance of the pooled trip always includes the distance between the two origins and that between the two destinations (both orders must be picked up before either of them is dropped off; otherwise the orders are effectively not pooled). 
The remaining distance depends on the sequence in which the jobs are picked up and dropped off, resulting in four possible routes. The minimum travel distance is then determined by evaluating the four routes and selecting the smallest travel distance.
Formally, let $\norm{\cdot}$ denote the Euclidean norm, the reward function is
%
%
\begin{align}
    \reward(\type,\type') = &  \norm{\xdestination-\xorigin} +  \norm{\xdestination'-\xorigin'}  \nonumber\\
    & -\left(\norm{\xorigin-\xorigin'} + \min\left\{\norm{\xdestination-\xorigin},\norm{\xdestination'-\xorigin},\norm{\xdestination-\xorigin'},\norm{\xdestination'-\xorigin'}\right\}+\norm{\xdestination-\xdestination'}\right). \label{eq:2D_reward_part_2}
\end{align}

Note that unlike the 1D common origin setting, the pooling reward in 2D space with heterogeneous origins is not always positive. As a result, we assume that all index-based greedy algorithms only pool jobs together when the reward is non-negative.

\paragraph{Potential.}
For a given job type $\type=(\xorigin,\xdestination)$, the maximum reward a platform can achieve by pooling it with another job is the distance of the job, $\norm{\xdestination - \xorigin}$.%
\footnote{
To see this, first observe that when a job $\type = (\xorigin,\xdestination)$ is pooled with another job of identical type, the travel distance saved is precisely $\norm{\xdestination - \xorigin}$.
On the other hand, let $m$ be the minimum in \eqref{eq:2D_reward_part_2}. Then, by triangle inequality $\|O'-D'\|\le \|O-O'\| + \|D-O\| + \|D-D'\|$, and thus $\reward(\type,\type')\le 2\|D-O\|-m \le \|D-O\|$, since $m\le \|D-O\|$.
To achieve an even higher pooling reward, it must be the case that the second part of the pooling reward \eqref{eq:2D_reward_part_2} is strictly smaller than $\norm{\xdestination' - \xorigin'}$. This implies that the total distance traveled to visit all four locations is smaller than the distance between two of them, which is impossible.
}  
As a consequence, the potential of a job $\type$ is given by $\potential(\type) = \norm{\xdestination - \xorigin}/2$. 
This is aligned with our 1D theoretical model, in which longer deliveries have higher potential reward from being pooled with other jobs in the future. 

\paragraph{Pooling window.} 
Instead of assuming that each job is available to be pooled for a fixed number of future arrivals, we take the actual timestamp of each job and assume that the jobs remain available for pooling for a fixed time window at the end of which it becomes critical.
Intuitively, the length of the window corresponds to the ``patience level''. 
This deviates from our theoretical model, but it is straightforward to see how our algorithm as well as the benchmarks we compare with can be generalized and applied.

\paragraph{Data Setting.} 

We focus our analysis on a three-hour lunch period (10:30am to 1:30pm), during which there are approximately $130$ orders per minute on average, and $\Njob\approx 24,000$ orders in total (see \Cref{fig:meituan_orders_per_hour} in \Cref{sec:market_dynamics} for an illustration of order volume over time).
Thus if the platform allows a time window of one minute for each order before it has to be dispatched, the average density would be $130$ jobs (although, these jobs could be at disparate locations such that delivery pooling may not yield any distance savings). 

\paragraph{Dual Estimation.} 
To simulate the \emph{hindsight-dual} algorithm $\dual$ for each instance associated to each day, we compute the shadow prices via the dual LP and then compute the ``online'' pooling outcomes using pooling reward minus these shadow prices as the index function.
To estimate historical average shadow prices for the $\averagedual$ algorithm, for each of the 8 days, we (i) use the remaining 7 days as the historical data, and (ii) discretize the 2D space into discrete types based on job origin and destination using H3, the hexagonal spatial indexing system developed by Uber \citep{uber_h3_2018}.\footnote{
The system supports sixteen resolution levels, each tessellating the earth using hexagons of a particular size.
For a given resolution, the set of all origin-destination (OD) hexagon pairs represents the partition of the type space $\typespace$, as described in \Cref{sec:sim_benchmarks}.
To compute the average shadow price for a particular OD pair, we compute the average of optimal dual variables associated with all historical jobs that share the same OD pair.
If there are no jobs for a given pair, we consider the hexagon pairs at coarser resolutions (moving one or more levels up in the H3 hierarchy, with each level increasing the hexagon size by a factor of 7) until there exist associated historical jobs.
There is a tradeoff between granularity and sparsity when we choose the baseline resolution level, as finer resolutions only average over trips with close-by origins and destinations, but may lack sufficient data for accurate estimation (see \Cref{fig:meituan_CDF_count_per_OD_pair} for the distribution of order volume by OD hexagon pairs).
After testing all available resolutions, we found that levels 8 or 9 perform the best, depending on the time window. The results presented in this section report the result corresponding to the best resolution for each time window.}

\subsubsection{Comparison with alternative opportunity costs.}

\begin{figure}[]
  \centering
  \subcaptionbox{Average regret (lower is better).%
    \label{fig:meituan_regret}}[0.49 \textwidth]{\includegraphics[width = \figWidth \textwidth]{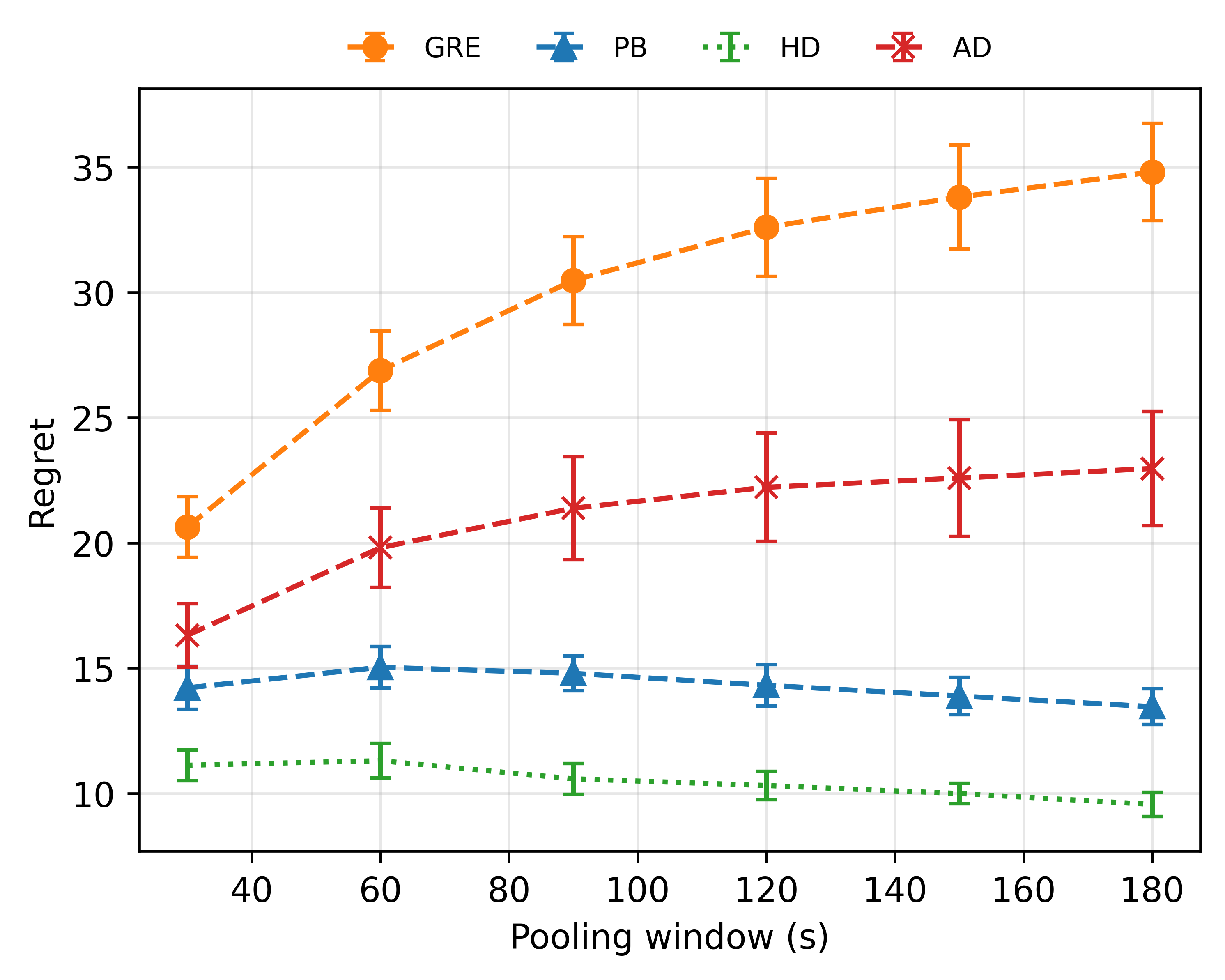}}
  \hfill
  \subcaptionbox{Average ratio (higher is better).%
    \label{fig:meituan_ratio}}[0.49 \textwidth]{\includegraphics[width = \figWidth \textwidth]{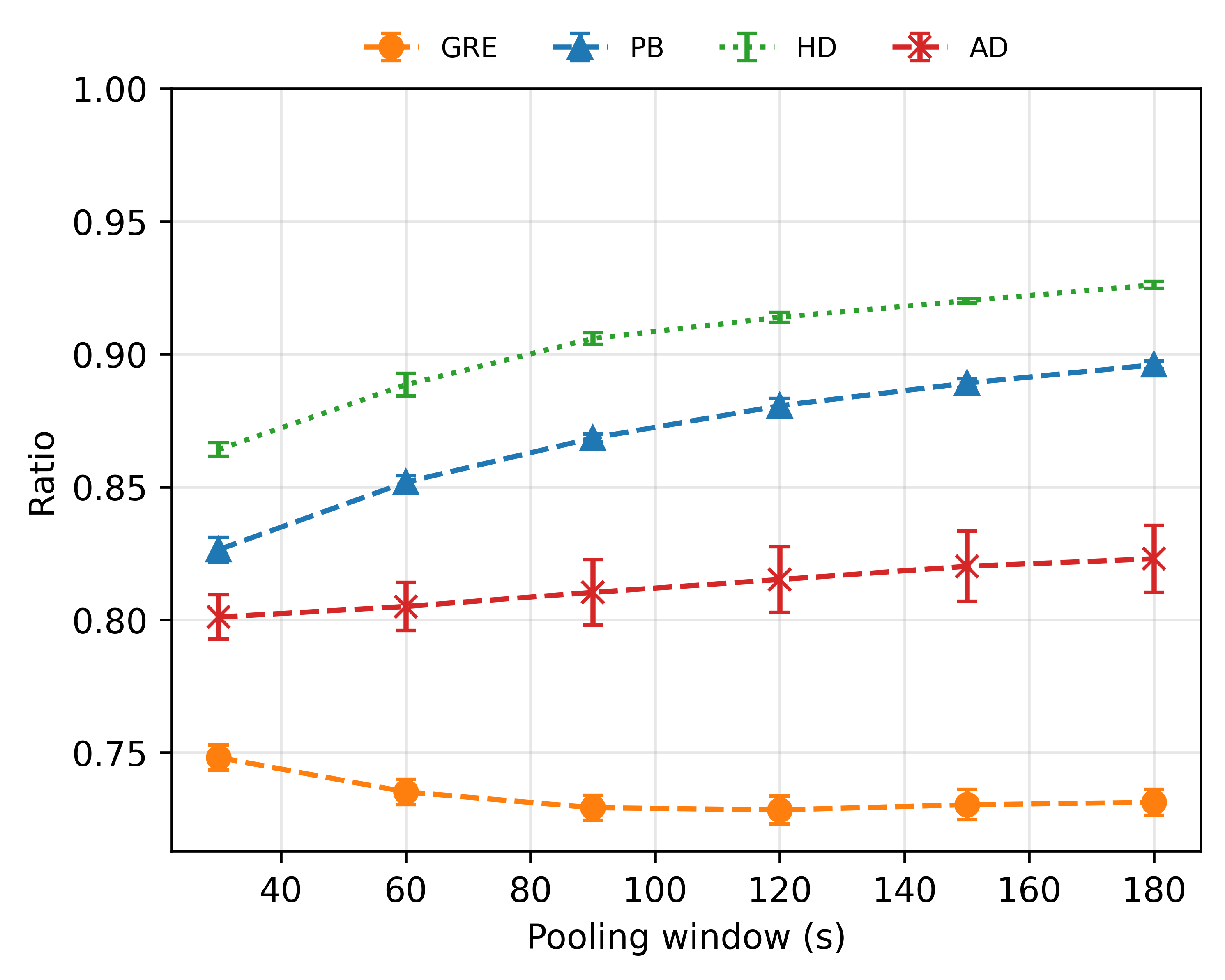}}
  \caption{Average regret and reward ratio, comparing with alternative opportunity costs, in Meituan data.}
  \label{fig:meituan}
\end{figure}
\Cref{fig:meituan} compares the index-based greedy heuristics with different shadow prices, as the pooling window increases from 30 seconds to 3 minutes.
We can see that $\PB$ outperforms both $\gre$ and $\averagedual$ (whereas $\dual$ is not a realistic algorithm). 
$\PB$ achieves around 90\% of the hindsight optimal reward if orders can wait $3$ minutes before dispatch.
This range is aligned with the timeframe for Meituan's dispatch system, where batches of orders are evaluated at time intervals of less than a minute, but assignments can be strategically delayed causing orders to be assigned to couriers between 3 to 5 minutes after order creation during peak lunch hours (see \Cref{fig:meituan_order_to_first_dispatch_how}).
Additionally, \Cref{fig:meituan_extra} indicates that with a $3$-minute window, $\PB$ pools over 90\% of jobs, and saves nearly 30\% of the travel distance compared to dispatching all jobs individually (recall that the theoretical upper bound is 50\%).

In contrast to the synthetic setting from \Cref{sec:sim_unif_1D}, having access to the hindsight information reflected in $\dual$ now leads to a meaningful improvement over $\PB$.
This appears to be driven by the spatially distributed instead of common job origins---additional simulations on 2D synthetic settings (\Cref{appx:additional_synthetic_results}), demonstrate that (i) $\dual$ performs only slightly better than $\PB$ in 2D space with common origin and destinations drawn uniformly at random from a unit square, and (ii) when both the origins and destinations uniformly drawn, $\dual$ outperforms $\PB$ in a way very similar to what we observe here with Meituan data.%

The weaker performance of $\averagedual$ relative to $\PB$ highlights the challenges of estimating the opportunity costs from historical data in realistic settings. 
Intuitively, taking historical data into consideration should be valuable in scenarios with highly non-uniform spatial distributions. In this case, however, the sparsity of the data for many if not most origin-destination pairs means that we are either averaging over a small number of observations or also considering jobs that are far in space.
$\PB$ bypasses this sparsity-granularity tradeoff by only considering the reward topology, i.e.\ the space of all possible types.

\subsubsection{Comparison with batching-based policies.} \label{sec:sim_meituan_batching}

\begin{figure}
  \centering
  \subcaptionbox{Average regret (lower is better).%
    \label{fig:meituan_batching_regret}}[0.49 \textwidth]{\includegraphics[width = \figWidth \textwidth]{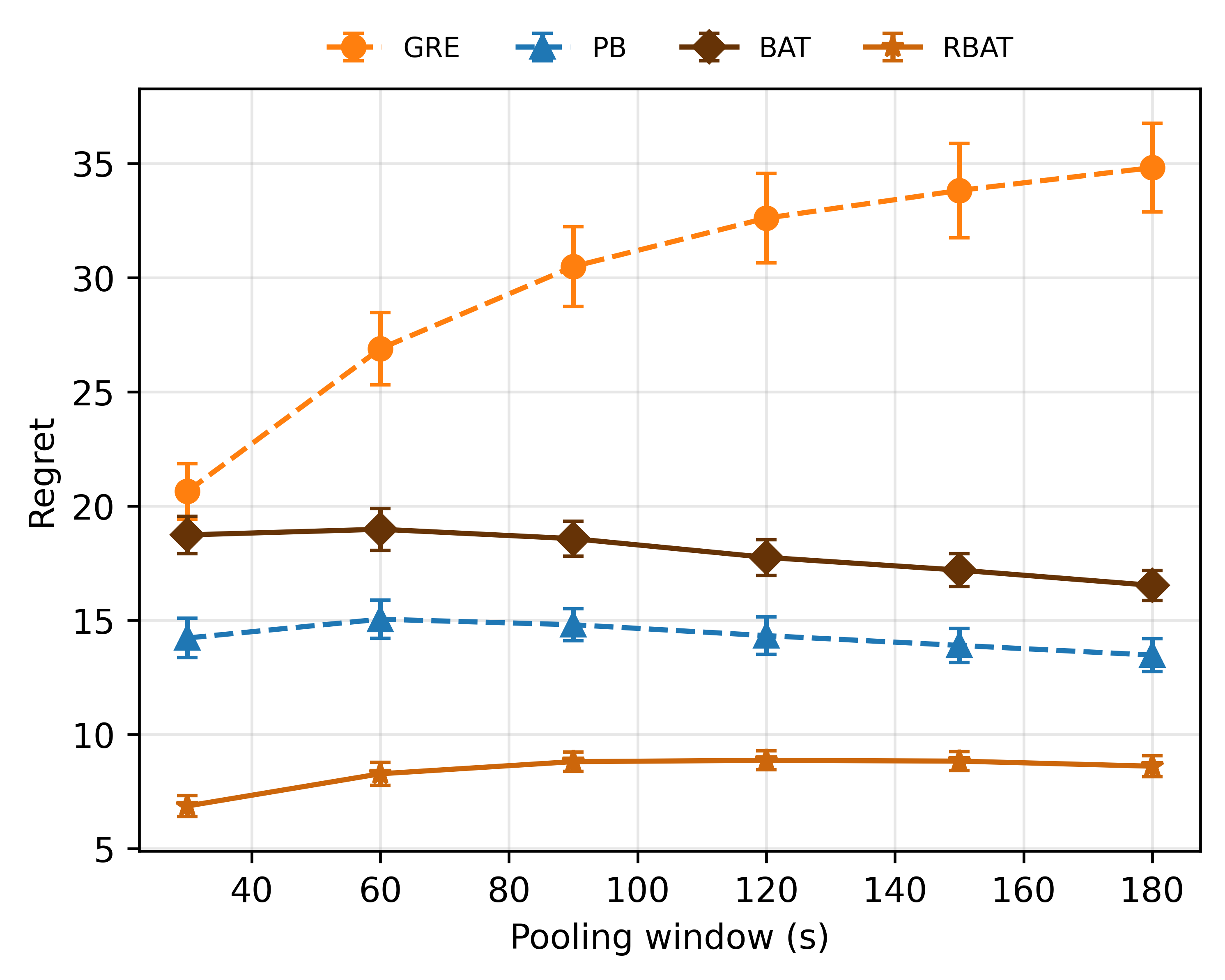}}
  \hfill
  \subcaptionbox{Average ratio (higher is better).%
    \label{fig:meituan_batching_ratio}}[0.49 \textwidth]{\includegraphics[width = \figWidth \textwidth]{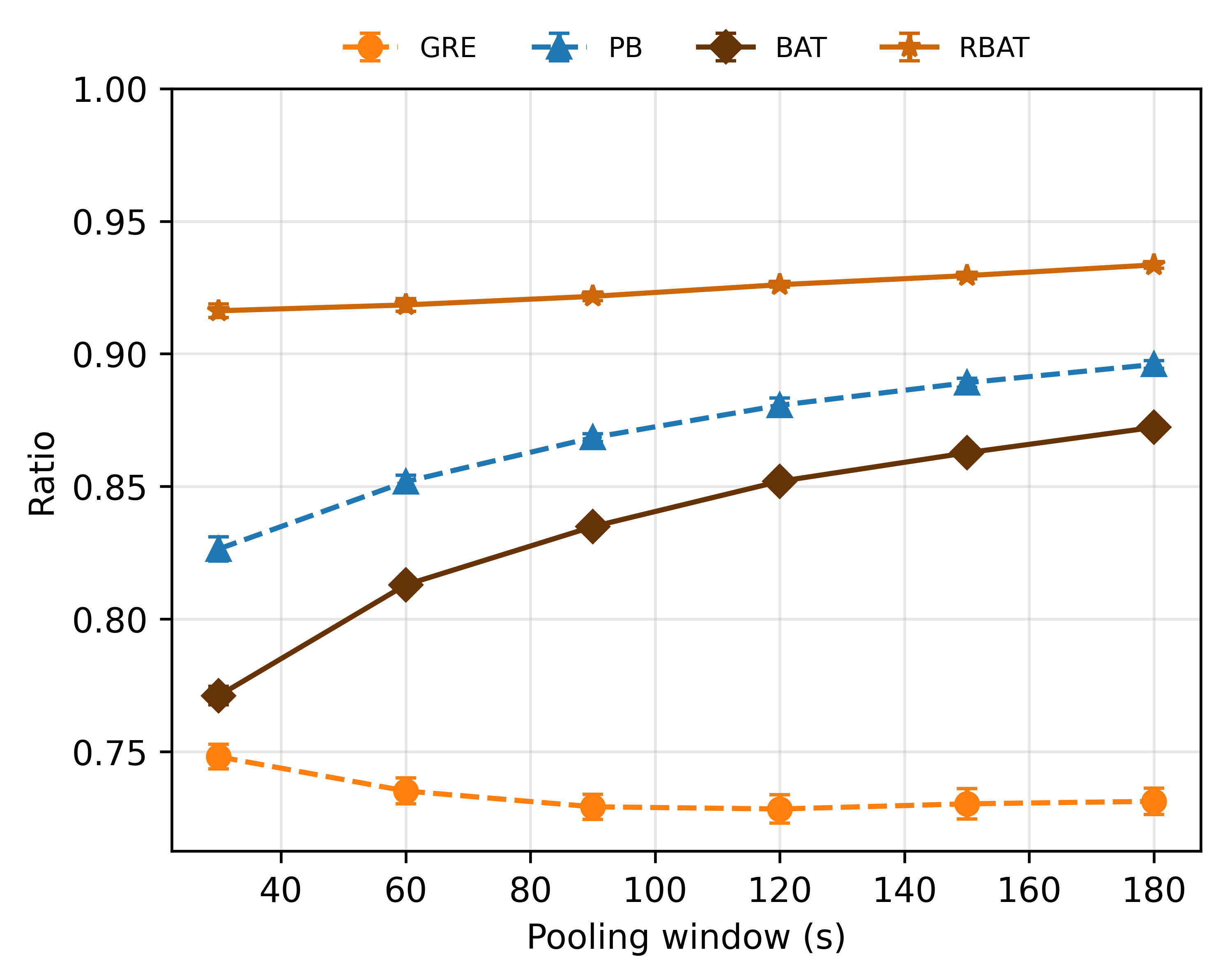}}
  \caption{Average regret and reward ratio, comparing with batching-based heuristics, in Meituan data.}
  \label{fig:meituan_batching}
\end{figure}

\begin{figure}
  \centering
  \subcaptionbox{Running time.%
    \label{fig:meituan_running_time}}[0.49 \textwidth]{\includegraphics[width = \figWidth \textwidth]{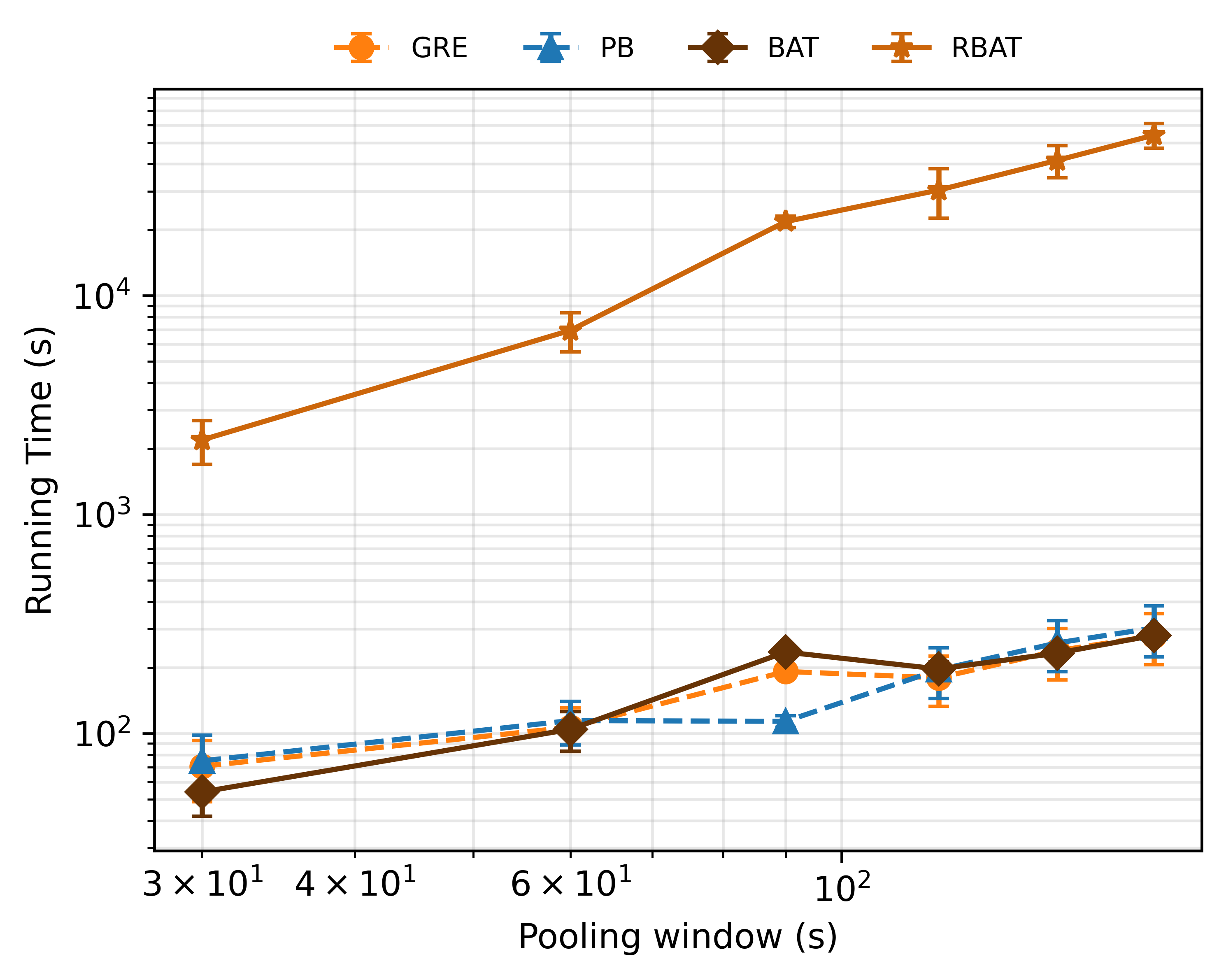}}
  \hfill
  \subcaptionbox{Match rate.%
    \label{fig:meituan_batching_match_rate}}[0.49 \textwidth]{\includegraphics[width = \figWidth \textwidth]{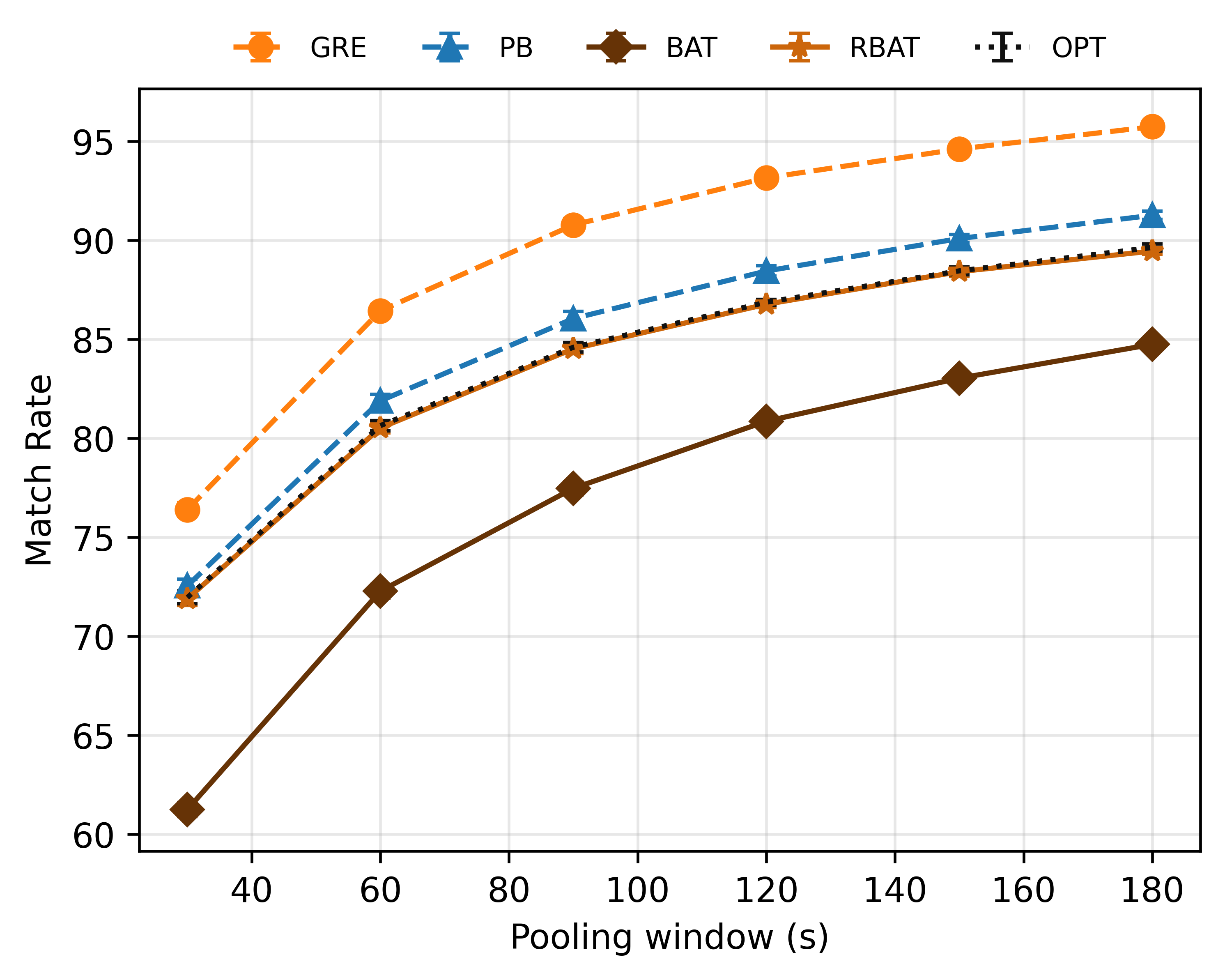}}
  \caption{Running time and match rate, comparing with batching-based heuristics, in Meituan data.}
  \label{fig:meituan_time_match}
\end{figure}

\Cref{fig:meituan_batching} compares our proposed algorithm with batching-based heuristics.
In contrast to the synthetic one-dimensional results, $\rbatching$ now outperforms $\PB$, highlighting the importance of properly taking into consideration the ``state'' (i.e. set of all currently-available jobs) in realistic settings. 
However, the value of explicitly capturing the state depends on spatial heterogeneity---additional 2D synthetic simulations showcase that $\rbatching$ is no better than $\PB$ when all jobs share a common origin (see \Cref{appx:2D_commom_origin}). %
$\batching$ remains at a disadvantage because, as it dispatches non-critical jobs prematurely and forgoes the possibility of more valuable future opportunities.

The superior performance of $\rbatching$, however, relies on re-solving an optimal match for the near-future every time any job becomes critical.
This results in an $100\times$ higher computational cost (\Cref{fig:meituan_running_time}), and limits the applicability of $\rbatching$ at a large platform like Meituan that observes over 100 million orders each day.
$\PB$, on the other hand, remains as fast as $\gre$ while closing a significant fraction of the performance gap relative to $\rbatching$.
Additionally, $\PB$ pools a larger fraction of jobs than $\rbatching$ (see \Cref{fig:meituan_batching_match_rate}). 
This is desirable in practice, since drivers who are offered just only one order from a platform may try to pool orders from competing platforms 
\citep[see e.g.][]{reddit2022grubhub,reddit2023doordash},
leading to poor service reliability for customers.

In \Cref{sec:rbat2}, we implement a more practical \emph{periodic} rolling batching that computes the optimal matching at fixed intervals (i.e., every 30 seconds) rather than every time a job becomes critical. 
Its performance matches that of $\PB$ when the pooling window is $1$ minute, and resides roughly half way between $\PB$ and the full $\rbatching$ for longer windows.
Nevertheless, for both versions of rolling batching, potential can be incorporated to further improve the performance, which we demonstrate next.

\subsubsection{Potential-augmented rolling batching.}
\label{sec:rbat_best}
\begin{figure}
  \centering
  \subcaptionbox{Average regret (lower is better).%
    \label{fig:meituan_batching_best_regret}}[0.49 \textwidth]{\includegraphics[width = \figWidth \textwidth]{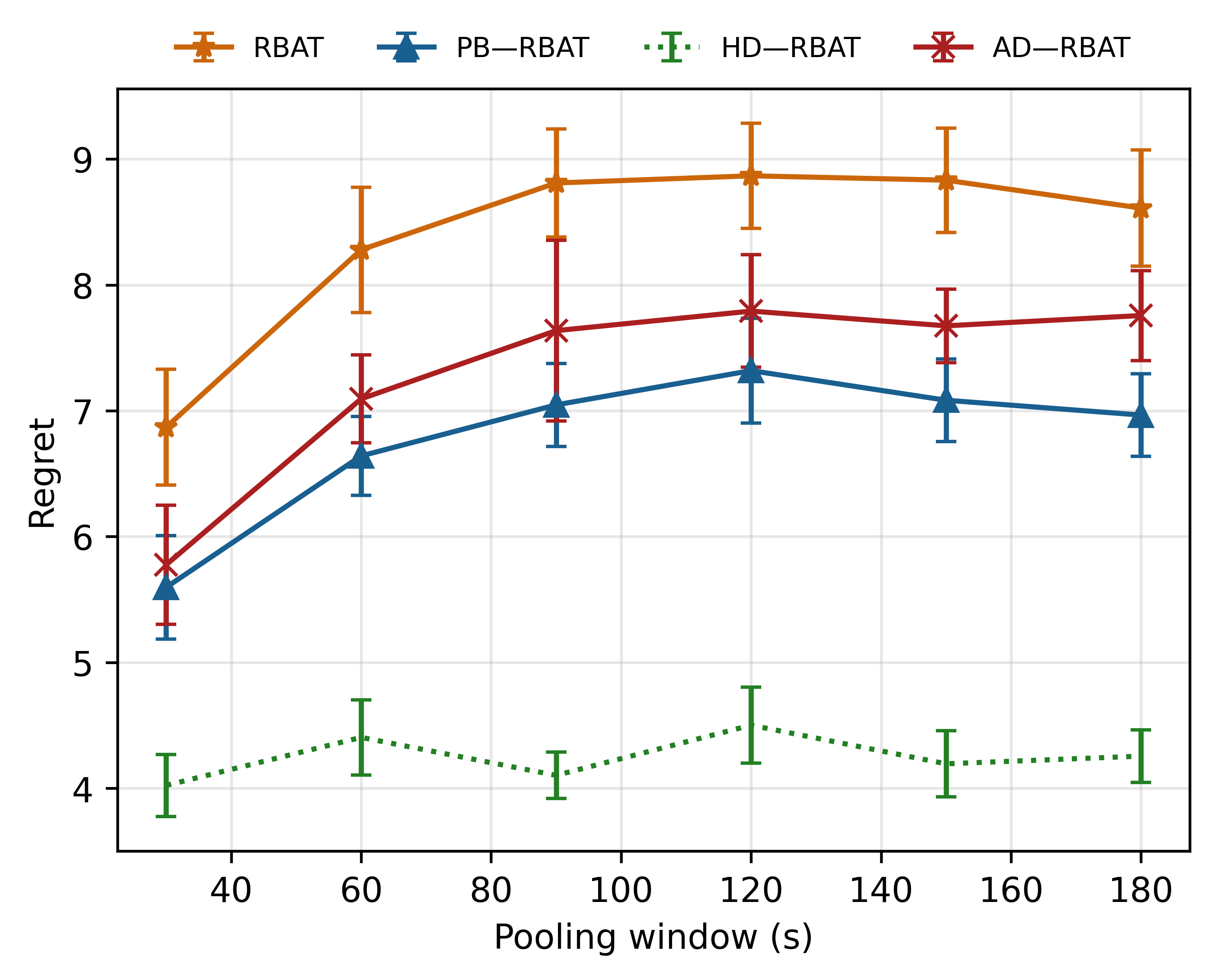}}
  \hfill
  \subcaptionbox{Average ratio (higher is better).%
    \label{fig:meituan_batching_best_ratio}}[0.49 \textwidth]{\includegraphics[width = \figWidth \textwidth]{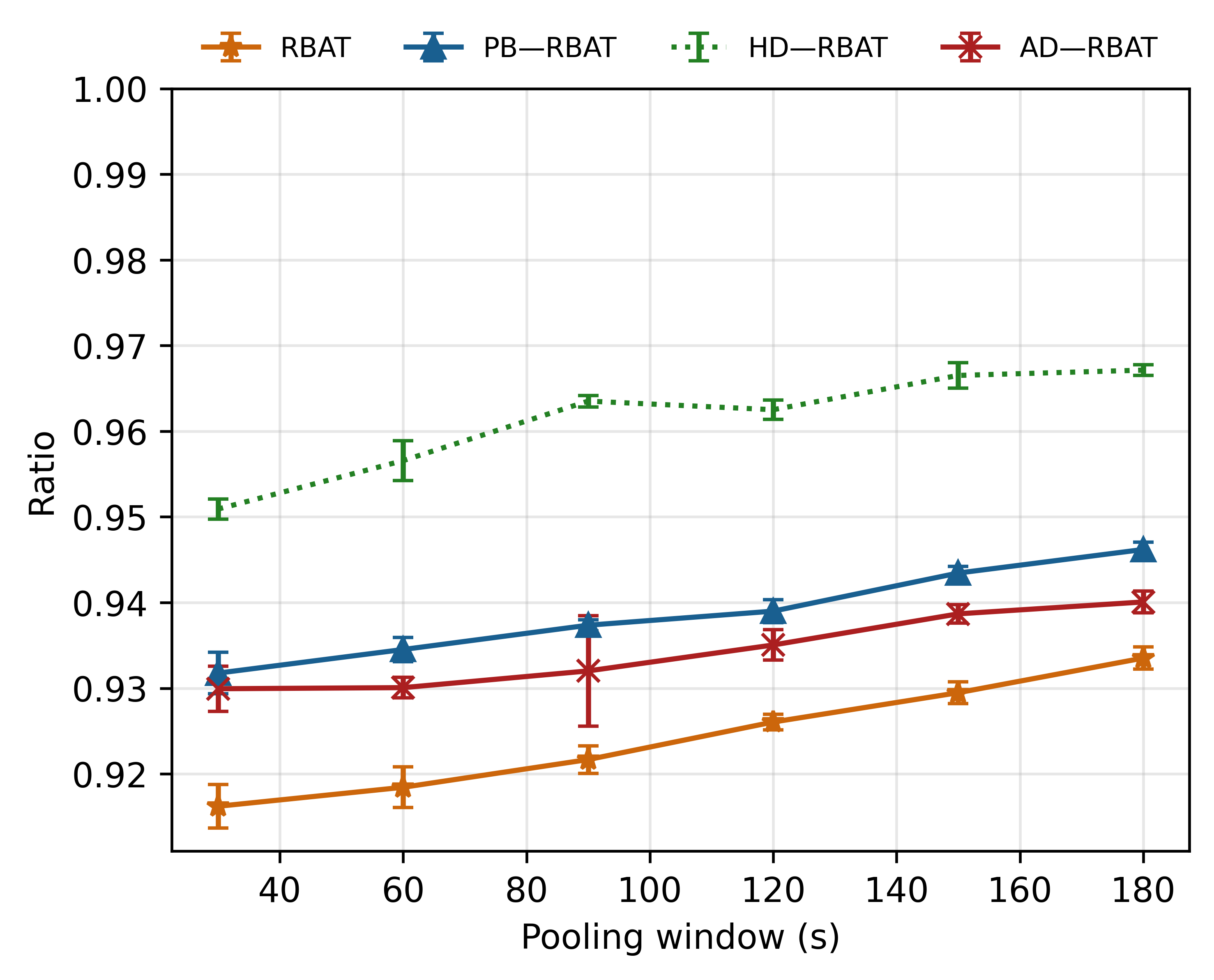}}
  \caption{Average regret and reward ratio, comparing with batching heuristics augmented with opportunity costs, in Meituan data.}
  \label{fig:meituan_best}
\end{figure}

The \emph{potential-augmented rolling batching} ($\PB-\rbatching$) algorithm is identical to $\rbatching$, except that rewards in the matching problem it solves are adjusted as
    $\Tilde{\reward}(\type_j,\type_k) = \reward(\type_j,\type_k) - \gamma\potential(\type_j)\indicator(j \text{ non-critical})
    - \gamma\potential(\type_k)\indicator(k \text{ non-critical})$,
where $\gamma\in[0,1]$ is a tuning parameter.
Intuitively, we subtract a fraction of the potential from the reward for jobs that are not yet critical.
$\dual-\rbatching$ and $\averagedual-\rbatching$ can be defined analogously, using the hindsight duals $\lambda_j$ and average duals $\Bar{\lambda}_j$ (as defined in \Cref{sec:sim_benchmarks}), respectively.
For each notion of shadow price, we tune $\gamma$ on a subset of the data (orders from an ``island'' which is disjoint from the rest of the city) for each pooling window, and then evaluate performance on the rest of the data (see \Cref{sec:opt_gamma} for details).
\Cref{fig:meituan_best} shows that incorporating shadow-price adjustments to $\rbatching$ yields modest but consistent improvements. 
Notably, $\PB-\rbatching$ achieves a $1$–$2\%$ increase in ratio performance compared to $\rbatching$, essentially at no extra computational cost.
Moreover, \Cref{fig:meituan_gamma_match_rate} shows that the optimal $\gamma$ for $\PB-\rbatching$ increases slowly with the pooling window, and the performance is not sensitive to $\gamma$ around the optimal coefficient.
Additionally, we show in \Cref{fig:meituan_prbat_best_regret} that augmented with potential, with a 180-second pooling window (very reasonable at Meituan, as we showed in \Cref{fig:meituan_order_to_first_dispatch_how}), periodic $\rbatching$ achieves the same outcome as that under full $\rbatching$, but does so at a run time that is two orders of magnitude lower.
These results highlight the potential as a simple yet robust proxy for future value, capable of enhancing the performance of complex policies or substantially reducing computation costs to achieve equivalent outcomes.

\section{Concluding Remarks and Future Work} \label{sec:conclusion}

In this work, we study dynamic non-bipartite matching in the context of delivery pooling.
Taking a half of each job's travel distance as its opportunity cost, the potential-based greedy algorithm ($\PB$) exploits the reward topology of this problem, and strategically holds longer jobs in the system for better future pooling opportunities.
We demonstrate the value of this insight both theoretically, via worst-case regret bounds, and empirically, using synthetic data and real data from the Meituan platform. 
Despite relying on no forecasts, $\PB$ consistently outperforms data-driven greedy heuristics and rivals the performance of computationally-intensive batching policies.
Moreover, the potential can also be incorporated into the batching frameworks to either further improve their performance or drastically reduce computation costs for equivalent outcomes.  
Our approach thus offers a simple, robust, and easy-to implement proxy for opportunity costs. 

We also provide intuition behind its success by showing through theory and experiments that various notions of opportunity cost are intimately related to our definition of potential, especially for markets with sufficient density.

We believe that the main insight of longer-distance jobs being more valuable to hold should apply even when pooling more than two orders is possible (see \citet{wei2023constant}), or when there is no hard deadline to dispatching any order but keeping the customers waiting is costly to the platform.
Moreover, our notion of potential (i.e., the highest possible matching reward) could be relevant on non-metric reward structures, where it is a measure of e.g.\ the popularity of a volunteer position (see \citet{manshadi2022online}).
Formalizing and generalizing these results is an interesting avenue for future research.
%


\ACKNOWLEDGMENT{%
The authors would like to thank
Nick Arnosti,
Itai Ashlagi,
Eric Balkanski,
Omar Besbes,
Gerard Cachon,
Francisco Castro,
Sergey Gitlin,
Alexandre Jacquillat,
Yash Kanoria,
Vahideh Manshadi,
Jake Marcinek,
Rad Niazadeh,
Scott Rodilitz,
Murat Oguz,
Daniela Saban,
Amin Saberi,
Alejandro Torielo,
Alfredo Torrico,
Yehua Wei,
Chiwei Yan and
Sophie Yu
for valuable comments and discussions.
The authors also thank Meituan and the organizers of the INFORMS TSL data-driven research challenge (2024), Hai Wang and Lei Zhao, for organizing the competition and providing the real-world data used in our empirical analysis.
We are also extremely grateful to Yiming He, R\'emi Torracinta, Junxiong Yin, Cinar Kilcioglu from the Uber Eats science team for very insightful discussions and for identifying an error in an initial implementation of our algorithms.}


\bibliographystyle{informs2014} 
\bibliography{DraftBib} 






  



%

\clearpage

\begin{APPENDIX}{}

\noindent{}We provide in \Cref{appx:proofs} proofs that are omitted from \Cref{sec:PB} of the paper.
Theoretical guarantees under alternative reward topologies are stated and proved in \Cref{appx:alternative_reward_topologies}. 
\Cref{sec: interpretation} shows an interpretation of potential in terms of the marginal value of jobs. 
Finally, we include in \Cref{sec:market_dynamics} high-level descriptions of market-dynamics observed in Meituan data, and in \Cref{sec: sim_results_extra} additional simulation results for settings studied in the body of the paper
as well as more general settings with 
non-uniform spatial distributions and different reward topologies.

%
%
\crefalias{section}{appendix}
\crefalias{subsection}{appendix}
\crefalias{subsubsection}{appendix}

\section{Proofs} \label{appx:proofs}


\subsection{Proof of \Cref{prop:greedy_linear_lower_bound}} \label{pf:greedy_linear_lower_bound}

Let $0<\eps<1/3$.
Consider an instance $\instance \in [0,1]^\Njob$ such that $\type_{n/2}<\type_{n/2-1}<\ldots<\type_{1}<\eps$ (low-type jobs) and $\type_{n/2+1}=\type_{n/2+2}=\ldots=\type_{\Njob}=1$ (high-type jobs). 
Note that because of \Cref{deliverygreedy}, when the first job becomes critical, $\gre$ chooses to match with a high-type job, i.e. $\matchof(1)\in \{n/2+1,\ldots,\Njob\}$, collecting $\reward(\type_1,\type_{\matchof(1)})\le \eps$ independent of the actual choice of $\matchof(1)$.
In particular, the second job is not matched and is then the next to become critical.
Since there are still high-type jobs to be matched, then by the same argument $\matchof(2)\in \{n/2+1,\ldots,\Njob\}$ and $\reward(\type_2,\type_{\matchof(2)})\le \eps$.
By induction, since there are as many high-type jobs as low-type jobs, every job $j \in \{1,\ldots,n/2\}$ is matched with $\matchof(j)\in \{n/2+1,\ldots,\Njob\}$ and $\reward(\type_j,\type_{\matchof(j)}) \le \eps$. Therefore, ${\gre}(\instance,\infty) \le \eps n/2$. 

On the other hand, $\reward(\type_{j},\type_{j+1}) = 1$ for all $j \in \{\Njob/2+1,\ldots,\Njob-1\}$. Thus, an optimal matching solution would always match all $\Njob/2$ high-type jobs together, collecting $\OPT(\instance, \infty) \ge \Njob/4$. 
Hence, 
%
\[ 
    \regret_\gre(\instance,\infty) = \OPT(\instance,\infty) - \gre(\instance, \infty) \ge \frac{(1-2\eps)}{4}\Njob, 
\]
independent of the tie-breaking rule. Since this inequality holds for all $0<\eps<1/3$, the proof is complete. \hfill \Halmos

\subsection{Proof of \Cref{prop:loglowerboundOffline}}
\label{pf:loglowerboundOffline}

We construct a sequence of instances $\instance(k)\in [0,1]^{\Njob(k)}$ with $\Njob(k)=2^{k+3} - 4$ for every integer $k\ge 0$, and show that the regret gain at each iteration, $\regret_\PB(\instance(k+1),\infty)-\regret_\PB(\instance(k),\infty)$ is at least $1/4$. For $k=0$, consider $\instance(0)=(1/2,0,1,0)$. We have that $\OPT(\instance(0),\infty) = 1/2$. In turn, under $\PB$, when the first job becomes critical, all available jobs are equally close, resulting in ties. Assume ties are broken so that the resulting matching is $\{(1,2),(3,4)\}$. Then, $\PB(\instance(0),\infty) = 0$, and thus $\regret_\PB(\instance(0),\infty) = 1/2$.
For $k\ge 1$, we inductively construct an instance $\instance(k)$ by carefully adding $2^{k+2}$ new jobs to be processed before the instance considered in the previous step. In particular, we add $2^{k}$ copies of $\instance(0)$, each scaled by $1/2^{k+1}$ and shifted in space so that $\PB$ matches within each copy. 

Formally, for $s\in\{0,\ldots,2^{k}-1\}$ and $r\in\{1,2,3,4\}$, define
\[ \instance(k)_{4s+r} = \frac{\instance(0)_r}{2^{k+1}} + \frac{s}{2^k}. \]
Lastly, define $\instance(k)_j = \instance(k-1)_{j-2^{k+2}} $ for $j \in \{ 2^{k+2}+1,\ldots,  \Njob(k) \}$.
By construction, for each $s\in \{0,\ldots,2^{k}-1\}$, $\PB$ matches job $4s+1$ with $4s+2$ since it is the closest in space (see \Cref{deliverypotential}), and then matches $4s+3$ with $4s+4$ for the same reason. After all these jobs are matched, the remaining jobs represent $\instance(k-1)$ exactly.
On the other hand, $\OPT$ is at least as good as processing $\OPT$ separately on each of the $2^k$ copies, and then on $\instance(k-1)$.
Hence, the regret gain $\regret_\PB(\instance(k),\infty) - \regret_\PB(\instance(k-1),\infty)$ is at least the sum of the regret of $\PB$ on each copy.
Moreover, note the reward function $\reward(\type,\type') = \min\{\type,\type'\}$ satisfies $\reward(\alpha+\type/\beta,\alpha\type'/\beta) = \reward(\type,\type')/\beta$ for any scalars $\alpha,\beta$ such that $\alpha+\type/\beta,\alpha\type'/\beta\in[0,1]$. 
Thus, the regret of $\PB$ on each copy is exactly $\regret_\PB(\instance(0),\infty)/{2^{k+1}}$, and then
$\regret_\PB(\instance(k),\infty) - \regret_\PB(\instance(k-1),\infty) \ge 2^k\frac{\regret_\PB(\instance(0),\infty)}{2^{k+1}} \ge \frac{1}{4}$, 
completing the proof.
\hfill \Halmos

\subsection{Proof of \Cref{prop:greedy_linear_lower_bound_dynamic}}
\label{pf:greedy_linear_lower_bound_dynamic}

Note that if $(\sojourn+1)$ is divisible by 4, then for any number of jobs $\Njob$ divisible by $(\sojourn+1)$, we have an exact partition $b=\Njob/(\sojourn+1)\in\{1,2,\ldots\}$ batches, each of size $(\sojourn+1)$.
We construct an instance $\instance\in[0,1]^\Njob$ that can be analyzed separately on each batch in the following sense.
For each $t=1,\ldots,b$, let $\instance^{(t)} = (\type_{j})_{j\in B_t}\in[0,1]^{(\sojourn+1)}$.
It is straightforward to check that $\OPT(\instance,\sojourn) \ge \sum_{t=1}^b \OPT(\instance^{(t)},\infty) $, since the matching solution induced by the offline instances on the right-hand side is feasible in the online setting.
Thus, if $\instance$ is such that $\gre(\instance,\sojourn) = \sum_{t=1}^b \gre(\instance^{(t)},\infty) $, then
\[ \OPT(\instance,\sojourn) - \gre(\instance,\sojourn) \ge \sum_{t=1}^b \regret_\gre(\instance^{(t)},\infty). \]
Moreover, we construct $\instance$ such that each $\instance^{(t)}\in[0,1]^{(\sojourn+1)}$ resembles the worst-case instance analyzed in \Cref{prop:greedy_linear_lower_bound}.
To this end, let $0<\eps<1$. Consider $\instance\in[0,1]^\Njob$ such that for each $t=0,\ldots,b-1$,
$\eps/2^{t+1}<\type_{(\sojourn+1)t+(\sojourn+1)/2}<\ldots<\type_{(\sojourn+1)t+1}<\eps/2^t$ (low-type jobs) and $\type_{(\sojourn+1)t+(\sojourn+1)/2+1}=\ldots=\type_{(\sojourn+1)(t+1)} = 1$ (high-type jobs).
By construction, when job $1$ becomes critical, the available jobs $k\in A(1)\subseteq B_1$ have types $\type_k < \type_1$ or $\type_k=1$. Then, by \Cref{deliverygreedy}, $\gre$ chooses to match with a high-type job $\matchof(1)\in B_1$.
As job $2$ becomes critical next, the new set of available jobs is $A(2)=A(1)\cup\{(\sojourn+1)+1\}\setminus\{\matchof(1)\}$, with $\type_{(\sojourn+1)+1}<\eps/2<\type_2$. Again, $\gre$ chooses to match with a high-type job $\matchof(2)\in B_2$.
By repeating this argument inductively, $\gre$ outputs $(C,\{\matchof(j)\}_{j\in C})$ such that $C = \bigcup_{t=1}^{b-1}\{(d+1)t + 1,\ldots, (d+1)t + \sojourn+1)t+(\sojourn+1)/2 \}$ and $\matchof(j) \in B_t\setminus C$ for all $j\in C\cap B_t$. 
Thus, $\gre(\instance,\sojourn) = \sum_{t=1}^b \gre(\instance^{(t)},\infty) $, and since $\instance^{(t)}\in[0,1]^{(\sojourn+1)}$ is specified as in the proof of \Cref{prop:greedy_linear_lower_bound}, then  $\regret_\PB(\instance^{(t)},\infty)\le (\sojourn+1)(1-2\eps)/4$. 
Hence, 
\[ \OPT(\instance,\sojourn) - \gre(\instance,\sojourn) \ge b\frac{(\sojourn+1)(1-2\eps)}{4} = \frac{\Njob}{4}(1-2\eps), \]
completing the proof.
\hfill \Halmos

\subsection{Proof of \Cref{prop:loglowerboundOnline}}
\label{pf:loglowerboundOnline}

Since $\Njob$ is divisible by $(\sojourn+1)$, we have an exact partition of the set of jobs into $b=n/(\sojourn+1)\in\{1,2,\ldots\}$ batches. 
As in \Cref{prop:greedy_linear_lower_bound_dynamic}, we construct an instance $\instance\in[0,1]^\Njob$ that can be analyzed separately on each batch, where we denote $\instance^{(t)}=(\type_j)_{j\in B_t}$ for $t=1,\ldots,b$.
By \Cref{prop:loglowerboundOffline}, there exists an instance $\instance^{(0)}\in[0,1]^{(\sojourn+1)}$ such that $\regret_\PB(\instance^{(0)},\infty) \ge (\log(\sojourn+5)-3)/4$.
Then, for each $t=1,\ldots,b$, let $\instance^{(t)}=\instance^{(0)}\cdot\frac13$ if $t$ is odd and $\instance^{(t)}=\instance^{(0)}\cdot\frac13+\frac23$ if $t$ is even.
Note that since each batch size is $\sojourn+1$ divisible by 4, it is possible to match within batch only. Moreover, by construction, every critical job has a matching candidate within batch, which is closer in space than any job on the next batch. Thus, because of \Cref{deliverypotential}, $\PB$ only matches within batch, and therefore $\PB(\instance,\sojourn) = \sum_{t=1}^b \PB(\instance^{(t)},\infty)$.
As in \Cref{prop:greedy_linear_lower_bound_dynamic}, we arrive at
$ \OPT(\instance,\sojourn) - \PB(\instance,\sojourn) \ge \sum_{t=1}^b \regret_\PB(\instance^{(t)},\infty) = b\frac{ \regret_\PB(\instance^{(0)},\infty)}{3} \ge \frac{\Njob}{3(\sojourn+1)}(\log(\sojourn+5)-3)/4, $
concluding the proof.
\hfill \Halmos

\section{Algorithmic Performance under Alternative Reward Topologies} \label{appx:alternative_reward_topologies}

In this section, we study the performance of both the naive greedy algorithm $\gre$ and potential-based greedy algorithm $\PB$, under the two alternative reward topologies defined in \eqref{eq:defn_reward_B} and \eqref{eq:defn_reward_C}.

\subsection{Reward Function $\reward(\type,\type')=1-|\type-\type'|$}
\phantomsection
\label{sec:reward2}
This reward function represents the goal to minimize the total distance between matched jobs.
In this case, for any $\type\in[0,1]$, the potential of a job of type $\type$ is $\potential(\type)=1/2$ constant across job types. Thus, the description of $\PB$ is equivalent to $\gre$, since their index functions differ only in an additive constant.

\begin{remark}\label{rem:rewardB}
    Under the 1-dimensional type space $\typespace=[0,1]$ and the reward function $\reward(\type,\type')=1-|\type-\type'|$, both the naive greedy and potential-based greedy algorithm generate the same output $(C,\matchof(\cdot))$ as described in \Cref{alg:dynamic}. In particular, both algorithms always match each critical job to an available job that is the closest in space.
\end{remark}

Moreover, we show that their performance can be reduced to the one of $\PB$ under reward function $\reward(\type,\type')=\min\{\type,\type'\}$.
In particular, we first show that when $\sojourn=\infty$, the regret is at least logarithmic in the number of jobs.
\begin{proposition}\label{prop:loglowerboundOffline_rewardB}     
    Under reward function $\reward(\type,\type')=1-|\type-\type'|$, if $\Njob=2^{k+3}-4$ for some integer $k\ge0$, then there exists an instance $\instance\in[0,1]^\Njob$ for which $\regret_\PB(\instance,\infty) = \regret_\gre(\instance,\infty) \ge (\log_2(\Njob+4)-3)/2 $.
\end{proposition}

\proof{Proof.}
    We prove the statement for $\gre$, since $\regret_\PB(\instance,\infty) = \regret_\gre(\instance,\infty)$ follows from \Cref{rem:rewardB}.
    Let $\Njob=2^{k+3}-4$, for $k\in\{0,1,\ldots\}$. From \Cref{prop:loglowerboundOffline}, there exists an instance $\instance\in[0,1]^\Njob$ such that the regret of $\PB$ under 
    $\reward(\type,\type')=\min\{\type,\type'\}$, which we denote by $R$ from here on, is at least $(\log_2(\Njob+4)-3)/4$.
    We show that under $\reward(\type,\type')=1-|\type-\type'|$, we have $\regret_\gre(\instance,\infty) \ge 2R$.
    First, note that since $|\type-\type'| = \type + \type' - 2\min\{\type,\type'\}$, we have 
    %
    \[ \gre(\instance,\infty) 
         = \sum_{j \in C} 1 - |\type_j-\type_{m(j)}|  
         = |C| - \sum_{j\in C} (\type_j +\type_{\matchof(j)})  + 2\sum_{j\in C} \min\{\type,\type'\}. 
    \]
    Similarly, recall that $\matchset_\OPT$ is the set of matches in the hindsight optimal solution. Then, 
    \[ \OPT(\instance,\infty) = |\matchset_\OPT| - \sum_{ (j,k)\in\matchset_\OPT} (\type_j +\type_{k})  + 2\sum_{(j,k)\in\matchset_\OPT} \min\{\type_j,\type_{k}\}. \]
    Since $\reward(\type,\type')\ge 0$ for any $\type,\type'\in \typespace$ and the total number of jobs $\Njob$ is even, both $\gre$ and $\OPT$ match every job. Hence, $|C|=|\matchset_\OPT|=n/2$, and moreover 
    \[ \sum_{(j,k)\in\matchset_\OPT} (\type_j +\type_{k}) = \sum_{j\in C} (\type_j +\type_{\matchof(j)}). \]
    Thus, 
    \begin{align*}
    &\OPT(\instance,\infty) - \gre(\instance,\infty) \\ 
    &\qquad \ge 2\left( \sum_{(j,k)\in\matchset_\OPT} \min\{\type_j,\type_{k}\} - \sum_{j\in C} \min\{\type_j,\type_{\matchof(j)}\} \right).
    \end{align*}
    We claim that the term in large parenthesis is exactly the regret of $\PB$ under reward topology $\reward(\type,\type')=\min\{\type,\type'\}$, which we denote by $R$.
    Indeed, from \Cref{rem:rewardB}, $(C,\matchof(\cdot))$ coincides with the resulting matching of $\PB$ under $\reward(\type,\type')=\min\{\type,\type'\}$, since it is specified to always match to the closest job (\Cref{deliverypotential}).
    On the other hand, since all jobs are matched, the set of matches $\matchset_\OPT$ must induce disjoint intervals, which coincides with the optimal matching solution under reward function $\reward(\type,\type') = \min\{\type,\type'\}$.
    Thus, by \Cref{prop:loglowerboundOffline}, we get $ \regret_\gre(\instance,\infty) \ge 2R \ge (\log_2(\Njob+4)-3)/2$, completing the proof.
\Halmos\endproof

We now extend this result to the online setting ($\sojourn<n$).

\begin{proposition} \label{prop:loglowerboundOnline_rewardB}
    Under reward topology $\reward(\type,\type') = 1-|\type-\type'|$, if $\sojourn+1 = 2^{k+3}-4$ for some integer $k\ge 0$, then for any number of jobs $\Njob$ divisible by $(\sojourn+1)$, there exists an instance $\instance \in [0,1]^\Njob$ for which $\regret_\PB(\instance,\sojourn) = \regret_\gre(\instance,\sojourn) \ge \frac{\Njob}{3(\sojourn+1)}(\log_2(\sojourn+5)-3)/2$.
\end{proposition}

\proof{Proof.}
    This proof is analogous to the proof of \Cref{prop:loglowerboundOnline}, since (i) there exists an offline instance that achieves the desired regret for $\sojourn+1=n$ (\Cref{prop:loglowerboundOffline_rewardB}), and (ii) the algorithm matches a critical job with the closest available job (\Cref{rem:rewardB}).
    \hfill \Halmos\endproof

Finally, we directly show the tight regret upper bound (up to constants) as an immediate consequence of the proof of \Cref{thm: dynamic}. Note that if $d+1=n$, then this yields the result for offline matching.

\begin{theorem}\label{thm:rewardB_dynamic}
Under reward topology $\reward(\type,\type') = 1-|\type-\type'|$, we have $\regret_\PB(\instance,\sojourn) = \regret_\gre(\instance,\sojourn) \le 1/2 + (\frac{\Njob}{d+1}+1)(1+\log(\sojourn+2))$ for any $\Njob$, and any instance $\instance\in[0,1]^\Njob$.
\end{theorem}

\proof{Proof.}
Let $\sojourn\ge 1$ and $(C,\matchof(\cdot))$ be the output of $\gre$ on $\instance$. Since $\potential(\type)=1/2$, for all $\type$, we have
\vspace{-0.5em}
\begin{align*}
\OPT(\instance,\sojourn)-\gre(\instance,\sojourn) 
& \le \sum_{j=1}^\Njob \potential(\type_j) - \gre(\instance,\sojourn)  \nonumber \\
& \le
\frac12 + \sum_{j \in C} |\type_j-\type_{\matchof(j)}|, 
\end{align*}
The second term is exactly the sum of distances between jobs matched by $\gre$. Recall that $(C,\matchof(\cdot))$ coincides with the matching output of $\PB$ under $\reward(\type,\type')=\min\{\type,\type'\}$, both are specified to always match with the closest job (\Cref{deliverypotential}, \Cref{rem:rewardB}). Thus, recalling \eqref{eq: distance_dynamic} in the proof of \Cref{thm: dynamic}, we have 
\[ \sum_{j \in C} |\type_j-\type_{\matchof(j)}| = \sum_{t=1}^b \sum_{j \in C\cap B_t} |\type_j-\type_{\matchof(j)}|\le \left(\frac{n}{d+1}+1 \right)\left(1+\log(d+2)\right),\]
and the result follows.
\hfill \Halmos\endproof

\subsection{Reward Function $\reward(\type,\type')=|\type-\type'|$}\label{sec: reward 3}
Under this reward topology, it is worst to match two jobs of the same type, contrasting the two settings studied in \Cref{sec:PB} and \Cref{sec:reward2}.
In this case, the potential of a job type $\type\in[0,1]$ is $\potential(\type)=\max\{\type,1-\type\}/2$.
We first show that \emph{any} index-based matching policy must suffer regret constant regret per job in the offline setting.
\begin{proposition}\label{prop:linearLB_offline_rewardC}
    Under reward topology $\reward(\type,\type')=|\type-\type'|$, when the number of jobs $\Njob$ is divisible by 4, for any index-based greedy matching algorithm $\ALG$, there exists an instance $\instance\in[0,1]^\Njob$ for which $\regret_\ALG(\instance,\infty)\ge c_\ALG \Njob$, for some $c_\ALG\in (0,1)$. In particular, $c_\gre = c_\PB = 1/4$.
\end{proposition}

\proof{Proof.}
Let $\Njob$ be divisible by 4.
Let $\ALG$ be an index-based greedy matching algorithm specified by index function $\indexf:[0,1]^2\to \R$.
Define $\type^{c} = \sup\{ \type\in[0,1] : \indexf(\type,1)\ge \indexf(\type,0) \}$.
We first construct an instance in the case $\type^{c}>0$ (e.g. $\type^{c}=1/2$ for $\gre$ and $\PB$), and analyze the alternative case separately in the end.
Consider an instance $\instance\in\typespace^\Njob$ such that $\type_1,\ldots,\type_{\Njob/4} = \type^{c}$, $\type_{{\Njob/4}+1},\ldots,\type_{{3\Njob/4}}=0$, and $\type_{{3\Njob/4}+1},\ldots,\type_{\Njob}=1$. 
It is straightforward to check that $\OPT(\instance,\infty) \ge (1+\type^c){\Njob/4}$.
On the other hand, $\ALG$ chooses to match every $j=1,\ldots,n/4$ to $\matchof(j)\in \{3n/4+1,\ldots,n\}$. In the presence of ties, such an output is achieved by some tie-breaking rule.
Then,
\[ \ALG(\instance,\infty) = 0 + \sum_{j=1}^{\Njob/4} \reward(\type_j,\type_{3\Njob/4+j}) = \frac{\Njob(1-\type^c)}{4}. \]
and thus $\regret_\ALG(\instance,\infty) = \OPT(\instance,\infty) - \ALG(\instance,\infty) \ge {{\Njob\type^{c}}}/{2},$ proving the statement with $c_\ALG = \type^{c}/2$.

Now, if $\type^{c}\le 0$, then $\indexf(\type,0) < \indexf(\type,1)$ for all $\type\in(0,1]$.
In particular, for any $\eps>0$, consider the instance $\type_1,\ldots,\type_{\Njob/4} = \eps$, $\type_{{\Njob/4}+1},\ldots,\type_{{3\Njob/4}}=1$, and $\type_{{3\Njob/4}+1},\ldots,\type_{\Njob}=0$.
Then, $\OPT(\instance,\infty) \ge (1+(1-\eps))n/4$, and $\ALG(\instance,\infty) = \eps n/4$.
Hence, $\regret_\ALG(\instance,\infty)\ge (1-\eps)/2$, and the statement is true with $c_\ALG = (1-\eps)/2$ for every $\eps>0$.
\Halmos\endproof

We now show that under $\PB$ and $\gre$, this lower bound extends to the online setting, losing a factor of 2.

\begin{proposition}\label{prop:linearLB_dynamic_rewardC}
    Under reward topology $\reward(\type,\type')=|\type-\type'|$, if $(\sojourn+1)$ is divisible by 4, then for any number of jobs $\Njob$ divisible by $2(\sojourn+1)$, there exists an instance $\instance \in [0,1]^\Njob$ for which $\regret_\gre(\instance,\sojourn)=\regret_\PB(\instance,\sojourn) \ge n/8$.
\end{proposition}

\proof{Proof.}
If $\Njob$ is divisible by $2(\sojourn+1)$, we have an exact partition of the set of jobs into $b=\Njob/(\sojourn+1)\in\{2,4,\ldots\}$ batches, defined as $B_{t+1} = \{t(\sojourn+1)+1,\ldots, (t+1)(\sojourn+1)\}$ for $t = 0, \ldots, b-1$.
We construct an instance $\instance\in[0,1]^\Njob$ for which $\gre$ (and $\PB$) ``resets'' every two batches, in the sense that all jobs in these two batches are matched among them.
To formally specify the construction, first fix $\eps\in(0,1/2)$. Define, for $t=0,\ldots,b/2-1$,
\[
\type_{2t(\sojourn+1)+j} =
\begin{cases}
\frac12,&j \le \frac{\sojourn+1}{4}, \\
\eps,&  \frac{\sojourn+1}{4} < j \le \frac{3(\sojourn+1)}{4}, \\
0,& \sojourn+1 < j \le \frac{3(\sojourn+1)}{2}, \\
1,&  \text{ otherwise }.
\end{cases}
\]
Consider the notation $\instance^{(0)}=(\type_j)_{j\le 2(\sojourn+1)}$.
It is easy to check that
\[ 
\OPT(\instance^{(0)},\infty) = \left(\frac12-\eps\right)\frac{\sojourn+1}{4} + (1-\eps)\frac{\sojourn+1}{4} + \frac{\sojourn+1}{2} =  \left(\frac{7-4\eps}{8}\right)(\sojourn+1),
\]
and then $\OPT(\instance,\sojourn) \ge \frac{b}{2}\OPT(\instance^{(0)},\infty) = \frac{7-4\eps}{16}\Njob $.
On the other hand, 
\[ 
\gre(\instance^{(0)},\infty) = \frac12\frac{\sojourn+1}{4} + \eps\frac{\sojourn+1}{4} + (1-\eps)\frac{\sojourn+1}{4} + \frac{\sojourn+1}{4} =  \frac58(\sojourn+1).
\]
We claim that $\gre(\instance,\sojourn) = \frac{b}{2}\gre(\instance^{(0)},\infty) = \frac{5}{16}\Njob $, and thus $\regret_\gre(\instance,\sojourn)\ge \frac{1-2\eps}{8}\Njob $.
Indeed, let $(C,\matchof(\cdot))$ be the output of $\gre$. Running $\gre$ on $\instance$ is as follows:
\begin{enumerate}
    \item For every job $j \le (\sojourn+1)/4$ of type $\type_j=1/2$, and then $3(\sojourn+1)/4 < \matchof(j) \le (\sojourn+1)$ with $\reward(\type_j,\type_{\matchof(j)})=1/2$.
    \item Then, for every job $\frac{\sojourn+1}{4} < j \le \frac{(\sojourn+1)}{2}$, the only available jobs are of type $\type\in\{0,\eps\}$, and then $ \sojourn+1 < \matchof(j) \le \frac{3(\sojourn+1)}{2}$ with $\reward(\type_j,\type_{\matchof(j)})=\eps$. 
    \item In turn, jobs $\frac{(\sojourn+1)}{2} < j \le \frac{3(\sojourn+1)}{4}$ get to observe jobs of type $\type=1$, and then $ \frac{3(\sojourn+1)}{2} < \matchof(j) \le 2(\sojourn+1)$ with $\reward(\type_j,\type_{\matchof(j)})=1-\eps$.
    \item Finally, the key observation is that the all remaining available jobs $j\le \frac{3(\sojourn+1)}{2}$ of type $\type_j=0$ observe jobs of type $1$, as well as jobs on a ``third'' batch. However, by construction every job $2(\sojourn+1) < k \le 2(\sojourn+1) + \frac{3(\sojourn+1)}{2} $ are of type $\type_k\in\{1/2,\eps\}$, and then each job $j\le \frac{3(\sojourn+1)}{2}$ is matched (within batch) to $ \frac{3(\sojourn+1)}{2} < \matchof(j) \le 2(\sojourn+1)$ with $\reward(\type_j,\type_{\matchof(j)})=1$.
\end{enumerate}
The proof for $\PB$ is analogous: the matching decisions coincide with $\gre$ on this instance. 
%
%
\Halmos\endproof


\section{Interpretation of Potential}\label{sec: interpretation}

When a job $j$ becomes critical, our potential-based greedy algorithm matches it to an available job $k$ maximizing $r(\theta_j,\theta_k)-p(\theta_k)$, where $p(\theta_k) = \frac12 \sup_{\theta\in\Theta} r(\theta_k,\theta)$ can be interpreted as the opportunity cost of matching job $k$.
This is an optimistic measure of opportunity cost because it assumes that job $k$ would otherwise be matched to an ``ideal'' type $\theta\in\Theta$ maximizing $r(\theta_k,\theta)$ (with half of this ideal reward $\sup_{\theta\in\Theta} r(\theta_k,\theta)$ attributed to job $k$).
We now prove that this ideal reward can indeed be achieved under asymptotically-large market thickness, for a stochastic model under our topology of interest.

\begin{definition}
Let $\instance \in \typespace^\Njob$.
For any job $j\in[\Njob]$, let $\instance^{-j}\in \typespace^{\Njob-1}$ be the same instance with the exception that job $j$ is not present. Meanwhile, let $\instance^{+j}\in \typespace^{\Njob+1}$ be the same instance with the exception that an additional copy of job $j$ is present.
Consider the following definitions.
\begin{enumerate}
\item Marginal Loss: $\marginalloss_j(\instance) = \OPT(\instance) - \OPT(\instance^{-j})$
\item Marginal Gain: $\marginalgain_j(\instance) = \OPT(\instance^{+j}) - \OPT(\instance)$
\end{enumerate}
\end{definition}

Note that definitions $\marginalloss_j(\instance),\marginalgain_j(\instance)$ are based solely on offline matching, and we will use them as our definitions of opportunity cost if the future was known.  One could alternatively use shadow prices from the LP relaxation of the offline matching problem, but we note that the LP is not integral.  In either case, there is no ideal definition of opportunity cost that is guaranteed to lead to the optimal offline solution in matching problems \citep[see][]{cohen2016invisible}.

We now establish the following \namecref{lem:marginal_as_interval} to help analyze the opportunity costs $\marginalloss_j(\instance),\marginalgain_j(\instance)$.

\begin{lemma}
\label{lem:marginal_as_interval}
Let $\typespace=[0,1]$ and $\reward(\type,\type') = \min\{\type,\type'\}$. Consider an instance $\instance\in\typespace^n$ and re-label the indices to satisfy $\type_1 \ge \type_2 \ge \ldots \ge \type_n$. Then,
$\marginalloss_j(\instance)
=\sum_{k\ge j,k\ \mathrm{even}}(\theta_k-\theta_{k+1})$, and $\marginalgain_j(\instance)
=\sum_{k\ge j,k\ \mathrm{odd}}(\theta_k-\theta_{k+1}),$
where we consider $\theta_{n+1}=0$.
\end{lemma}
\proof{Proof.}
    Let $\instance \in \typespace^\Njob$ such that $\type_1 \ge \type_2 \ge \ldots \ge \type_n$. 
    Then, $\OPT(\instance) =\sum_{k\ \mathrm{even}}\theta_k$, and moreover 
    \begin{align*}
    \OPT(\instance^{-j}) &=\sum_{k<j, k\ \mathrm{even}}\theta_k+\sum_{k>j, k\ \mathrm{odd}}\theta_k
    \\ \OPT(\instance^{+j}) &=\theta_j+\OPT(\instance^{-j})
    \end{align*}
    Therefore, 
    %
    \begin{align*}
    \marginalloss_j(\instance)
    &=\sum_{k\ge j,k\ \mathrm{even}}\theta_k
    -\sum_{k>j,k\ \mathrm{odd}}\theta_k
    =\sum_{k\ge j,k\ \mathrm{even}}(\theta_k-\theta_{k+1})
    \\ \marginalgain_j(\instance)
    &=\theta_j-\sum_{k\ge j,k\ \mathrm{even}}(\theta_k-\theta_{k+1})
    =\sum_{k\ge j,k\ \mathrm{odd}}(\theta_k-\theta_{k+1})
    \end{align*}
    completing the proof.
\Halmos\endproof

Note that because $\OPT(\instance^{+j}) = \OPT(\instance^{-j}) + \type_j$ and $\potential(\type_j) = \reward(\type_j,\type_j)/2$ for this reward function, we immediately get the following \namecref{cor: marginal_average}.

\begin{corollary}\label{cor: marginal_average}
        If $\typespace=[0,1]$ and $\reward(\type,\type') = \min\{\type,\type'\}$, then for all jobs $j$,
    \[ \potential(\type_j) = \frac{\marginalloss_j(\instance) + \marginalgain_j(\instance)}{2}. \]
\end{corollary}

We are now ready to state our main result about the interpretation of potential, that the true opportunity costs $\marginalloss_j(\instance),\marginalgain_j(\instance)$ concentrate around $p(\theta_j)$ in a random uniform instance, assuming the market is sufficiently thick.  We without loss consider job $j=1$ and fix its type $\theta_1$.

\begin{proposition}
\label{thm:interpretation}
Let $\typespace=[0,1]$ and $\reward(\type,\type') = \min\{\type,\type'\}$.
Fix $\theta_1\in\typespace$ and suppose $\theta_2,\ldots,\theta_n$ are drawn IID from the uniform distribution over $[0,1]$, forming a random instance $\instance\in\typespace^n$, for some $n\ge 2$. Then, both $\mathbb{E}[\marginalloss_1(\instance)]$ and $\mathbb{E}[\marginalgain_1(\instance)]$ are within $O\left(\frac{1}{n}\right)$ of $\potential(\type_1)$ and moreover $\mathrm{Var}(\marginalloss_1(\instance))=\mathrm{Var}(\marginalgain_1(\instance)) = O\left(\frac{1}{n}\right)$.
\end{proposition}
\proof{Proof.}
We prove the statement only for $\marginalloss_1(\instance)$, and the analogous result for $\marginalgain_1(\instance)$ follows from \Cref{cor: marginal_average}.
If $\type_1=0$, then $\marginalloss_1(\instance)=0$ for any $\instance$, coinciding with $\potential(\type_1)=0$. Then, for the remainder of the proof, assume that $\type_1>0$.
Consider the random variable $N=|\{j:\theta_j<\theta_1\}|$, which counts the number of points between $0$ and $\theta_1$ and has distribution $\text{Binom}(\Njob-1,\type_1)$.
These $N$ points divide the interval $[0,\theta_1]$ into $N+1$ intervals with total length $\type_1$.
From \Cref{lem:marginal_as_interval}, the marginal loss $\marginalloss_1(\instance)$ is determined by computing the total length of a subset of these intervals. 
It is known that if $N$ random variables are drawn independently from $\text{Unif}[0,1]$, then the joint distribution of the induced interval lengths is $\text{Dirichlet}(1,1,\ldots,1)$ with $N+1$ parameters all equal to 1 (i.e., drawn uniformly from the simplex).
In particular, since the Dirichlet distribution is symmetric, the sum of any $k$ of these intervals is equal in distribution to the $k$-th smallest sample ($k$-th order statistic) of the $N$ uniform random variables, whose distribution is known to be $\text{Beta}(k,N+1-k)$.
Thus, conditional on $N$, with $N\ge 1$, since the distribution of each of the $N$ jobs to the left of $\type_1$ is $\text{Unif}[0,\type_1]$, the distribution of $\marginalloss_1(\instance)/\type_1$ is $\text{Beta}(k_L,N+1-k_L)$,
where $k_L$ is either $\lceil\frac{N+1}2\rceil$ or $\floor{\frac{N+1}2}$.
To be precise, for $N\ge 1$, $k_L=\floor {N/2}+1=\lceil\frac{N+1}2\rceil$ if $n$ is even, and $k_L=\ceil{N/2}=\floor{\frac{N+1}2}$ if $n$ is odd.
Moreover, if $N=0$, then $\marginalloss_1(\instance)=\type_1$ if $n$ is even, and $\marginalgain_1(\instance)=0$ if $n$ is odd.
Hence, 
$\mathbb{E}[\marginalloss_1(\type) \mid N] = \type_1\frac{k_L}{N+1}$.
for all $N\in\{0,\ldots,n-1\}$.
Since $|k_L-(N+1)/2|\le 1$, then $|\mathbb{E}[\marginalloss_1(\type) \mid N]-\type_1/2|\le \type_1/(N+1)$, thus

\begin{align*}
&\left|\mathbb{E}[\marginalloss_1(\type)]-\frac{\type_1}{2}\right| 
\le \type_1\mathbb{E}\left[\frac{1}{N+1}\right] \\
&\qquad = \type_1\sum_{k=0}^{n-1} \frac{1}{k+1}\binom{n-1}{k}\type_1^k(1-\type_1)^{n-1-k} \\
&\qquad = \frac{\type_1}{n}\sum_{k=0}^{n-1} \binom{n}{k+1}\type_1^k(1-\type_1)^{n-1-k} \\
&\qquad = \frac{1}{n}\sum_{k=1}^{n} \binom{n}{k}\type_1^k(1-\type_1)^{n-k} \\
&\qquad = \frac{1-(1-\type_1)^{n}}{n} = O\left(\frac{1}{n}\right).
\end{align*}
This completes the proof of the statement about $\mathbb{E}[\marginalloss_1(\instance)]$.

For the statement about variance, we know $\mathrm{Var}(\marginalloss_1(\type)) = \mathbb{E}[\mathrm{Var}(\marginalloss_1(\type) \mid N)] + \mathrm{Var}(\mathbb{E}[\marginalloss_1(\type)\mid N])$ by the law of total variance. For the first term, we have $\mathrm{Var}(\marginalloss_1(\type) \mid N) = \type_1^2\frac{k_L(N+1-k_L)}{(N+1)^2(N+2)}\indicator\{N\ge 1\}\le \frac{\type_1^2}{N+1}$, and then from the previous argument $\mathbb{E}[\mathrm{Var}(\marginalloss_1(\type) \mid N)]=O(1/n)$.
For the second term, 
\begin{align*}
\mathrm{Var}(\mathbb{E}[\marginalloss_1(\type)\mid N]) 
&= \mathrm{Var}\left(\mathbb{E}[\marginalloss_1(\type)\mid N] - \frac{\type_1}{2}\right) \\
&\le \mathbb{E} \left[\left(\mathbb{E}[\marginalloss_1(\type)\mid N] - \frac{\type_1}{2}\right)^2\right] \\
&\le \mathbb{E}\left[\frac{\type_1^2}{(N+1)^2}\right] \\
&\le \type_1^2\mathbb{E}\left[\frac{1}{N+1}\right] = O\left(\frac{1}{n}\right).
\end{align*}
Thus, $\mathrm{Var}(\marginalloss_1(\instance)) = O\left(1/n\right)$, completing the proof.
\Halmos\endproof

\subsection{Linear Relaxation and Dual Variables}\label{sec: LP}

\begin{figure}
  \centering
  \subcaptionbox{Average Total Reward.%
    \label{fig:reward}}[0.49 \textwidth]{\includegraphics[width = \figWidth \textwidth]{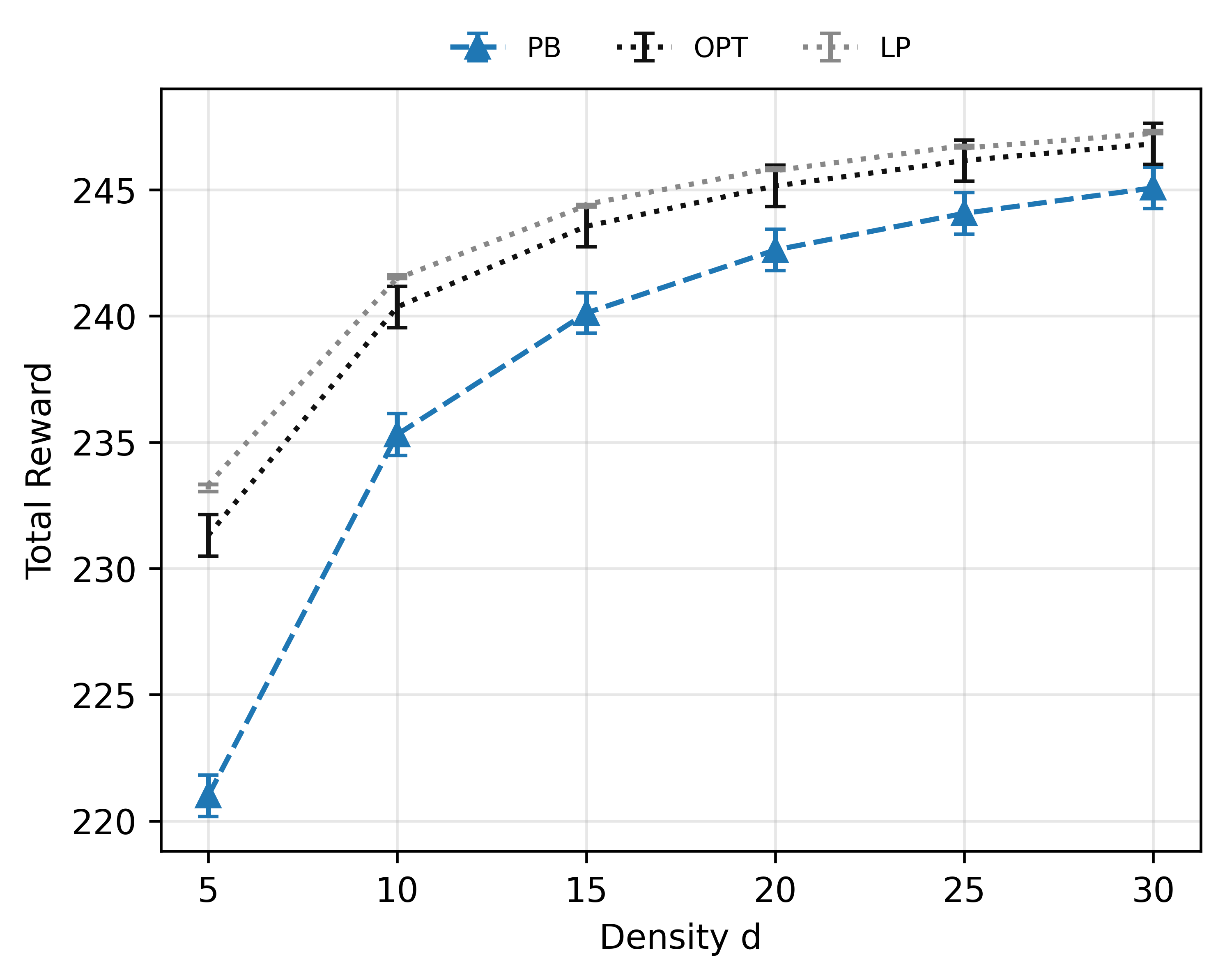}}
  \hfill
  \subcaptionbox{Fraction of LP value achieved by $\OPT$.%
    \label{fig:LP_Gap}}[0.49 \textwidth]{\includegraphics[width = \figWidth \textwidth]{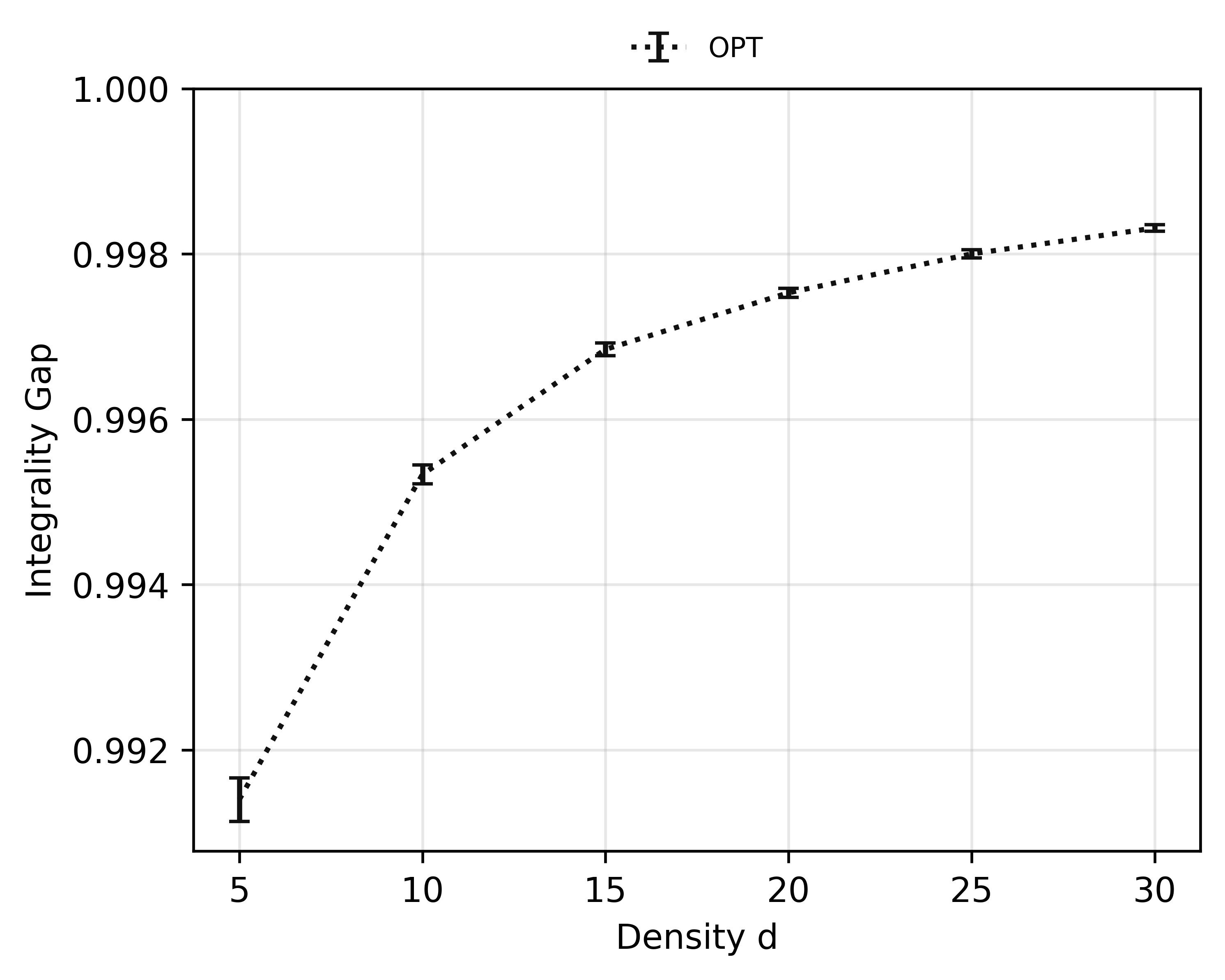}}
  \caption{Comparison with LP value.
  }
  \label{fig: LP_gap}
\end{figure}

In this section, we provide details on the LP formulation and dual variables used for the $\dual$ and $\averagedual$ benchmarks. 
Given market density $\sojourn\ge 1$ and full information of the sequence of arrivals $\instance\in\typespace$, we can define a linear program from (\ref{eq: OPT}) by relaxing the integrality constraints.
It is known that the LP relaxation is integral only for bipartite graphs. However, we observe in \Cref{fig: LP_gap} that the integrality gap is small, and thus the value of the LP can be reasonably used as a benchmark.

Also, note that the potential can always be used as a feasible (but not necessarily optimal) solution to the dual LP
%
%
%
\begin{mini}
{\lambda}{ \sum_{j\in[\Njob]} \lambda_{j} }
{\label{eq: Dual-LP}}{}
\addConstraint{ \lambda_j + \lambda_k}{ \ge \reward(\type_j,\type_k),\quad}{ j\neq k, |j-k|\le \sojourn}
\addConstraint{ \lambda_{j} }{\ge 0 ,}{ j\in [\Njob].}
\end{mini}

Indeed, $\reward\ge 0$ and $\reward(\type,\type')\le \frac{1}{2}\reward(\type,\type')+\frac{1}{2}\reward(\type,\type') \le \potential(\type)+\potential(\type')$, for any $j,k$, verifying the dual feasibility constraints.



When we extract the shadow prices from the 400 instances 
in the historical data in \Cref{sec:sim_unif_1D}, we can see that these concentrate around potential, as density increases (\Cref{fig: dual_converge}). This is aligned with our results in \Cref{sec: interpretation}.

\newcommand{\scatterplotWidth}{0.45}

\begin{figure}
  \centering
  \subcaptionbox{Density $\sojourn=5$.%
    \label{fig:dual_conv_d5}}[0.4 \textwidth]{\includegraphics[width = \scatterplotWidth \textwidth]{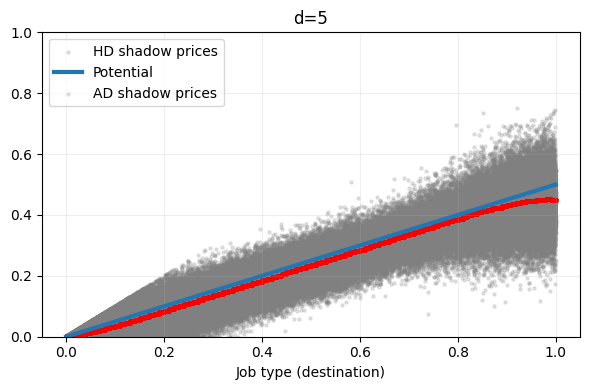}}
  \hspace{2em}
  \subcaptionbox{Density $\sojourn=30$.%
    \label{fig:dual_conv_d30}}[0.4 \textwidth]{\includegraphics[width = \scatterplotWidth \textwidth]{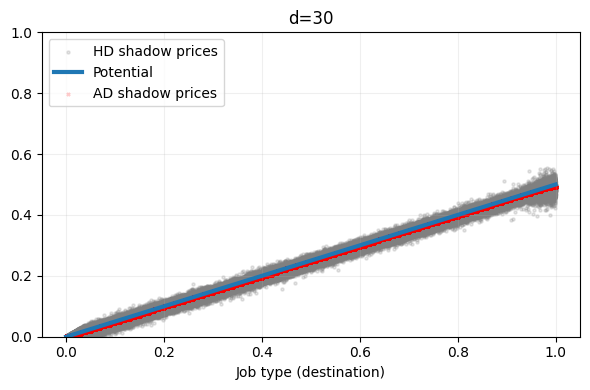}}
  \caption{Scatter plot of shadow prices, average, and potential.
  }
  \label{fig: dual_converge}
\end{figure}

\section{Meituan Data - Market Dynamics} \label{sec:market_dynamics}

In this appendix, we briefly describe the Meituan dataset made public by the INFORMS TSL Data-Driven Challenge, and provide high-level descriptions of the market-dynamics.

\subsection{Temporal Dynamics} \label{appx:meituan_temporal_dynamics}

\Cref{fig:meituan_orders_per_hour} provides the average number of orders per hour by \emph{hour-of-week} (HOW). 
HOW 0 corresponds to midnight to 1am on Mondays, and 
HOW 1 corresponds to 1am to 2am on Mondays, and so on. 
The gray bars indicate the peak lunch hours (10:30am-1:30pm); 
the green bars indicate peak dinner hours (5pm-8pm).
We can see that the platform observes highest order volume during lunch time, averaging around 200 order per minute between noon and 1pm.

\newcommand{\FigWidthHOW}{0.85}

\begin{figure}
    \centering
    \includegraphics[width=\FigWidthHOW\linewidth]{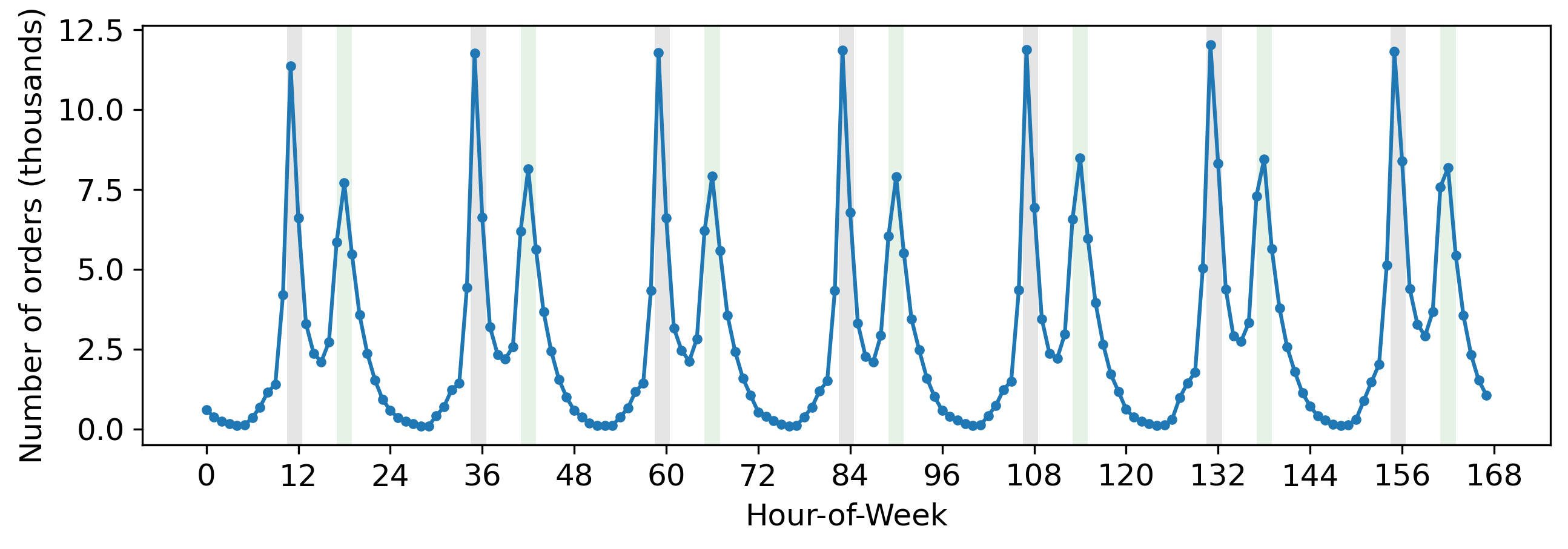}
    \vspace{-0.5em}
    \caption{Average number of orders by hour-of-week in Meituan data.}
    \label{fig:meituan_orders_per_hour}
\end{figure}

\Cref{fig:meituan_pooled_orders_per_hour} in \Cref{sec:intro} 
illustrates the fraction of orders placed during each hour-of-week that is pooled with at least one other order. 
This is computed by combining the order-level data (which contains the timestamp of each order and the order ID) and the ``wave level'' data, where each wave corresponds to a list of orders that were pooled and assigned to the same driver \citep{tsl_meituan_2024}. Each wave may contain more than two orders, though some orders may have been dropped off by the driver before some other orders were picked up.   
The orders that are not pooled with any other order correspond to waves with only one order, and \Cref{fig:meituan_pooled_orders_per_hour} illustrates 1 minus this fraction of orders that are not pooled.



Finally, \Cref{fig:meituan_order_to_first_dispatch_how} illustrates the amount of time that elapses, between platform\_order\_time, i.e. when each order is placed by the customer, and first dispatch\_time, i.e. when the order was offered to a delivery driver for the first time.
\begin{figure}
    \centering    \includegraphics[width=\FigWidthHOW\linewidth]{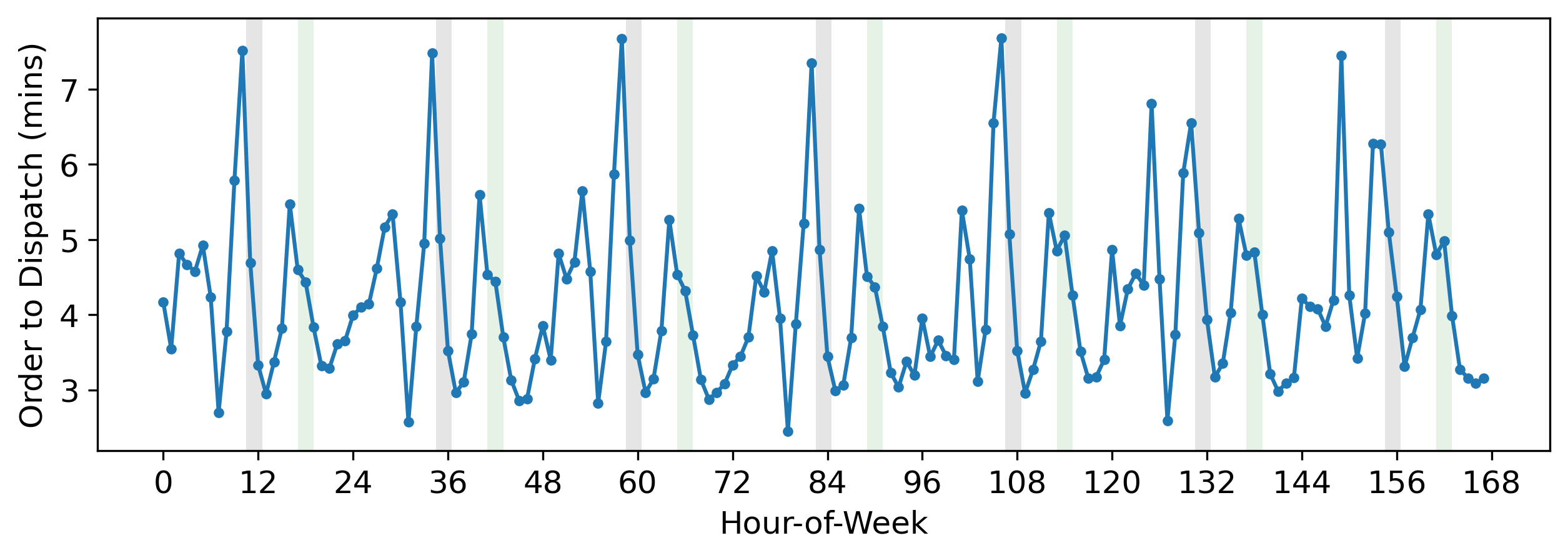}
    \vspace{-0.5em}
    \caption{
    The average amount of time (minutes) that elapses between (i) when an order is placed by the customer, and (ii) when the order is dispatched to a delivery driver for the first time.
    }
    \label{fig:meituan_order_to_first_dispatch_how}
\end{figure}
Intuitively, this is the amount of time it takes the platform to start dispatching each order, and the average is around 4 minutes across all orders. 
Despite the fact that it takes an estimated $12$ minutes for the restaurant to prepare the orders, the platform starts to dispatch the orders shortly after they are placed. This is potentially due to the fact that it takes time for drivers to get to the restaurants 
, so orders are dispatched before they are ready in order to reduce the total wait time experienced by the customers.

\subsection{Spatial Distribution} \label{appx:meituan_spatial_distribution}

We now visualize the distribution of orders in space, and provide the distribution of order volume by origin-destination hexagon pairs. We use resolution-9 hexagons from the h3 package, which is the resolution that performs the best for $\averagedual$ when the matching window is at least $2$ minutes. The average size of each hexagon is  $0.1 \mathrm{km}^2$, and the average edge length is around $0.2 \mathrm{km}~$\citep{uber_h3_2018}.

First, \Cref{fig:meituan_heatmaps} provides heat-maps of trip volume by order origin and destination, respectively. We can see that the order origins are more concentrated than destinations, which is aligned with the more concentrated locations of restaurants in comparison to residential areas and office buildings. 
Note that both distributions are highly skewed. For hexagons with at least one originating order, the median, average, and maximum number of orders are 330, 960, and 17151, respectively. For hexagons that are the destination of at least one order, the median, average, and maximum order count are 71, 327 and 5167, respectively.

\begin{figure}[t!]
  \centering
  \subcaptionbox{By Origin.%
    \label{fig:meituan_count_by_orig}}[0.49 \textwidth]{\includegraphics[width = \figWidth \textwidth]{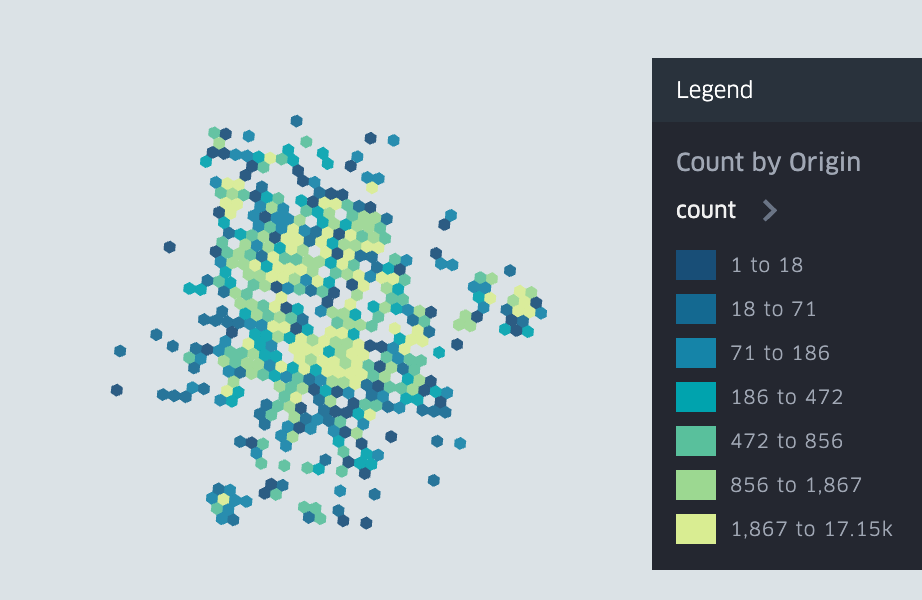}}
  \hfill
  \subcaptionbox{By Destination.%
    \label{fig:meituan_count_by_dest}}[0.49 \textwidth]{\includegraphics[width = \figWidth \textwidth]{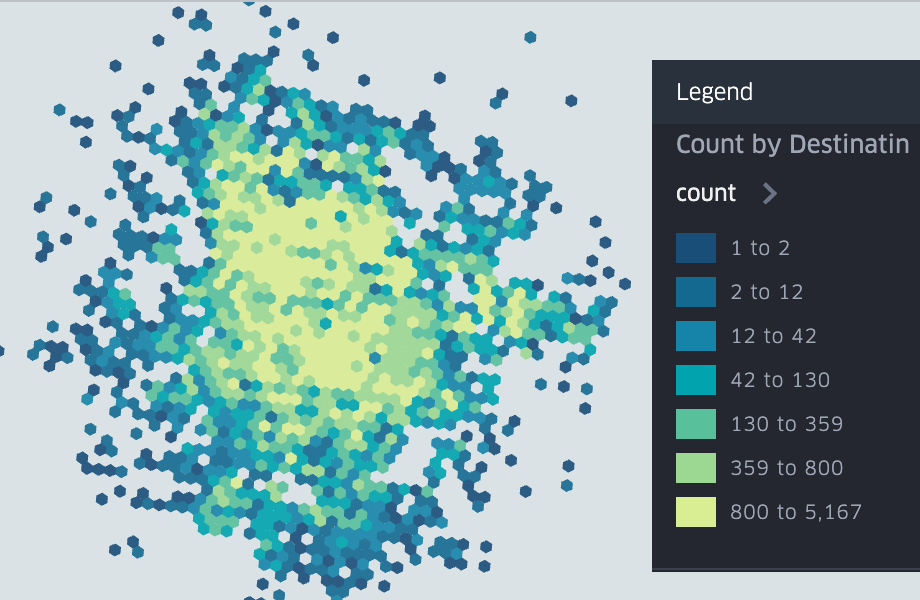}}
  \caption{Number of orders by order origin hexagon (left) and order destination hexagon (right).
  }
  \label{fig:meituan_heatmaps}
\end{figure}

Finally, \Cref{fig:meituan_CDF_count_per_OD_pair} presents the distribution order volume by origin and destination hexagon pairs. From the unweighted CDF on the left, we can see that the vast majority of OD pairs have fewer than 100 orders. Out of all OD hexagon pairs with at least one order, the median, average, and maximum number of orders are 3, 8.22, and 808, respectively.
The CDF on the right is weighted by order count.
Overall, roughly 14\% of orders are associated with OD pairs with at least 100 orders, and 72.8\% of orders are from OD pairs with at least 10 orders.
\begin{figure}
    \centering
    \includegraphics[width=0.75\linewidth]{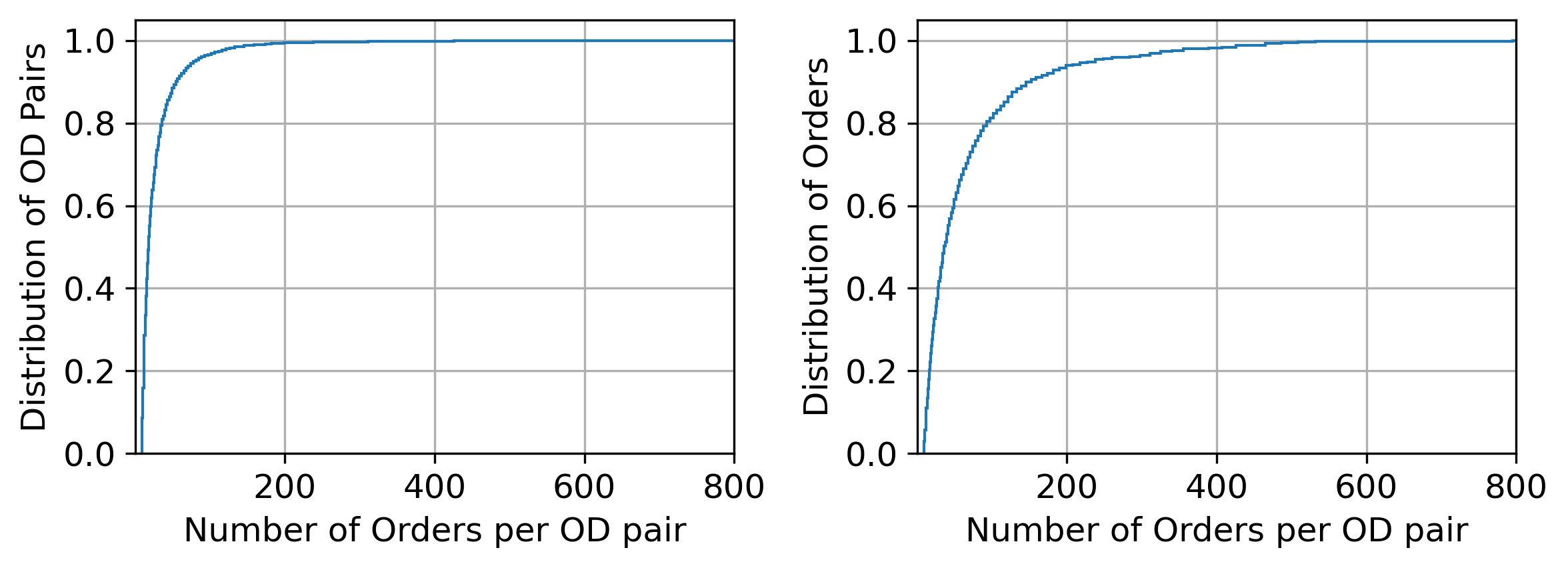}
    \caption{
    Distribution (CDF) of the number of orders per origin-destination hexagon pair (resolution-9), unweighted (left) and weighted by the number of orders (right). 
    }
    \label{fig:meituan_CDF_count_per_OD_pair}
\end{figure}

\section{Additional Simulation Results}\label{sec: sim_results_extra}

We provide in this section additional details and numerical results for settings studied in \Cref{sec:numerical_experiments} of the paper 
as well as more general synthetic environments, non-uniform spatial distributions, and different reward topologies.
We also provide results on a more practical version of the \textit{rolling batching} algorithm (which solves for the optimal solution over available jobs every 30 seconds, instead of every time a job becomes critical) and demonstrate that this algorithm also benefits from 
taking potential into consideration to account for the opportunity costs in the (more distant) future.


\subsection{Match Rate and Saving Fraction}
\label{sec:match_rate_2}  


We provide in this section two additional performance metrics: the \newterm{match rate}, i.e. the fraction of jobs that were pooled instead of dispatched on their own, and  the \newterm{saving fraction}, i.e. the fraction of the total distance that is reduced by pooling, relative to the total travel distance without any pooling~\citep[see][]{aouad2020dynamic}.

Both regret and reward ratio are performance metrics that compare pooling algorithm relative to the hindsight optimal pooling outcome. 
The saving fraction illustrates the benefit of delivery pooling in comparison to the total distance traveled, and is upper bounded by 0.5 (since reward from pooling a job cannot exceed its distance).
The match rate, as shown in \Cref{fig:1D_unif_match_rate,fig:meituan_match_rate}, highlights an additional desirable property of $\PB$ and, more generally, of \emph{index-based greedy matching} algorithms: they pool as many jobs as possible.
%

%

\subsubsection{Uniform one-dimensional case.} \label{sec: 1D_extra}

\Cref{fig:1D_extra} compares the match rate and the saving fraction achieved for the one-dimensional synthetic environment studied in \Cref{sec:sim_unif_1D}.
%
In this setting, all jobs share the same origin at $0$ and destinations are drawn uniformly at random from $[0,1]$. 
\Cref{fig:1D_unif_saving_fraction} shows that the reduction in travel distance can be remarkably high for this setting, with $\OPT$ saving more than 46\% even at low density levels.
%

The match rate in \Cref{fig:1D_unif_match_rate} illustrates that all \emph{index-based greedy matching} algorithms match every one of the $\Njob = 1000$ jobs.
This is because (i) the reward from matching any two jobs is non-negative in this 1D setting, and (ii) the total number of jobs is even.
On the other hand, $\batching$ matches every job only if $\sojourn+1$ is even, since every time it computes an optimal matching, the \textit{batch size} is $\sojourn+1$ and it dispatches \textit{all} available jobs in pooled trips. 
%
%

\begin{figure}[]
  \centering
  \subcaptionbox{Average match rate (higher is better).%
    \label{fig:1D_unif_match_rate}}[0.49 \textwidth]{\includegraphics[width = \figWidth \textwidth]{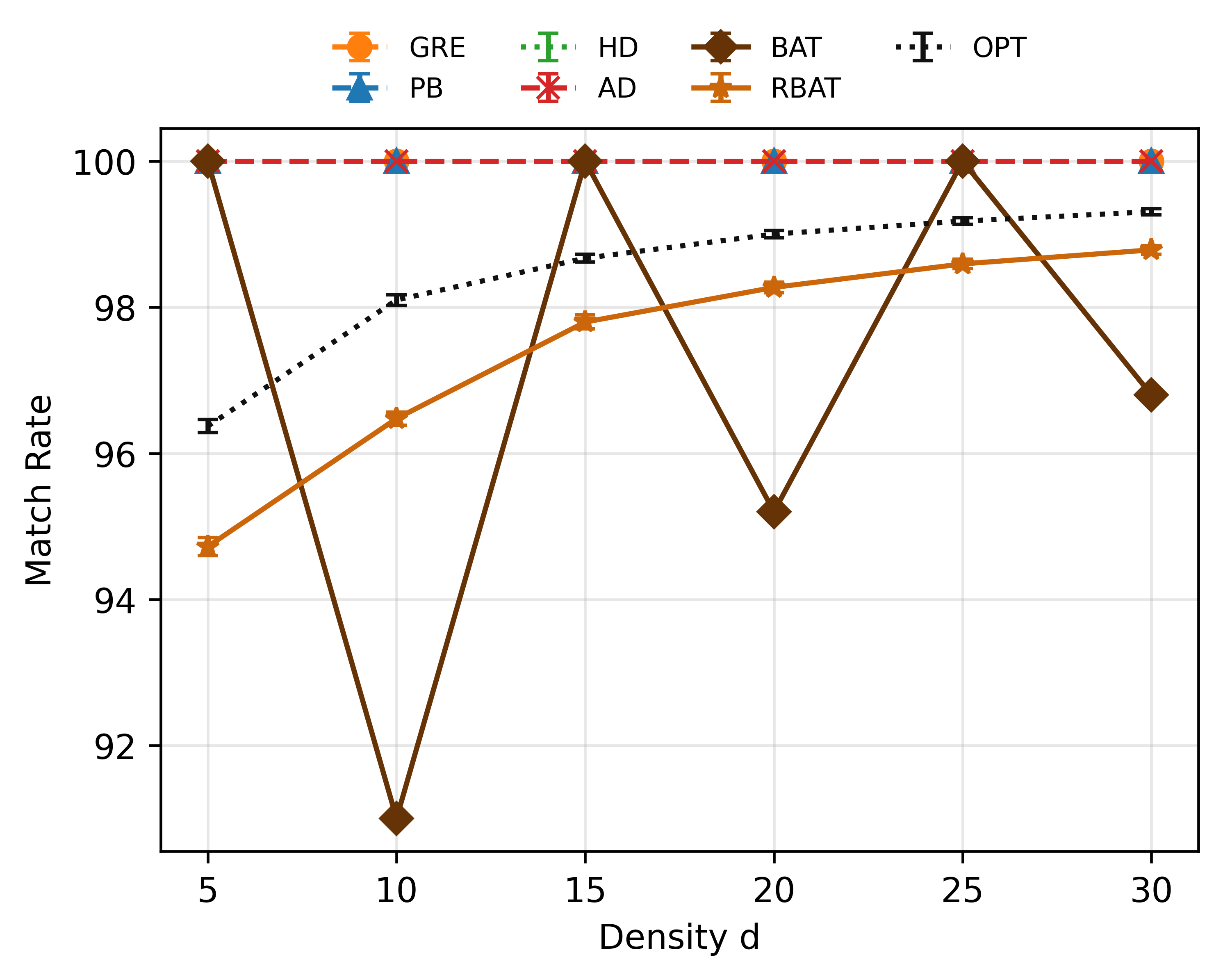}}
  \hfill
  \subcaptionbox{Average saving fraction (higher is better).%
    \label{fig:1D_unif_saving_fraction}}[0.49 \textwidth]{\includegraphics[width = \figWidth \textwidth]{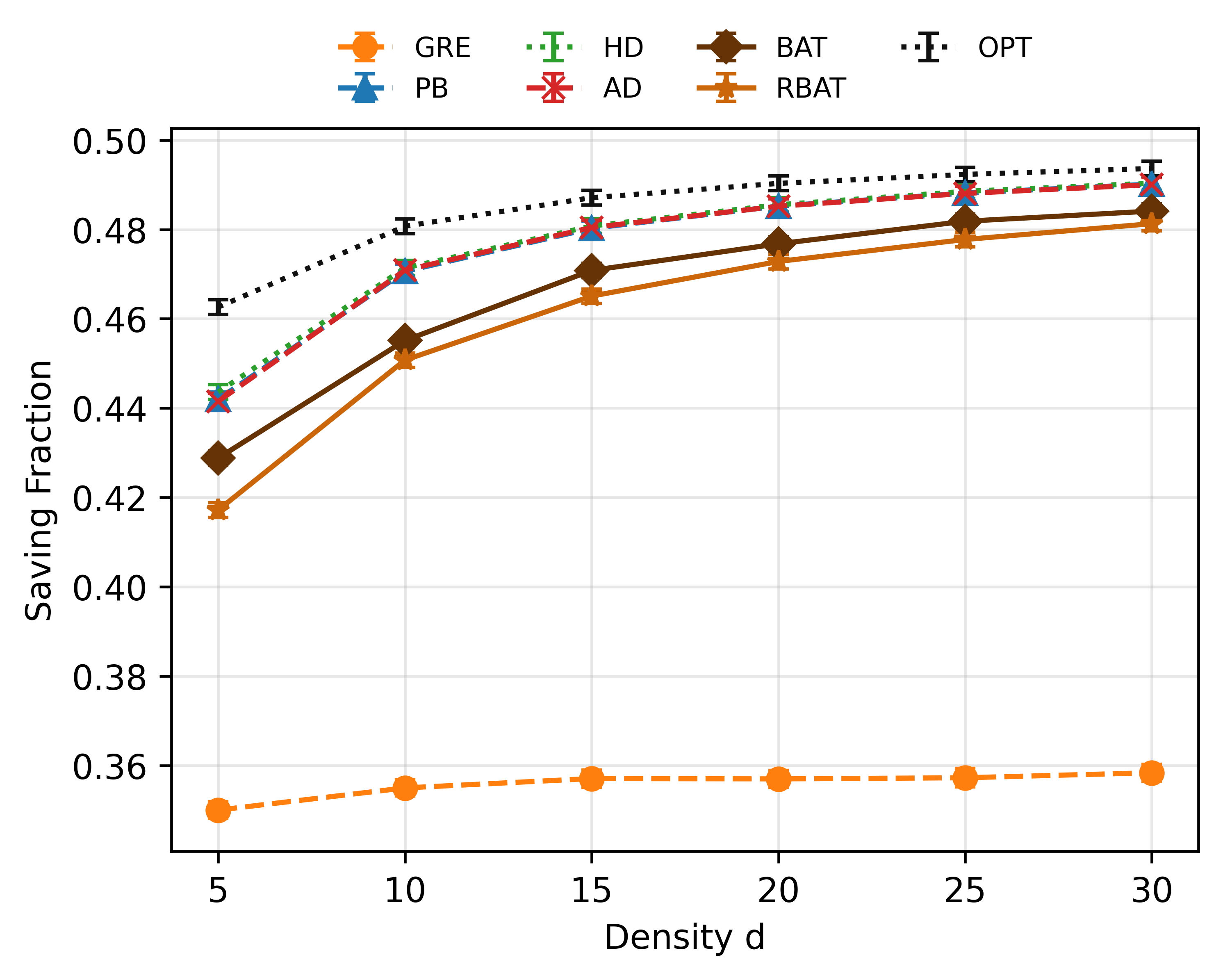}}
  \caption{Match rate and saving fraction in random 1D instances.}
  \label{fig:1D_extra}
\end{figure}

\subsubsection{Meituan data.} \label{sec: meituan_extra}
\begin{figure}
  \centering
  \subcaptionbox{Average match rate (higher is better).%
    \label{fig:meituan_match_rate}}[0.49 \textwidth]{\includegraphics[width = \figWidth \textwidth]{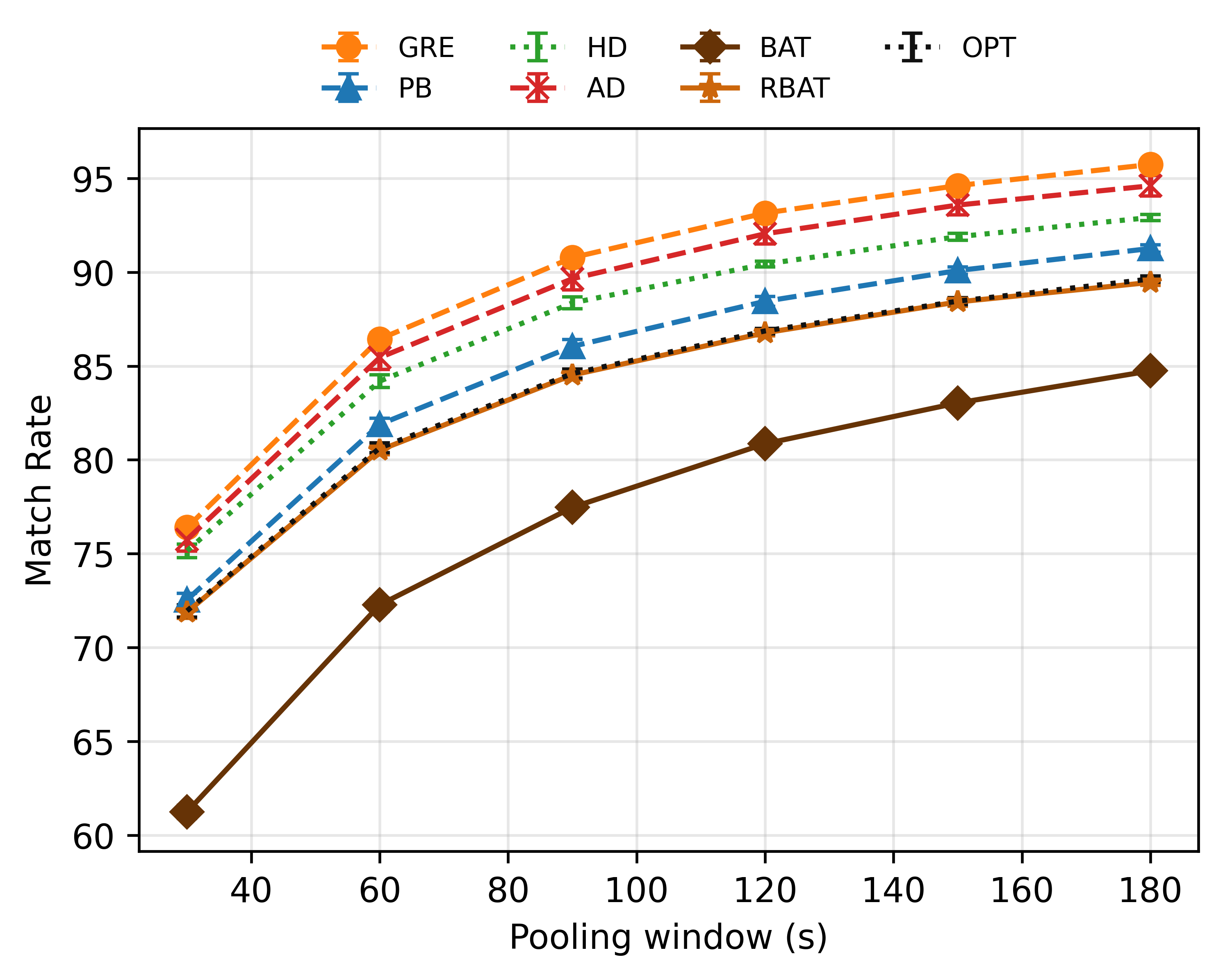}}
  \hfill
  \subcaptionbox{Average saving fraction (higher is better). %
    \label{fig:meituan_saving_fraction}}[0.49 \textwidth]{\includegraphics[width = \figWidth \textwidth]{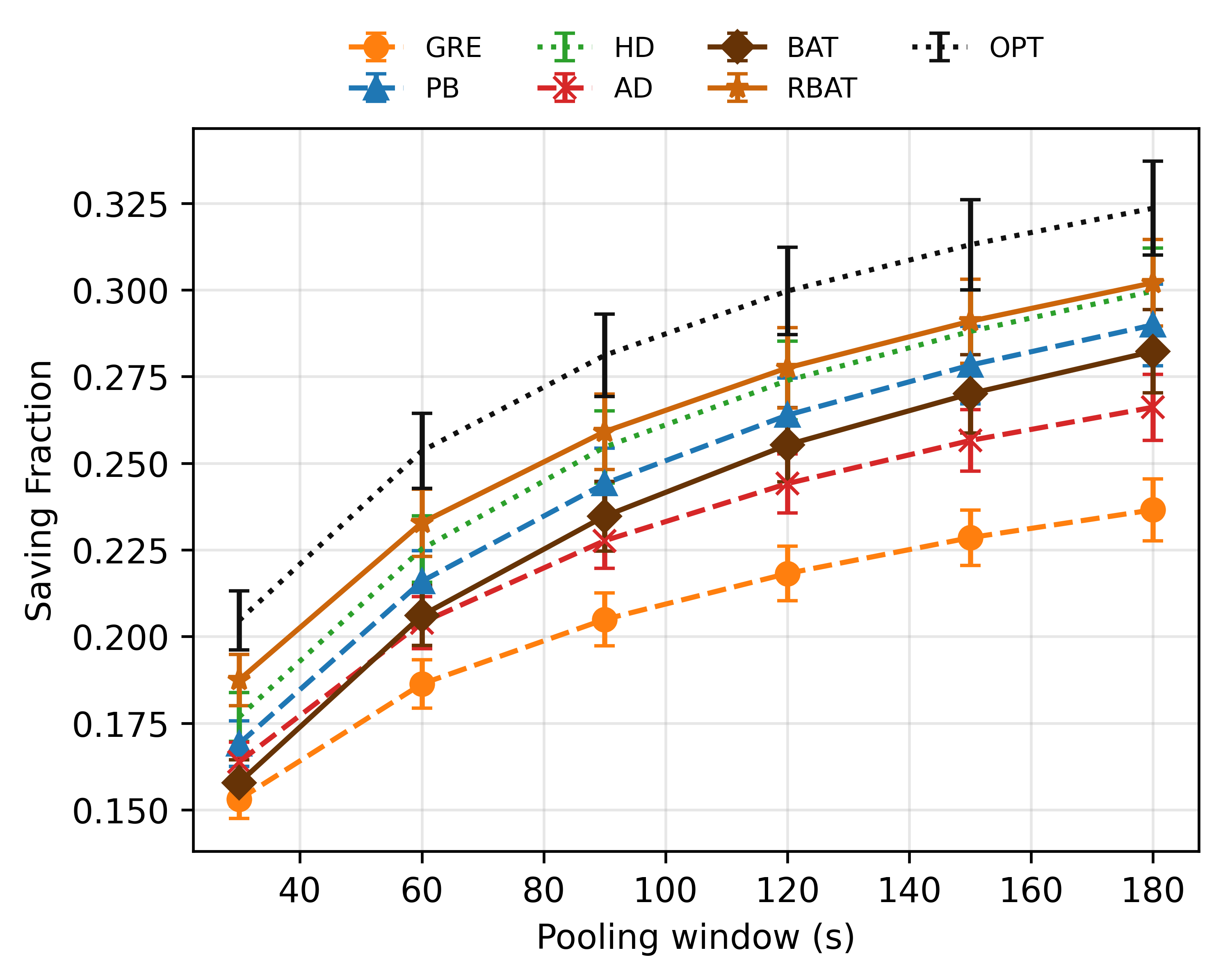}}
    \hfill
  %
  \caption{Match rate and saving fraction for Meituan Data.}
  \label{fig:meituan_extra}
\end{figure}

\Cref{fig:meituan_extra} provides a comparison of the match rate and the saving fraction for the setting presented in \Cref{sec:sim_meituan}.
Recall that as we extend our definition of pooling reward to allow for 2D locations with heterogeneous origins, it is no longer guaranteed to be non-negative.
Consequently, both the matching rate and saving fraction of all algorithms are lower than in the one-dimensional setting.
We see from \Cref{fig:meituan_match_rate} 
that greedy algorithms ($\gre$, $\PB$, $\dual$, $\averagedual$) consistently achieve higher match rates in comparison to $\batching$, $\rbatching$, and $\OPT$, suggesting a trade-off between the pooling reward and the match rate in the real-world setting.
We can also see from \Cref{fig:meituan_saving_fraction} that if the platform allows a time window of at least $1$ minute for each order before it has to be dispatched, 
$\PB$ achieves a reduction in distance traveled by over 20\%. At $3$ minutes, this fraction increases to roughly 30\%, still far from the 50\% theoretical upper bound but the aggregate savings from 100 million orders a day is substantial.

\subsection{Parameter Tuning for Potential-augmented Rolling Batching} \label{sec:opt_gamma}

We now provide additional details on potential-augmented rolling batching policies introduced in \Cref{sec:rbat_best}.
For each pooling window and each notion of shadow price (potential, hindsight duals, and average duals), we run the augmented batching algorithms for all $\gamma$ in $\{0, 0.1, 0.2, \dots, 1\}$ on a small subset of the data corresponding to a geographically isolated ``island'' within the city. 
We then choose the $\gamma$ that achieve the best performance for each setting, and evaluate the performance on the rest of the data (which is disjoint from the tuning set).

\begin{figure}
  \centering
  \subcaptionbox{Parameter tuning.%
    \label{fig:meituan_best_gamma}}[0.49 \textwidth]{\includegraphics[width = \figWidth \textwidth]{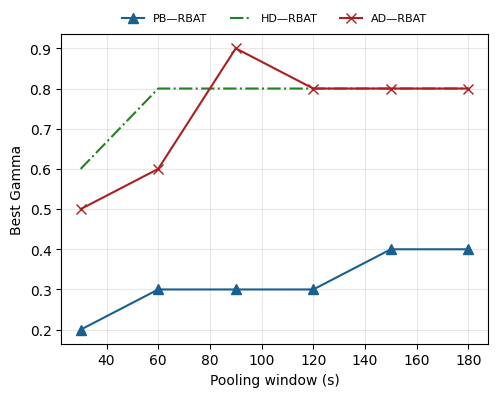}}
  \hfill
  \subcaptionbox{Average match rate (higher is better).%
    \label{fig:meituan_best_match_rate}}[0.49 \textwidth]{\includegraphics[width = \figWidth \textwidth]{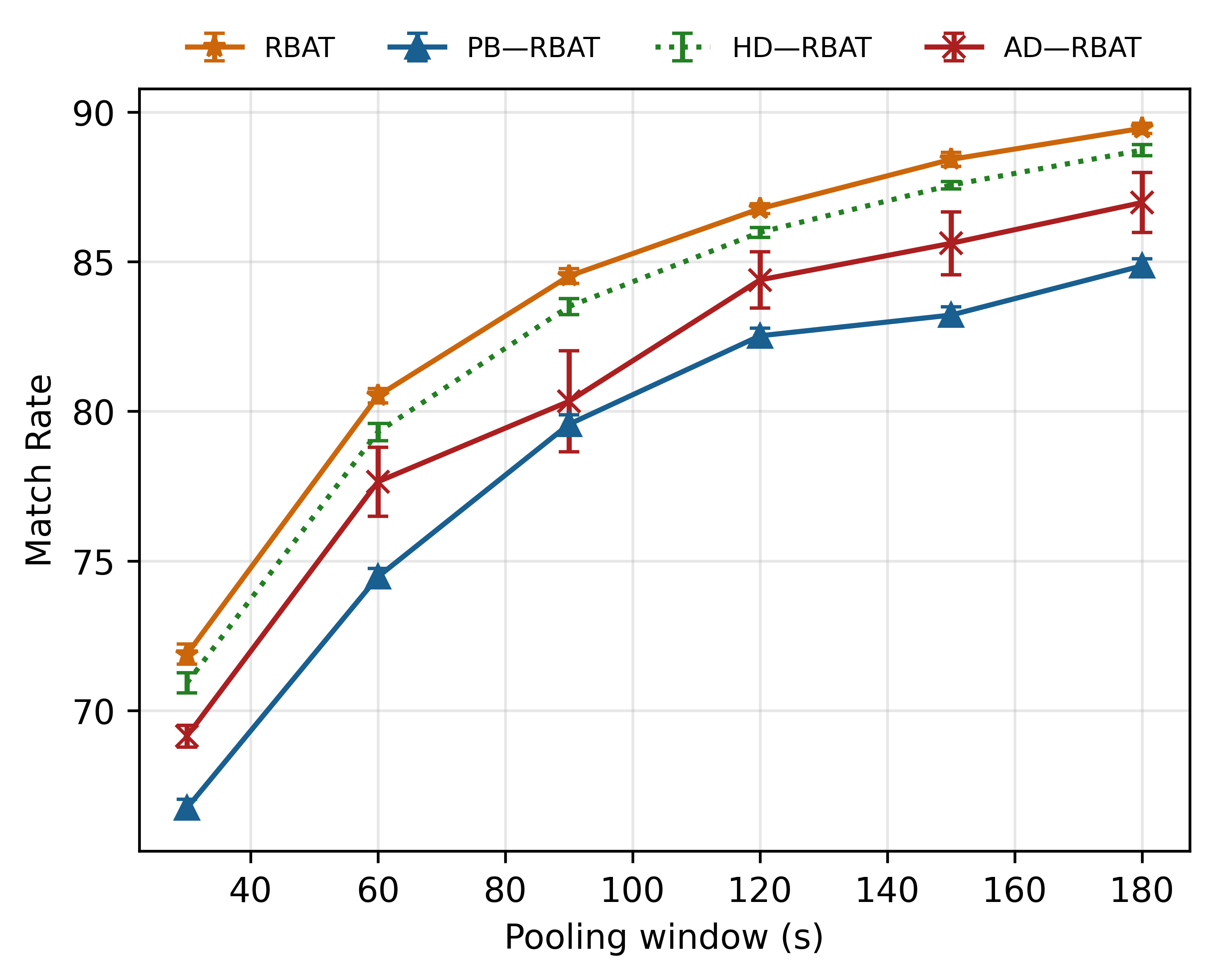}}
  \caption{Parameter tuning and match rate for batching-based heuristics augmented with opportunity costs in Meituan data.}
  \label{fig:meituan_gamma_match_rate}
\end{figure}

\Cref{fig:meituan_best_gamma} illustrates the optimal values of $\gamma$ selected under each shadow-price specification.
We can see that the optimal coefficient for potential-augmented $\rbatching$ is not very sensitive to the pooling window. 
Additionally, we have also observed that the performance of the augmented rolling batching algorithms is not sensitive to $\gamma$ around the optimal choice.

Moreover, we see from \Cref{fig:meituan_best_match_rate} that out of all ``realistic'' shadow prices (i.e. excluding $\dual$),  higher pooling reward corresponds to a lower match rate. This highlights the same trad-off between cost savings and match rate as observed in Section~\ref{sec:sim_meituan_batching}, and the coefficient $\gamma$ effectively functions as the tuning parameter for this trade-off for $\PB$-rolling-batching (i.e., lower than optimal $\gamma$ leads to a to higher match rate at the expense of lower pooling rewards).

\subsection{Periodic Rolling Batching}
\label{sec:rbat2}

The $\rbatching$ algorithm described in \Cref{sec:sim_benchmarks} recomputes the optimal pooling 
outcome whenever any job becomes critical, solving the matching problem over all available jobs and then dispatching only the critical job (and its match, if any).
As discussed Section~\ref{sec:sim_meituan_batching}, this achieves high pooling reward, but the intense computational need renders the algorithm practically infeasible.
In practice, batching heuristics are often executed periodically.
To capture this possibility, we study in this section the \textit{periodic rolling batching} ($\periodicrbatching$) algorithm in which the optimization is solved every 30 seconds.
At each such `decision epoch', the algorithm computes the optimal pooling outcome based on the currently available jobs,
and dispatches any job (and their match) whose deadline occurs before the next decision epoch.

\begin{figure}
  \centering
  \subcaptionbox{Average regret (lower is better).%
    \label{fig:meituan_prbat_regret}}[0.49 \textwidth]{\includegraphics[width = \figWidth \textwidth]{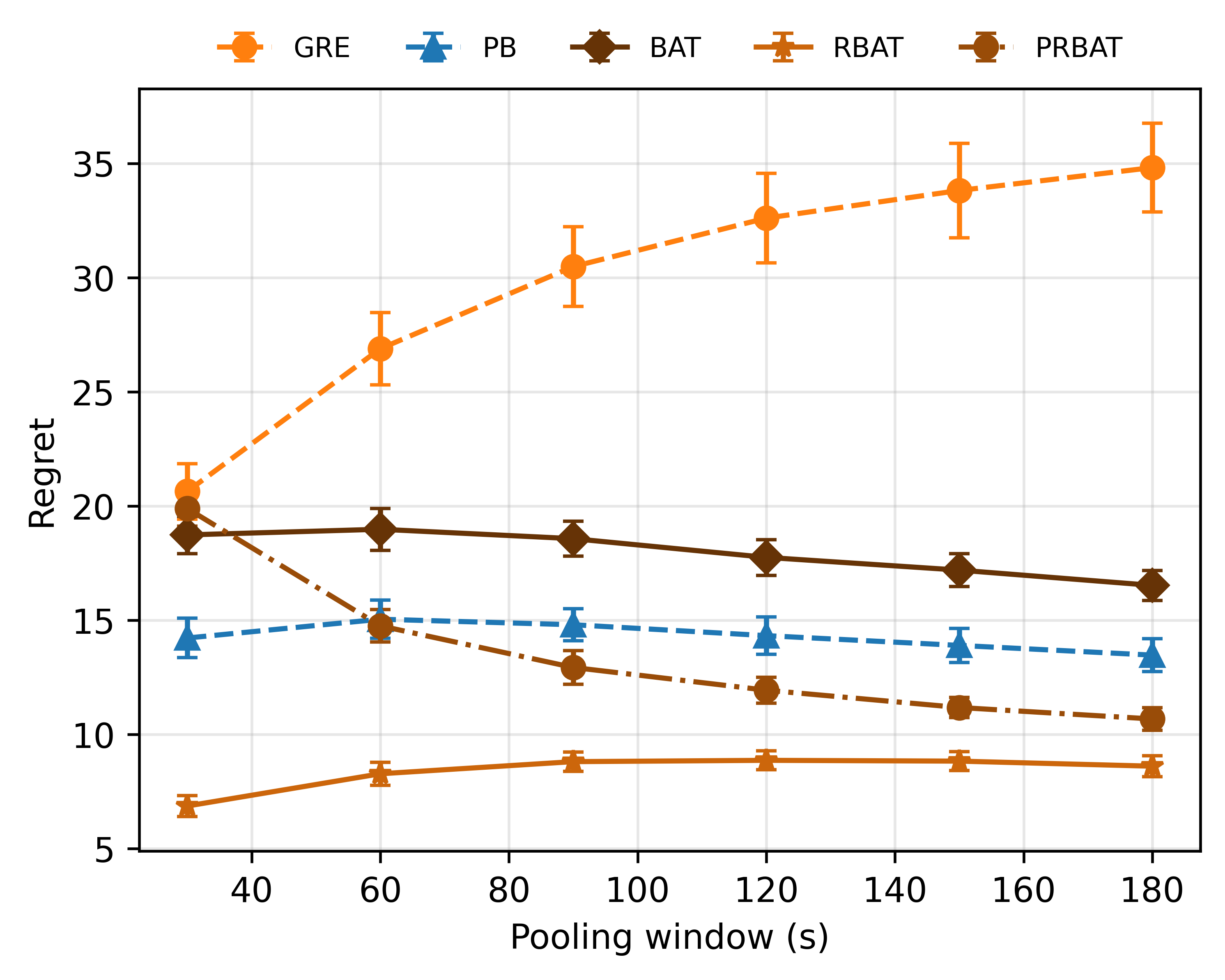}}
  \hfill
  \subcaptionbox{Average ratio (higher is better).%
    \label{fig:meituan_prbat_ratio}}[0.49 \textwidth]{\includegraphics[width = \figWidth \textwidth]{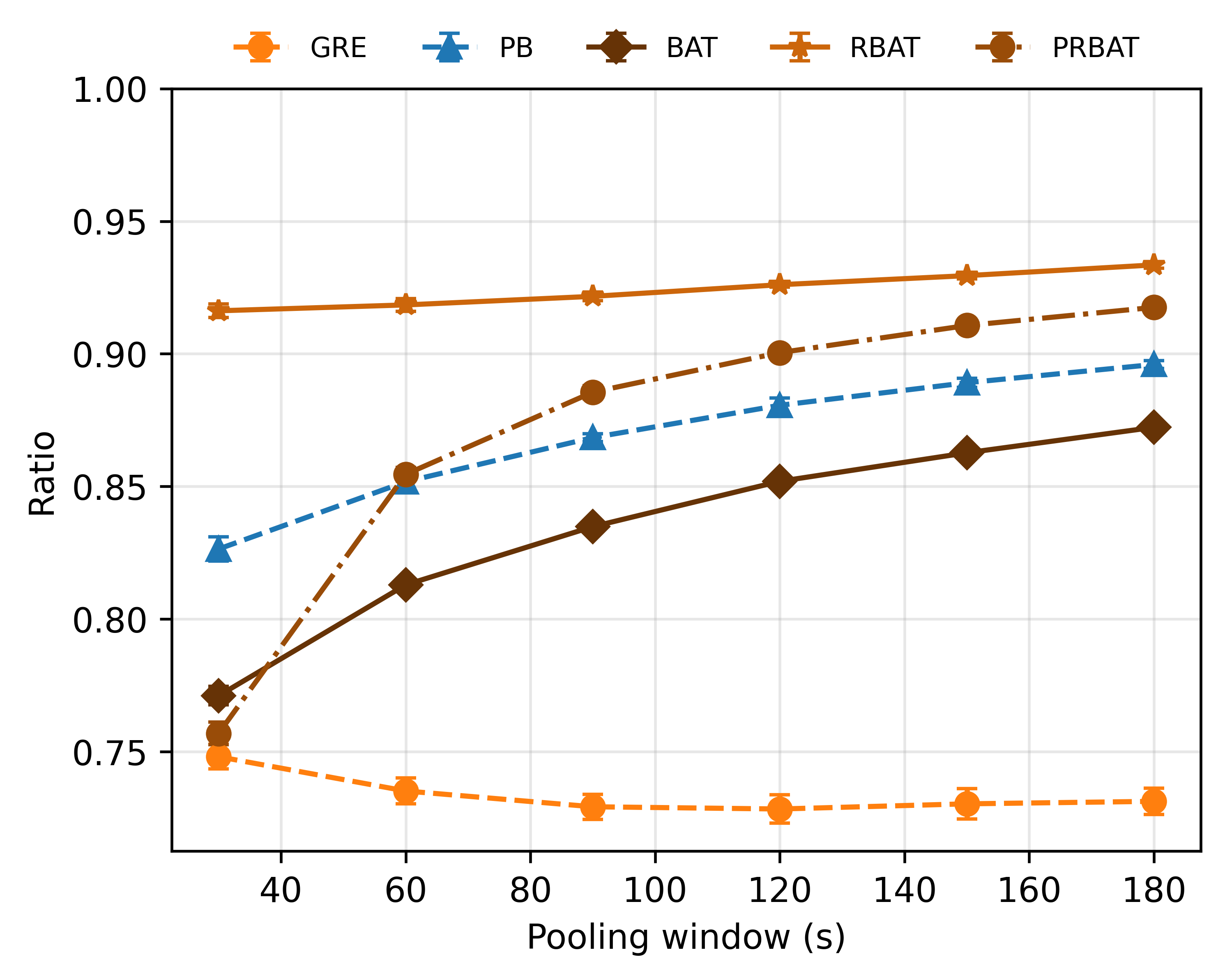}}
  \caption{Average regret and reward ratio, comparing with batching-based heuristics (including $\periodicrbatching$), in Meituan data.}
  \label{fig:meituan_prbat}
\end{figure}

\Cref{fig:meituan_prbat} reports the performance of periodic rolling batching.
With a pooling window is of 30 seconds, $\periodicrbatching$ is closer in behavior to full batching ($\batching$). 
In this regime, the algorithm often dispatches a relatively large fraction of non-critical orders, limiting future pooling opportunities.
As the pooling window increases, the performance of $\periodicrbatching$ improves substantially, outperforming $\PB$ at 60 seconds and halving the gap between $\PB$ and $\rbatching$ at 180 seconds.
The match rate of $\periodicrbatching$ interpolates smoothly between that of $\batching$ at short intervals and that of $\rbatching$ at longer intervals.
Moreover, we see from \Cref{fig:meituan_prbat_running_time} that the computational cost of $\periodicrbatching$ is substantially lower than that of $\rbatching$, and only modestly higher than that of greedy policies. 
Thus, periodic rolling batching offers a practically feasible trade-off between performance and runtime.

\begin{figure}
  \centering
  \subcaptionbox{Running time.%
    \label{fig:meituan_prbat_running_time}}[0.49 \textwidth]{\includegraphics[width = \figWidth \textwidth]{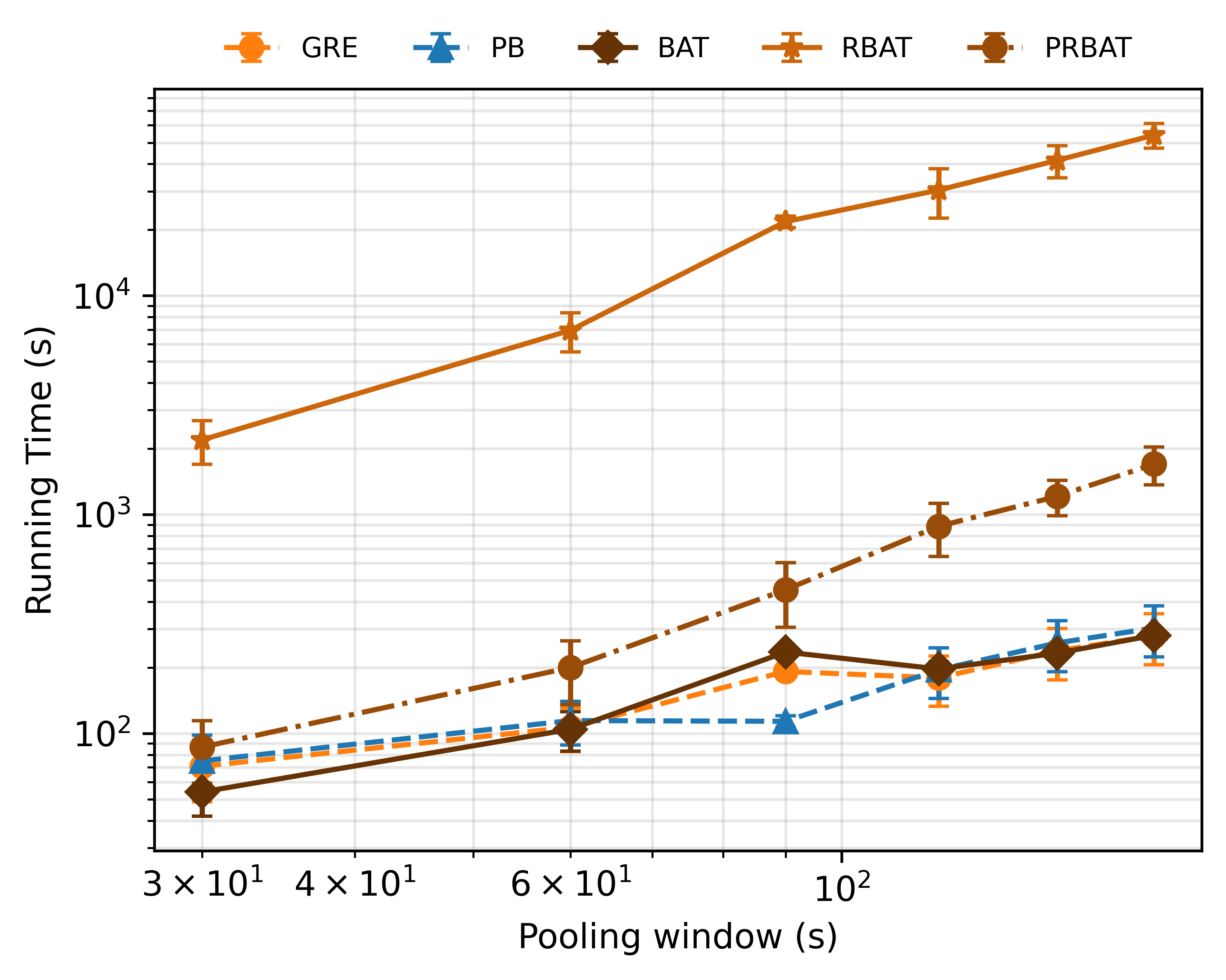}}
  \hfill
  \subcaptionbox{Match rate.%
    \label{fig:meituan_prbat_match_rate}}[0.49 \textwidth]{\includegraphics[width = \figWidth \textwidth]{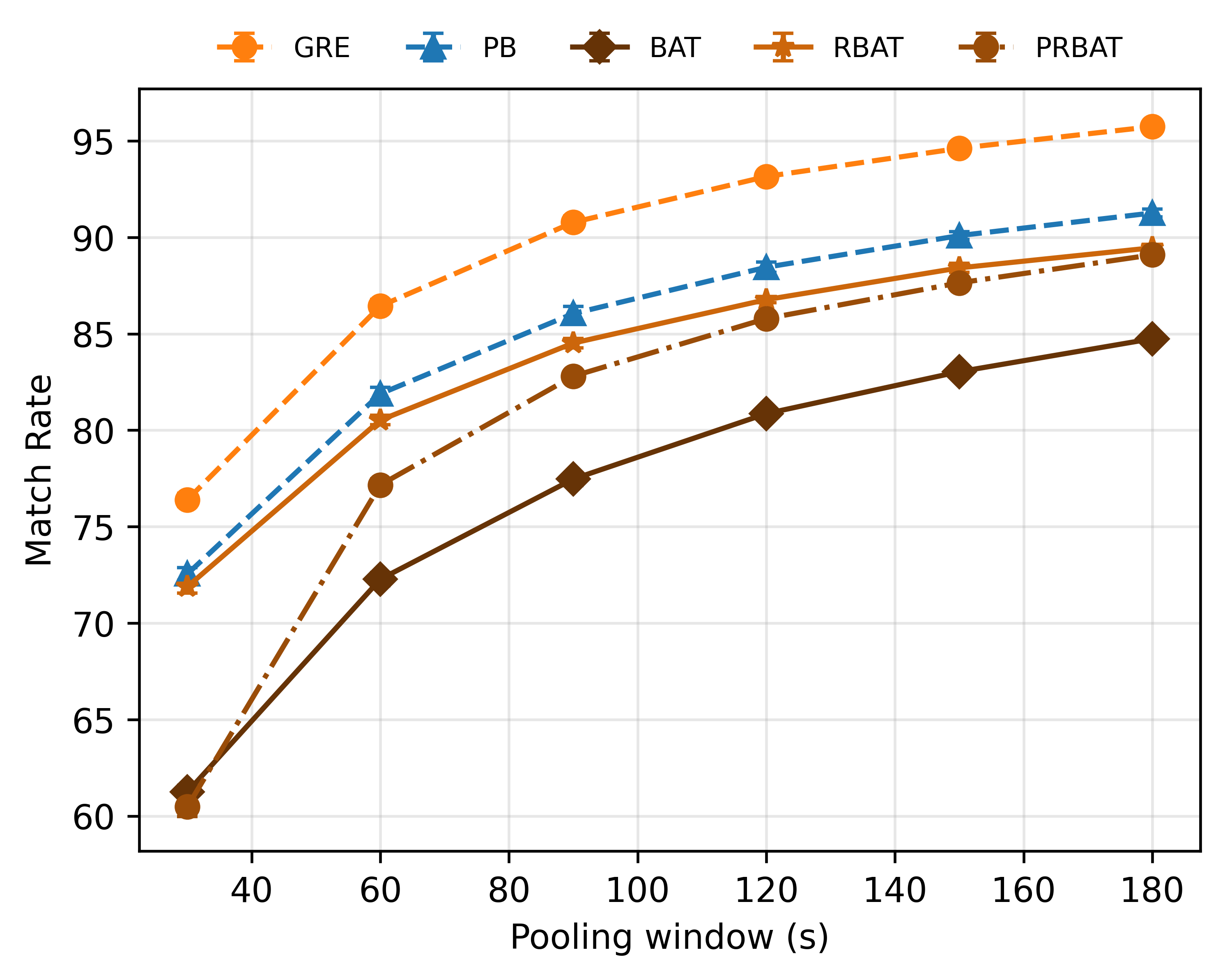}}
  \caption{Running time and match rate, comparing with batching-based heuristics (including $\periodicrbatching$), in Meituan data.}
  \label{fig:meituan_prbat_time_match}
\end{figure}

\paragraph{Potential-Augmented Periodic Rolling Batching.}

Finally, we augment the periodic rolling-batching algorithm with the potential and other shadow prices (in the same way as our analysis in \Cref{sec:rbat_best} and \Cref{sec:opt_gamma}), and report the performance under the optimal $\gamma$ for each setting in \Cref{fig:meituan_prbat_best} and \Cref{fig:meituan_prbat_gamma_match_rate}.

Recall from \Cref{fig:meituan_prbat} that periodic rolling-batching $\periodicrbatching$ with no shadow price augmentation closes roughly a half of the gap between $\PB$ and $\rbatching$ when the pooling is at least 120 seconds. 
From \Cref{fig:meituan_prbat_best}, we can see that augmenting $\periodicrbatching$ with the potential, $\PB-\periodicrbatching$ now achieves the same pooling reward as that under the full rolling batching benchmark for the 180-second pooling window. 
Importantly, it does so with a run time that is two orders of magnitude lower than that of $\rbatching$ (the run time of $\PB-\periodicrbatching$ is the same as that of $\periodicrbatching$, as shown in \Cref{fig:meituan_prbat_running_time}).
This result reinforces the central message of the paper: the potential provides an effective proxy for future value, allowing simpler and more computationally efficient policies to recover much of the performance of more complex optimization-based benchmarks.

\begin{figure}
  \centering
  \subcaptionbox{Average regret (lower is better).%
    \label{fig:meituan_prbat_best_regret}}[0.49 \textwidth]{\includegraphics[width = \figWidth \textwidth]{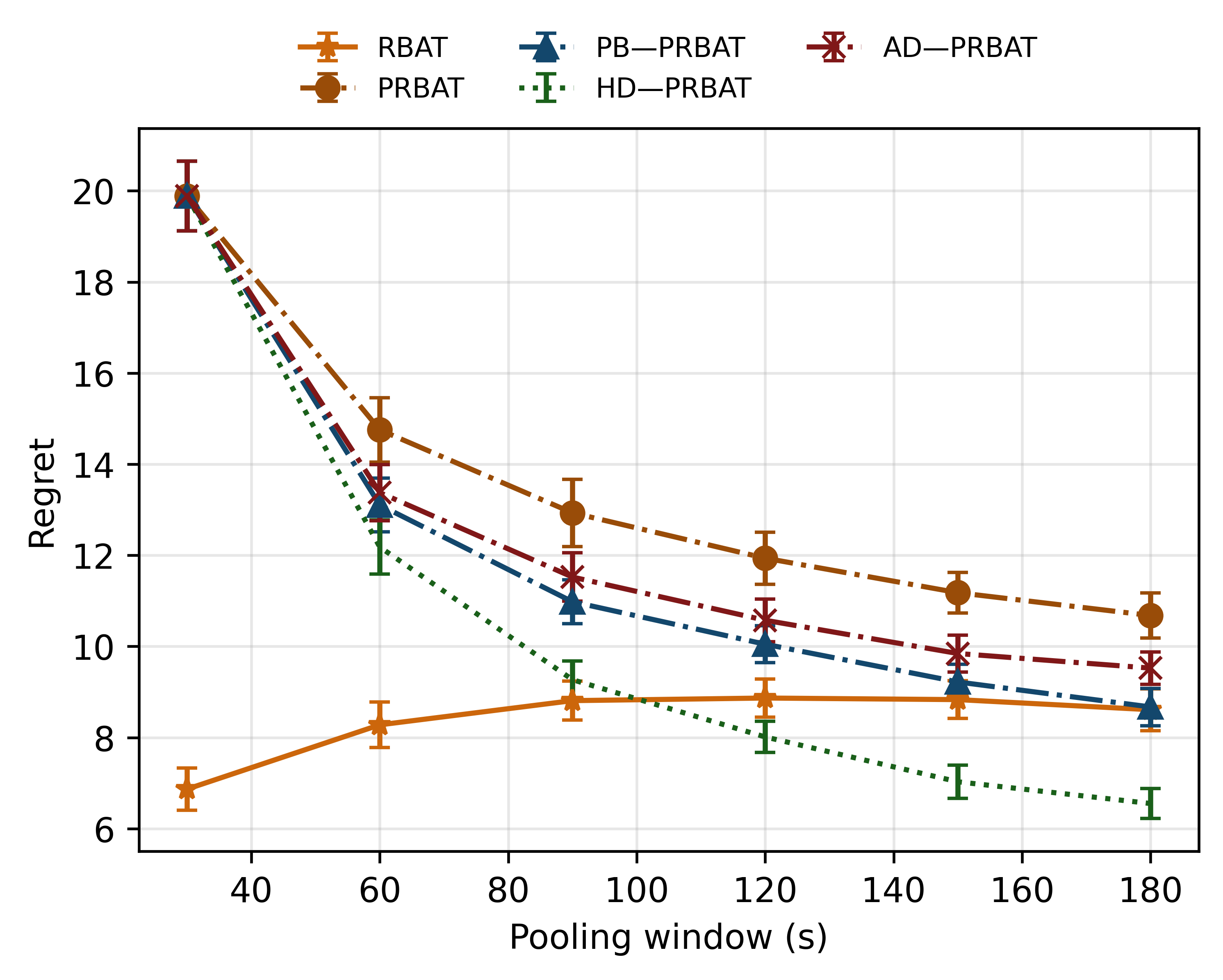}}
  \hfill
  \subcaptionbox{Average ratio (higher is better).%
    \label{fig:meituan_prbat_best_ratio}}[0.49 \textwidth]{\includegraphics[width = \figWidth \textwidth]{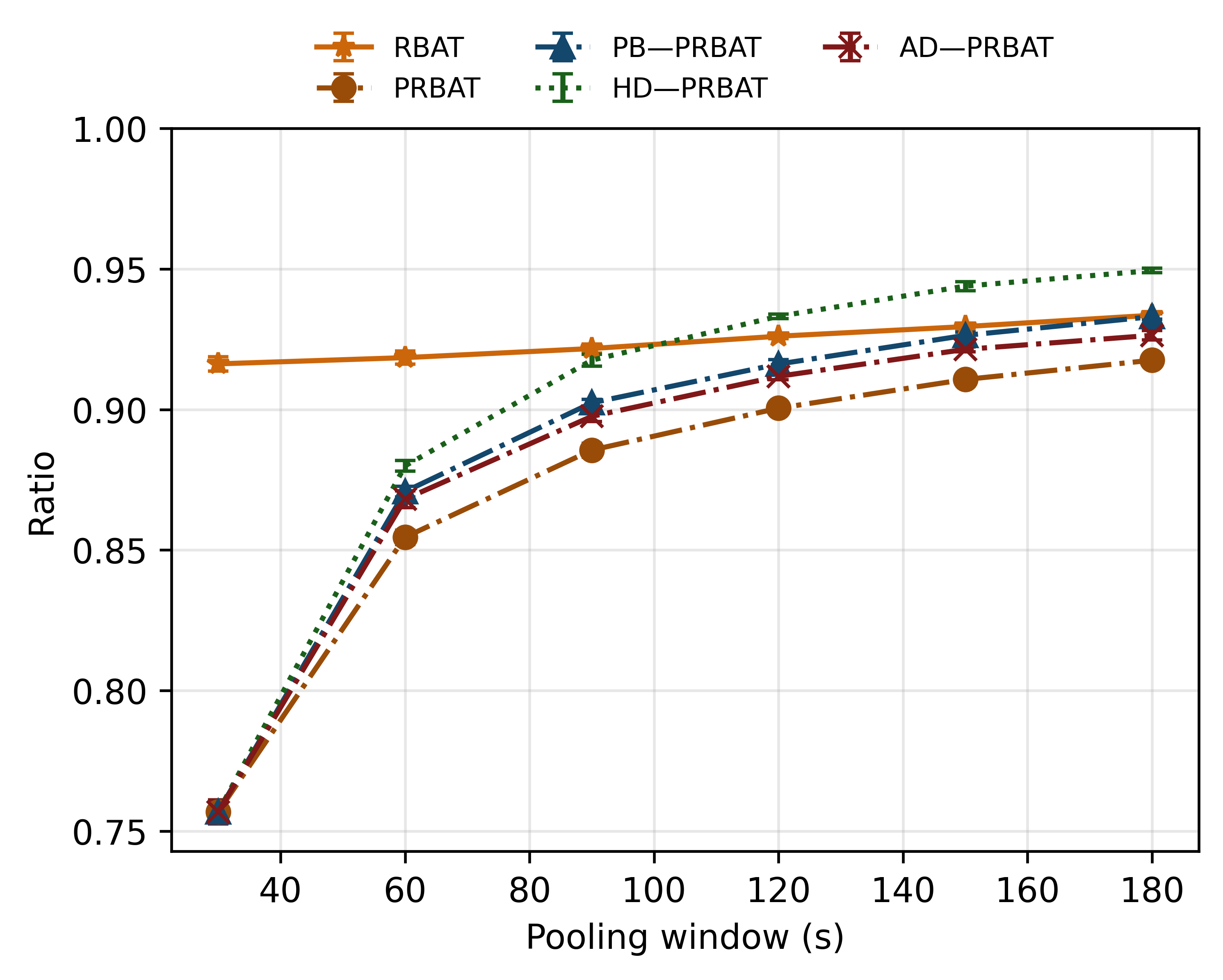}}
  \caption{Average running time and reward ratio, comparing with batching-based heuristics augmented with opportunity costs (including $\periodicrbatching$), in Meituan data.}
  \label{fig:meituan_prbat_best}
\end{figure}

\begin{figure}
  \centering
  \subcaptionbox{Parameter tuning.%
    \label{fig:meituan_prbat_best_gamma}}[0.49 \textwidth]{\includegraphics[width = \figWidth \textwidth]{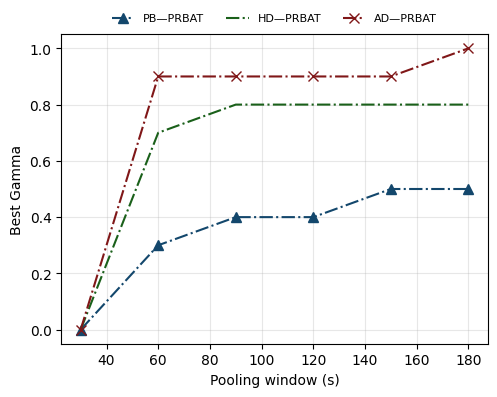}}
  \hfill
  \subcaptionbox{Average match rate (higher is better).%
    \label{fig:meituan_prbat_best_match_rate}}[0.49 \textwidth]{\includegraphics[width = \figWidth \textwidth]{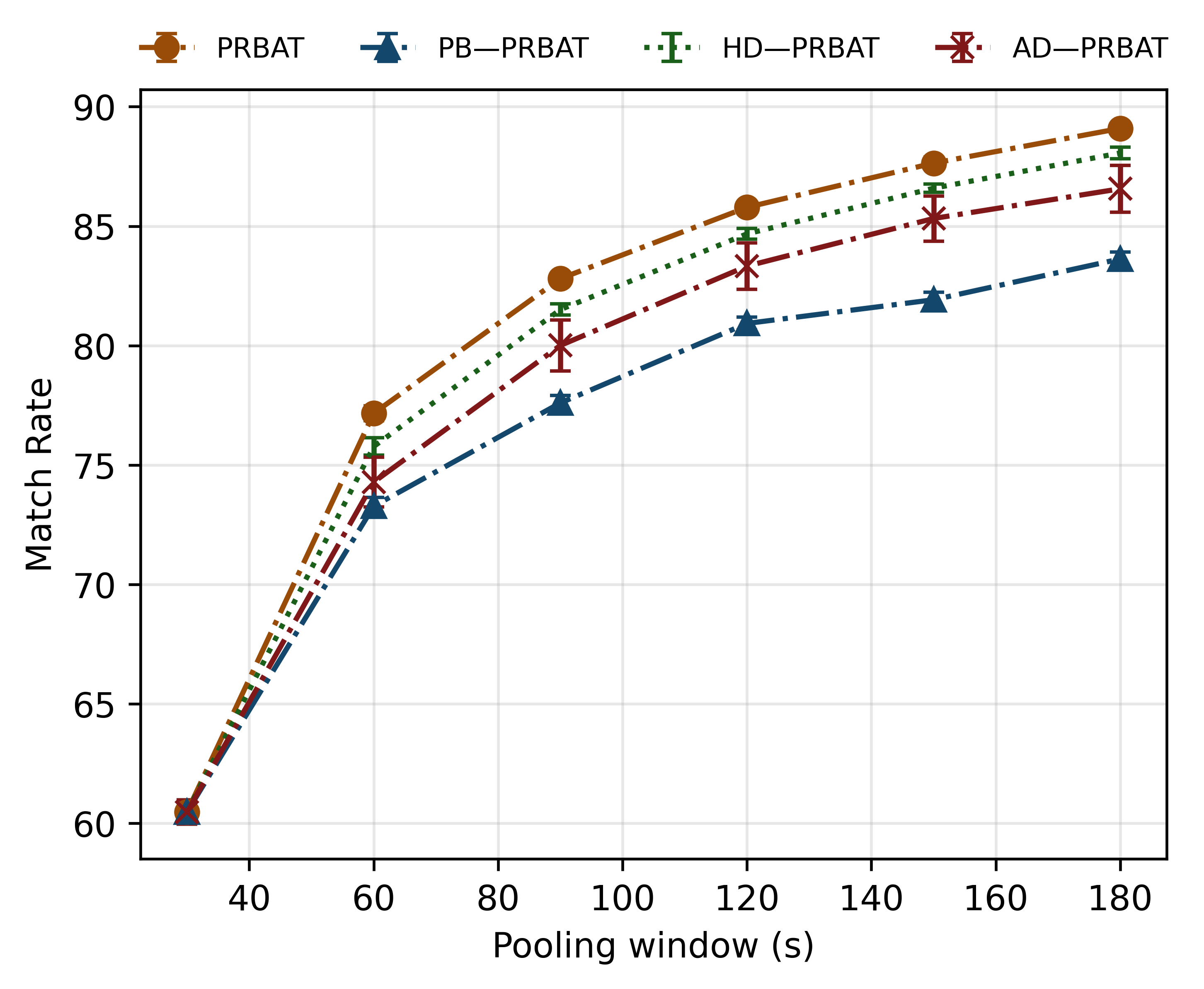}}
  \caption{Parameter tuning and match rate for batching-based heuristics augmented with opportunity costs in Meituan data.}
  \label{fig:meituan_prbat_gamma_match_rate}
\end{figure}


\subsection{Additional Synthetic Simulations} \label{appx:additional_synthetic_results}

\subsubsection{1D Common origin w/ non-uniform destinations.}\label{sec: nonunif_1D}
In this section, we show that our experimental results in one-dimensional synthetic data are robust to non-uniform distribution of types. 
We consider an analogous setting from \Cref{sec:sim_unif_1D}, but now each job type is drawn IID from a distribution Beta(0.5, 2). Overall, the results are consistent (see \Cref{fig:1D_nonunif}).

\begin{figure}
  \centering
  \subcaptionbox{Average regret (lower is better).%
    \label{fig:1D_nonunif_regret}}[0.49 \textwidth]{\includegraphics[width = \figWidth \textwidth]{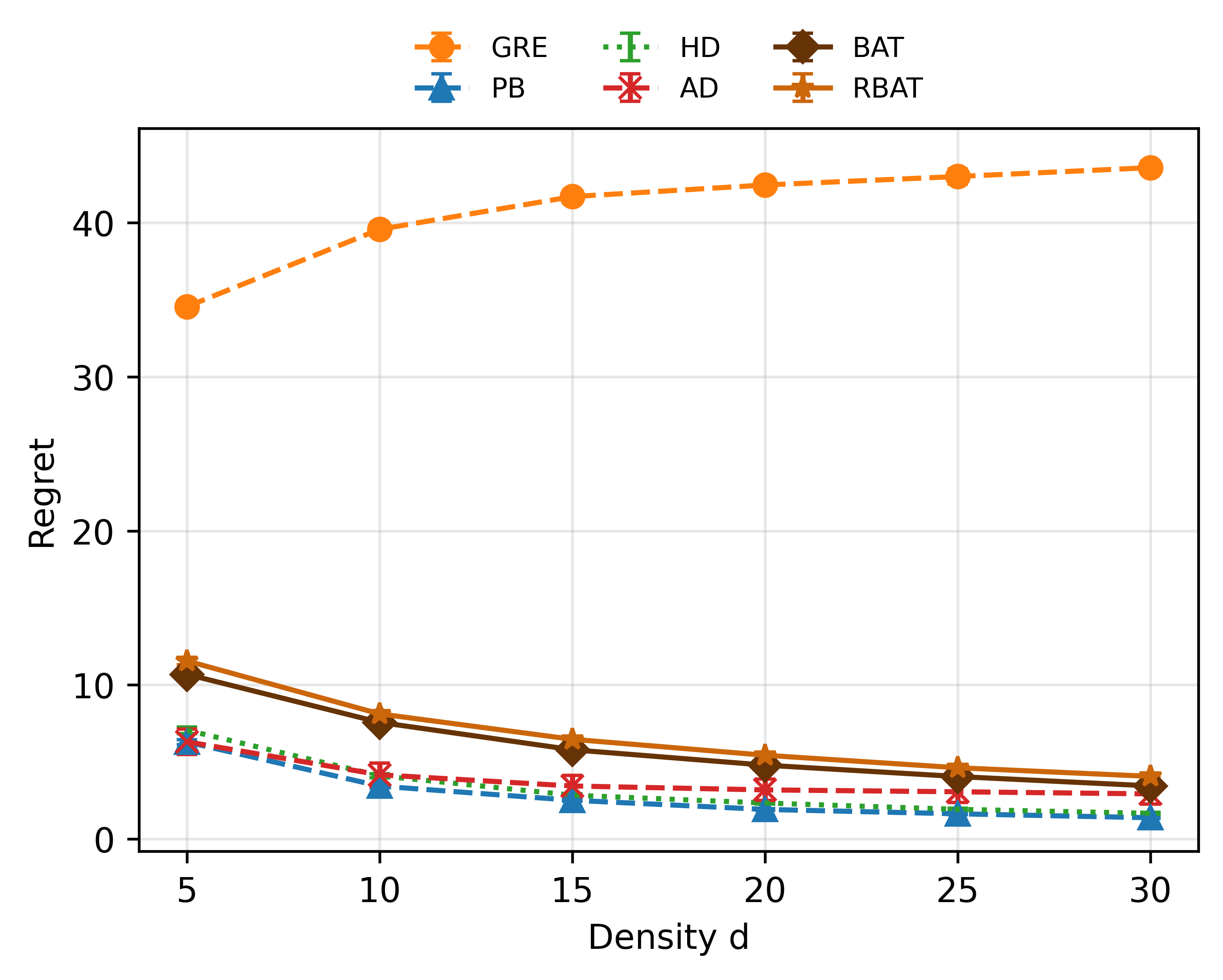}}
  \hfill
  \subcaptionbox{Average ratio (higher is better).%
    \label{fig:1D_nonunif_ratio}}[0.49 \textwidth]{\includegraphics[width = \figWidth \textwidth]{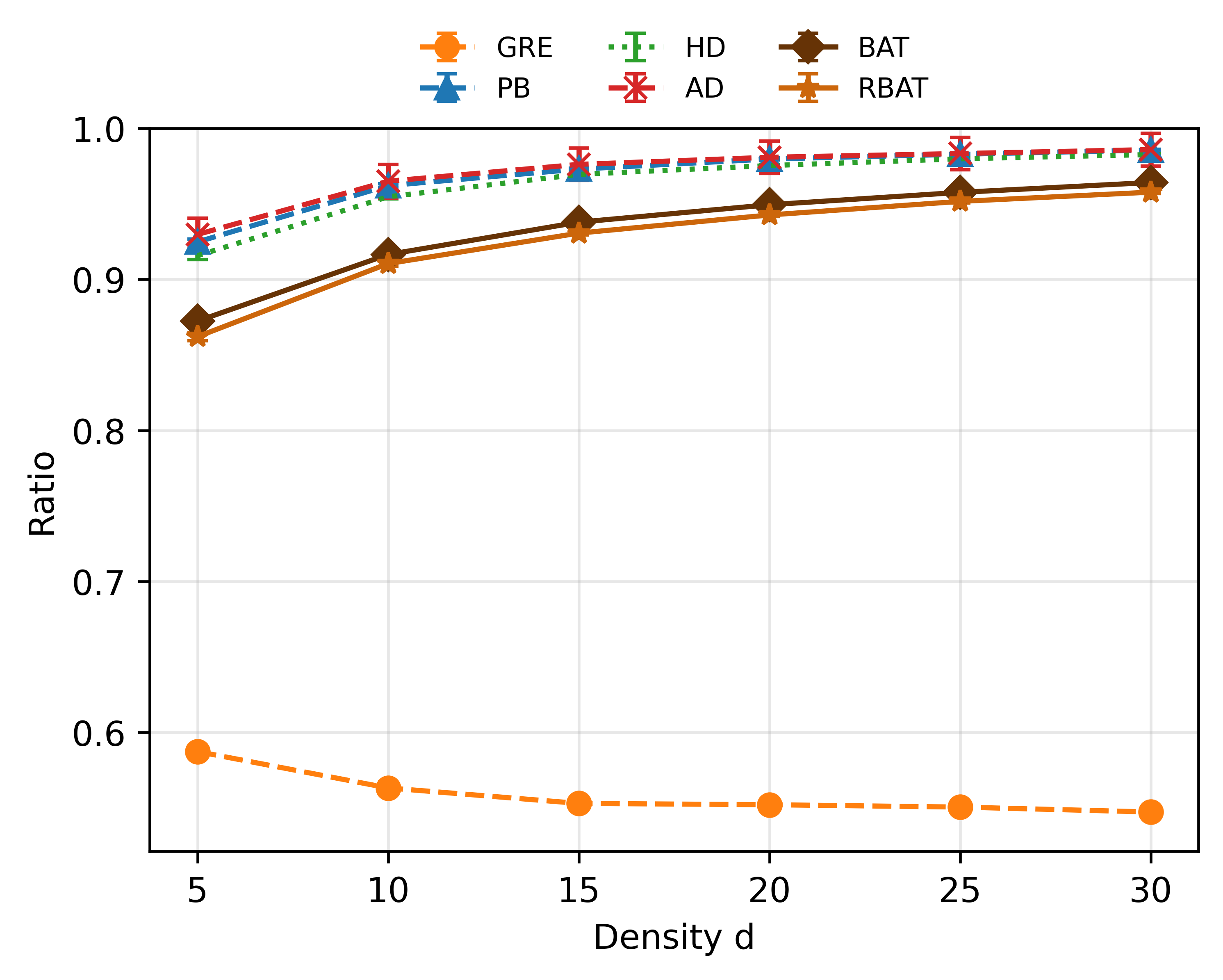}}
  \caption{Average regret and reward ratio for 1D, common origin, where destinations are drawn IID from Beta(0.5, 2).}
  \label{fig:1D_nonunif}
\end{figure}

\subsubsection{2D Common origin.} \label{appx:2D_commom_origin}

We next consider a two-dimensional extension of the common-origin setting studied in the main text, where job destinations are distributed uniformly over $[0,1]^2$ and all jobs have a common pick-up location at $(0,0)$. Overall, the relative performance patterns are similar to those observed in the one-dimensional case (see \Cref{fig:2D}).
Importantly, the batching-based heuristics do not outperform $\dual$ in this setting.
The main difference from the 1D setting is that the hindsight-dual benchmark ($\dual$) 
now achieves slightly better performance than $\PB$ across the range of densities considered.
Since $\dual$ relies on full hindsight information and is not realistically implementable, $\PB$ remains the strongest practical heuristic among the evaluated policies.

\begin{figure}
  \centering
  \subcaptionbox{Average regret (lower is better).%
    \label{fig:2D_regret}}[0.49 \textwidth]{\includegraphics[width = \figWidth \textwidth]{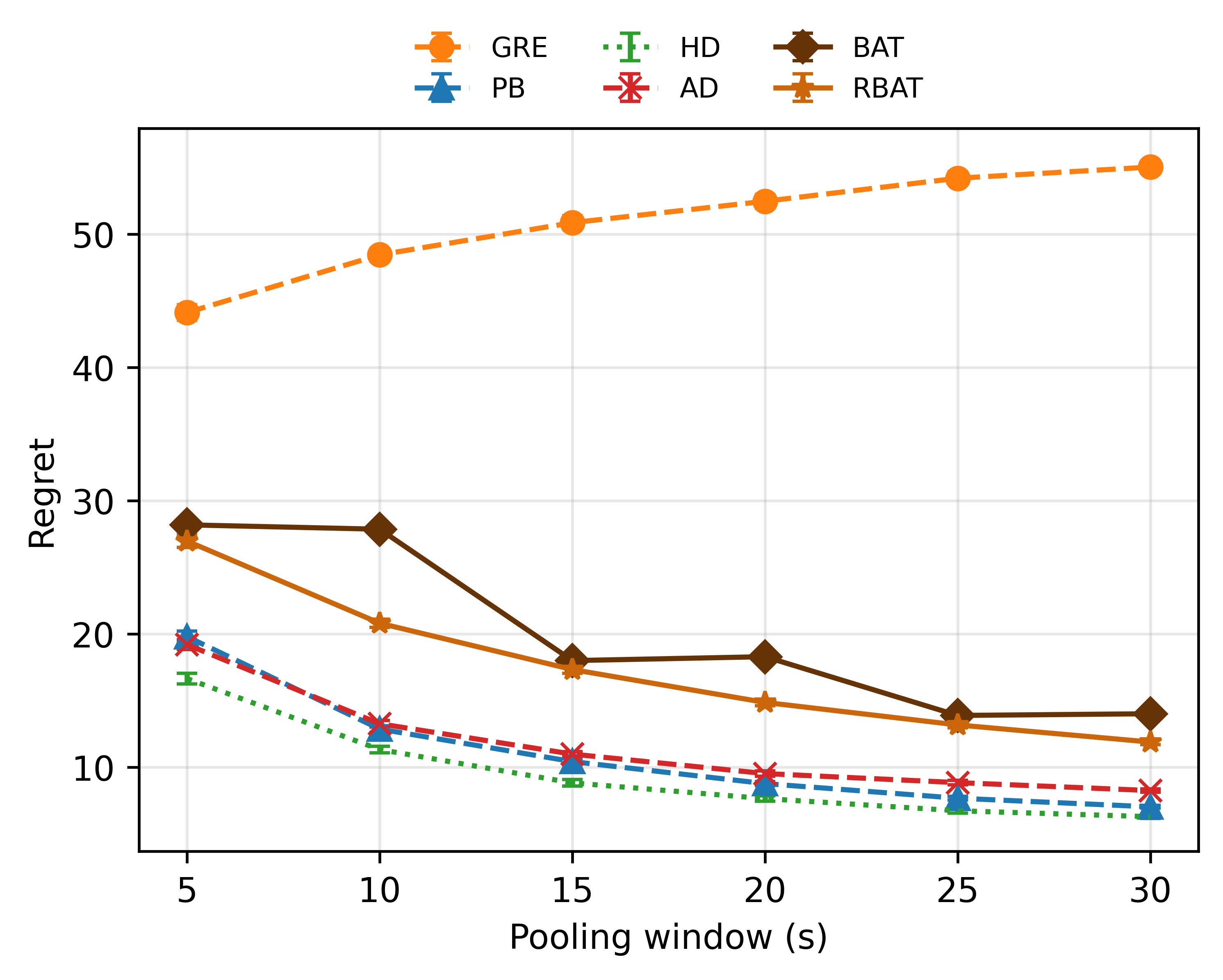}}
  \hfill
  \subcaptionbox{Average ratio (higher is better).%
    \label{fig:2D_ratio}}[0.49 \textwidth]{\includegraphics[width = \figWidth \textwidth]{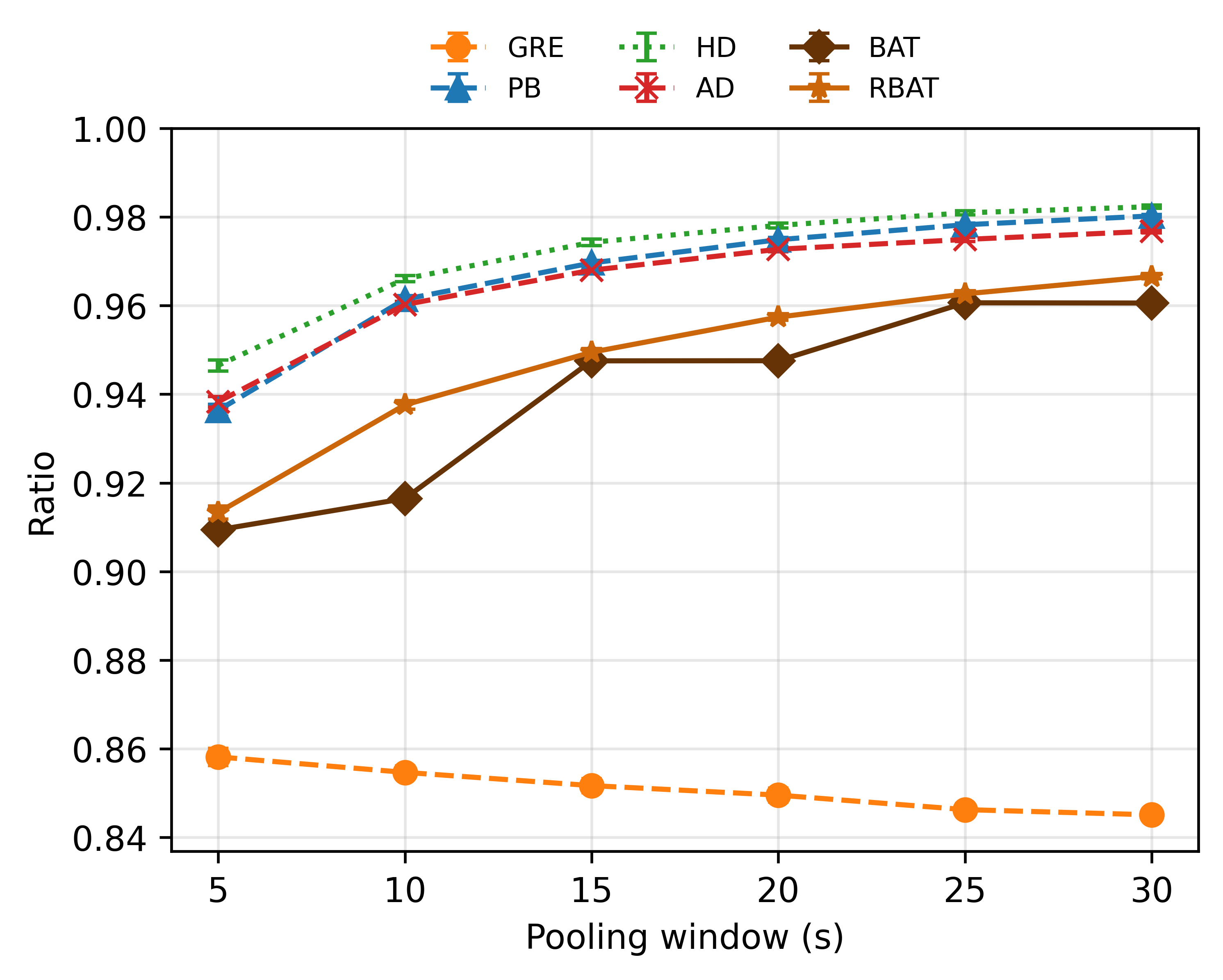}}
  \caption{Average regret and reward ratio for 2D space with common origin and uniformly distributed destinations.}
  \label{fig:2D}
\end{figure}

\subsubsection{2D Heterogeneous origin.}
\label{appx:2D_heterogeneous_origin}

We now consider two-dimensional locations with heterogeneous origins. 
Specifically, for each order, both the origin and the destination are drawn from a uniform distribution in the unit square.
As shown in \Cref{fig:2Dhet}, the qualitative behavior in this environment differs from
the common-origin case. 
In particular, $\rbatching$ outperforms all other
policies across the range of densities considered, mirroring the behavior observed in the Meituan data.
This improvement reflects the increased value of explicitly considering the ``current state'' (i.e. the set of all available jobs) when origins are heterogeneous, since the spatial ``proximity'' between jobs becomes more complex, and the effective density (of jobs that can benefit from being pooled together) decreases.

Nonetheless, $\PB$ continues to outperform all realistic greedy baselines and remains competitive with batching-based heuristics at substantially lower computational cost. 
These results highlight the trade-off between performance and computational complexity in heterogeneous spatial environments, and showcase the role of batching-based methods when richer spatial structure is present.

\begin{figure}
  \centering
  \subcaptionbox{Average regret (lower is better).%
    \label{fig:2Dhet_regret}}[0.49 \textwidth]{\includegraphics[width = \figWidth \textwidth]{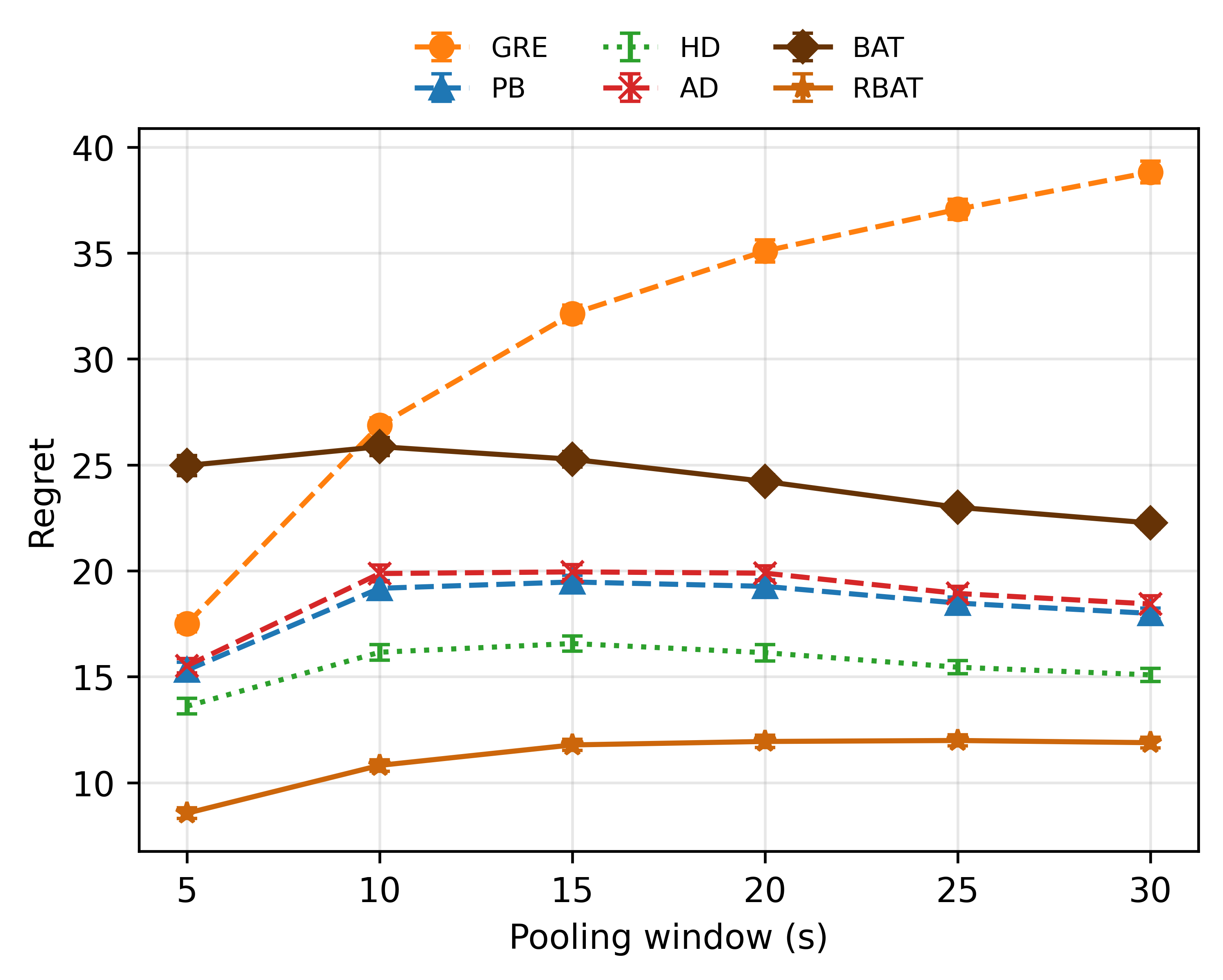}}
  \hfill
  \subcaptionbox{Average ratio (higher is better).%
    \label{fig:2Dhet_ratio}}[0.49 \textwidth]{\includegraphics[width = \figWidth \textwidth]{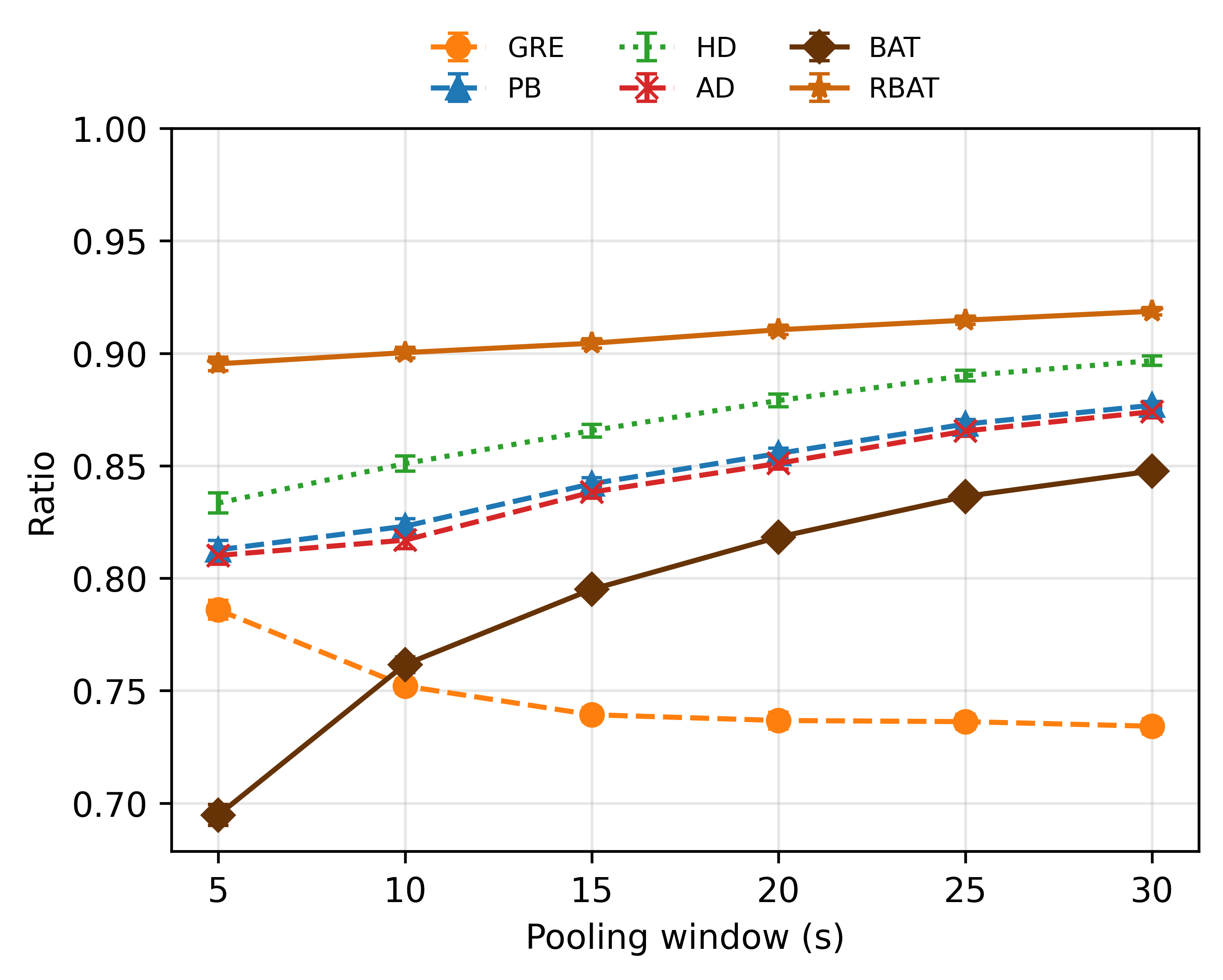}}
  \caption{Average regret and reward ratio for 2D space with uniformly distributed origins and destinations.}
  \label{fig:2Dhet}
\end{figure}




\subsection{Different Reward Topology} \label{sec:sec:sim_reward3}

In this appendix, we present numerical results under the two alternative reward topologies defined in \eqref{eq:defn_reward_B} and \eqref{eq:defn_reward_C}.
This illustrates the importance of reward topology in determining the performance of algorithms.
As in \Cref{sec:sim_unif_1D}, we generate instances of size $\Njob=1000$, where each type is drawn IID from a uniform distribution on $[0,1]$.
Overall, we observe that $\PB$ performs on par with the best realistic benchmarks (i.e., outside $\dual$ and $\OPT$). This suggests that $\PB$ could be a leading heuristic even beyond the reward function $r(\type,\type')=\min\{\type,\type'\}$ from delivery pooling, although here we are still restricting to continuous distributions over (one-dimensional) metric spaces.

\subsubsection{Reward function $\reward(\type,\type')=1-|\type-\type'|$.} 

Recall that in this case the potential of a job is $1/2$, independent of the job type, and thus both $\PB$ and $\gre$ are the same algorithm (see \Cref{sec:reward2}).
\Cref{fig:R2} shows the superior performance of greedy-like algorithms compared to batching-based heuristics.
Moreover, as in \Cref{sec:sim_unif_1D}, forecast-aware heuristics are no better than $\PB$. 

\begin{figure}[H]
  \centering
  \subcaptionbox{Average regret (lower is better).%
    \label{fig:R2_regret}}[0.49 \textwidth]{\includegraphics[width = \figWidth \textwidth]{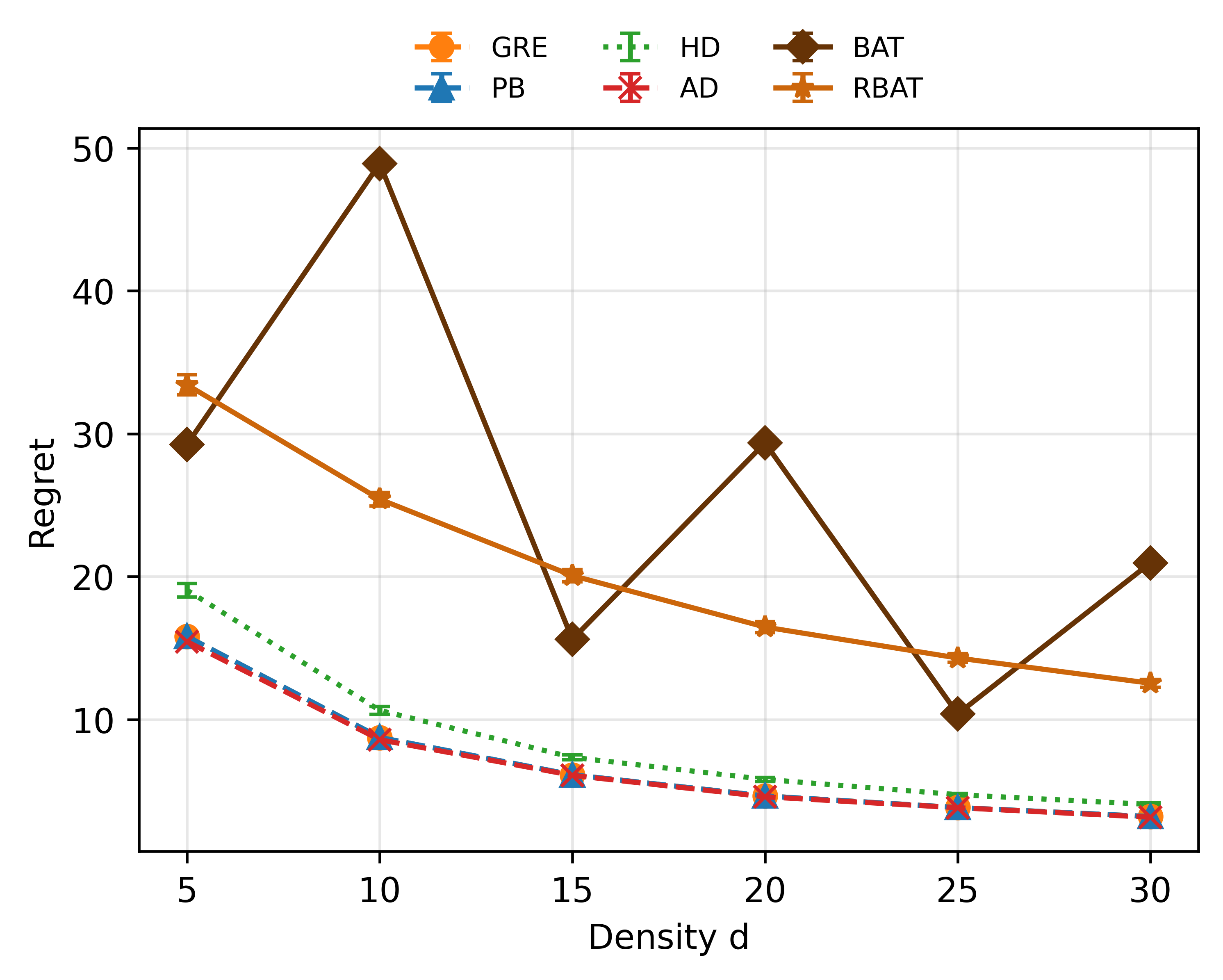}}
  \hfill
  \subcaptionbox{Average ratio (higher is better).%
    \label{fig:R2_ratio}}[0.49 \textwidth]{\includegraphics[width = \figWidth \textwidth]{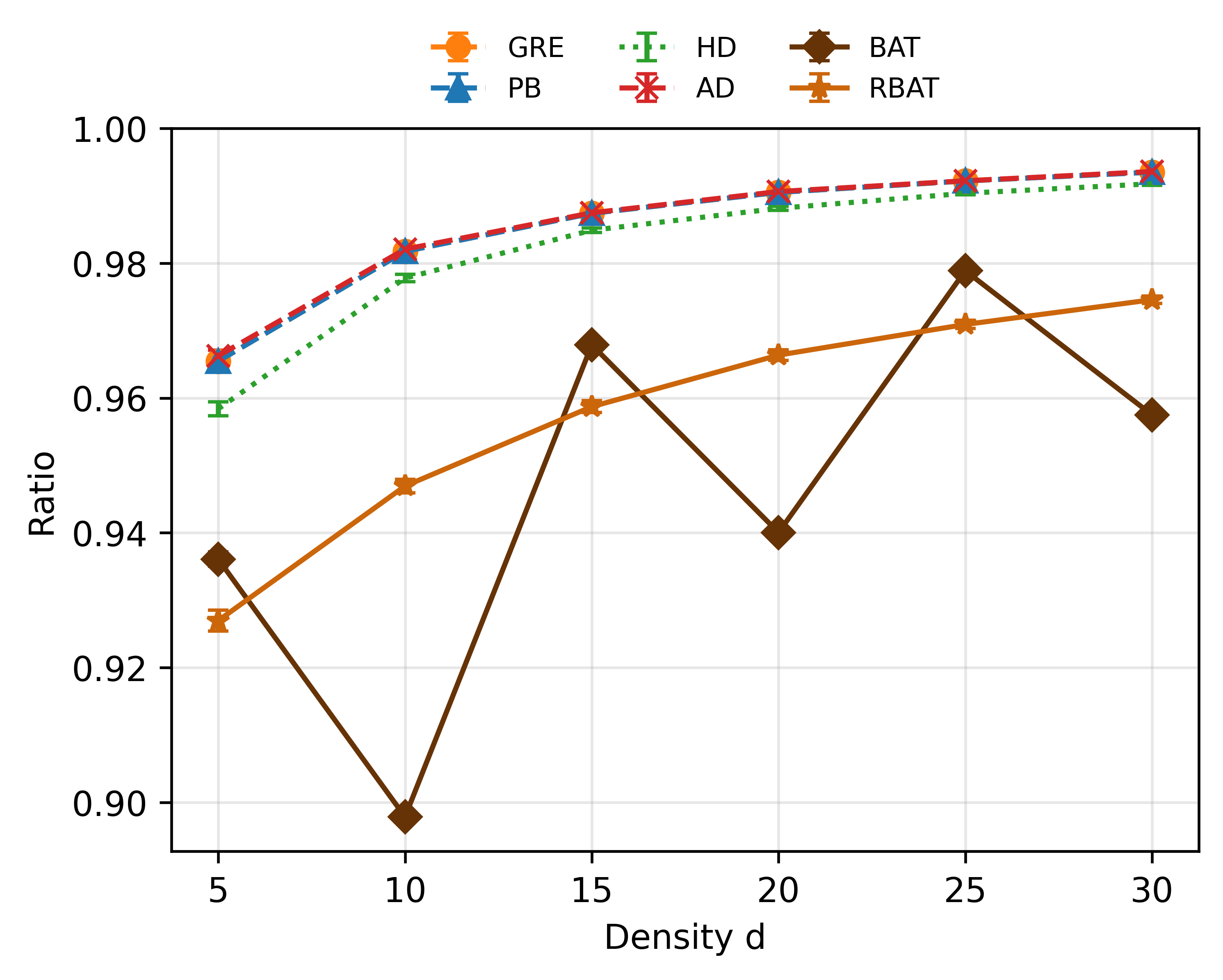}}
  \caption{Average regret and reward ratio in random 1D instances under $\reward(\type,\type')=1-|\type-\type'|$.}
  \label{fig:R2}
\end{figure}

\subsubsection{Reward function $\reward(\type,\type')=|\type-\type'|$.}

Consider $\typespace=[0,1]$ and $\reward(\type,\type') = |\type - \type'| $ (analyzed in \Cref{sec: reward 3}).
\Cref{fig:R3} highlights that greedy-like algorithms perform better than batching-based heuristics, with $\PB$ outperforming every practical matching algorithm (i.e., aside from $\dual$ and $\OPT$ that use hindsight information).

\begin{figure}[H]
  \centering
  \subcaptionbox{Average regret (lower is better).%
    \label{fig:R3_regret}}[0.49 \textwidth]{\includegraphics[width = \figWidth \textwidth]{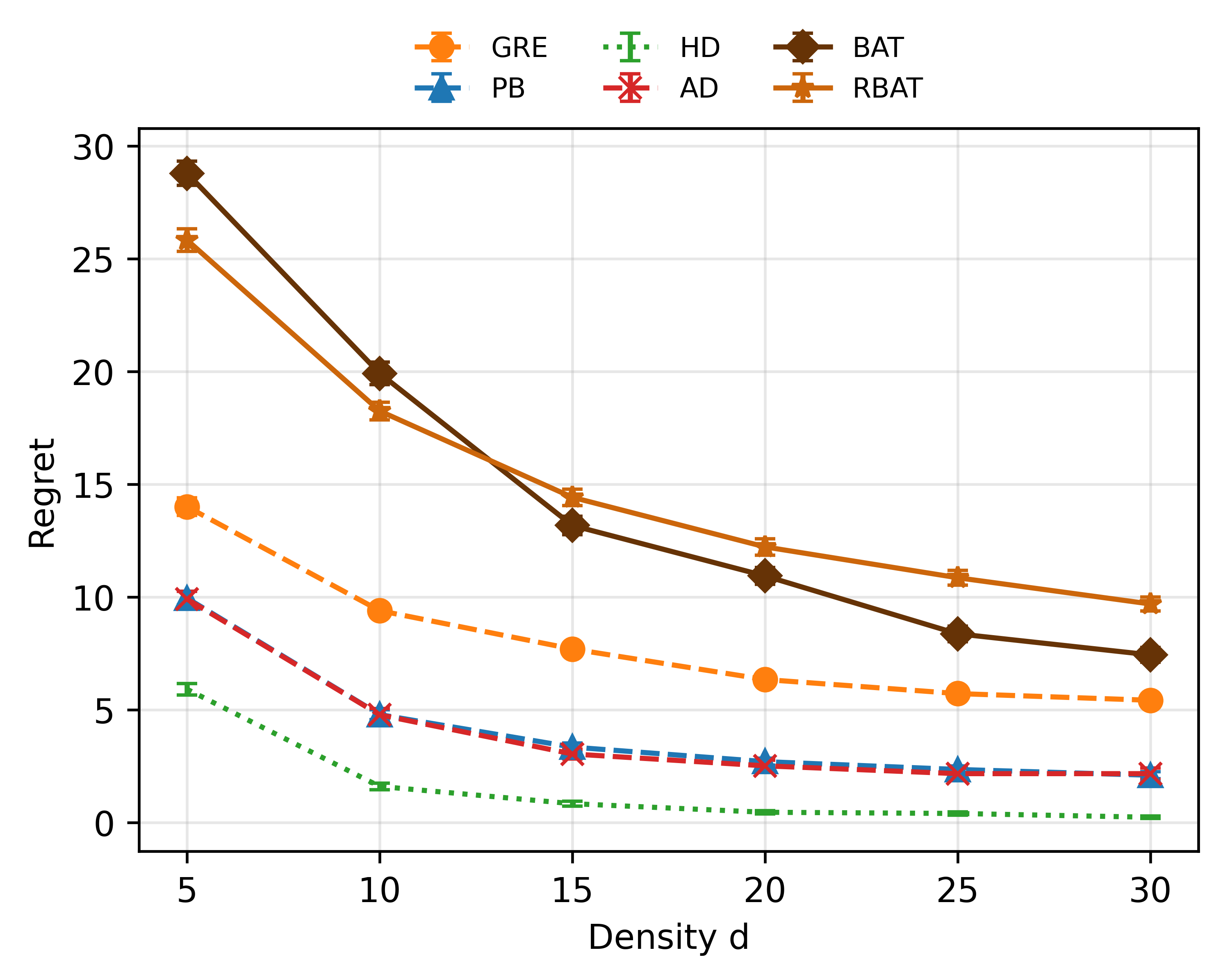}}
  \hfill
  \subcaptionbox{Average ratio (higher is better).%
    \label{fig:R3_ratio}}[0.49 \textwidth]{\includegraphics[width = \figWidth \textwidth]{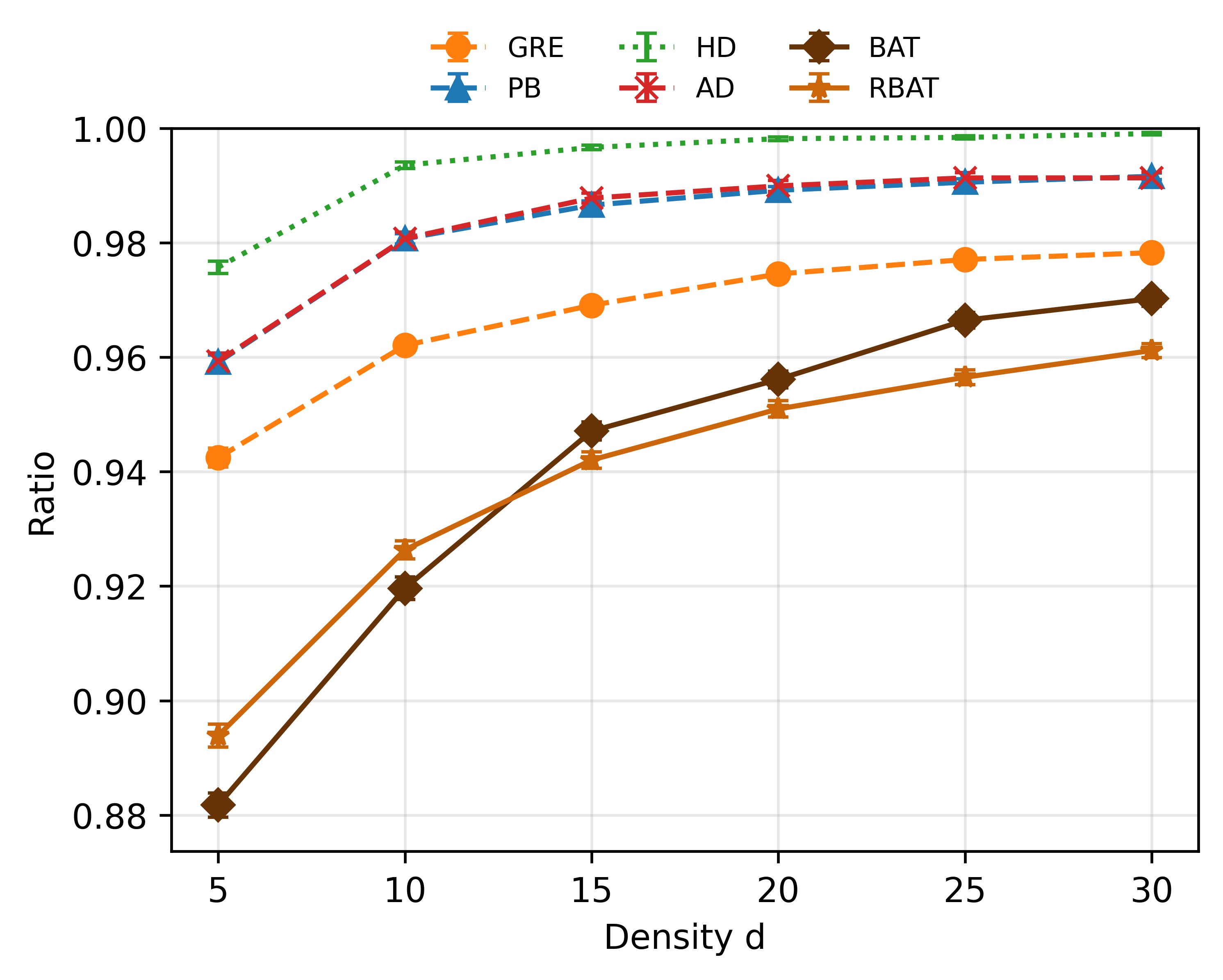}}
  \caption{Average regret and reward ratio in random 1D instances under $\reward(\type,\type')=|\type-\type'|$.} 
  \label{fig:R3}
\end{figure}

\end{APPENDIX}

\end{document}